\documentclass[11pt,leqno]{article}
\usepackage{latexsym}
\usepackage{epsfig}
\usepackage{color}
\usepackage{subfigure}
\usepackage{amssymb,amsmath}

\numberwithin{equation}{section}

\newtheorem{theorem}{Theorem}[section]
\newtheorem{lemma}[theorem]{Lemma}
\newtheorem{proposition}[theorem]{Proposition}

\newtheorem{definition}{Definition}[section]
\newtheorem{corollary}[theorem]{Corollary}
\newtheorem{example}{Example}[section]

\def\remark #1{\noindent{\bf Remark:} #1\\}
\def\example #1{\noindent{\bf Example:} #1\\}

\long\def\claim #1 #2{\bigskip\noindent{\bf Claim {#1}} {\it #2}\bigskip}
\def\xclaim #1 #2{\noindent{\bf Claim {#1}} {\it #2}\bigskip}
\newenvironment{proof}{\noindent{\bf Proof:}}{\hfill $\Box $\\}

\long\def\claim #1 {\bigskip\noindent{\bf Claim {#1}} \bigskip}

\newcommand{\ecproof}{\hfill $\diamondsuit$\\}

\renewcommand{\thetheorem}{\arabic{section}.\arabic{theorem}}
\def\sqr#1#2{{\vcenter{\vbox{\hrule height .#2pt
              \hbox{\vrule width .#2pt height#1pt \kern#1pt
              \vrule width .#2pt} \hrule height .#2pt}}}}

\def\ncas #1 {\noindent {\bf Case #1.}\ }

\def\bipart #1 #2{\bigskip \noindent {\bf #1} {\it #2}}
\def\xbipart #1 #2{\noindent {\bf #1} {\it #2}}
\def\iipart #1 #2{\bigskip \noindent {\it #1} {\it #2}}
\def\xiipart #1 #2{\noindent {\it #1} {\it #2}}
\def\brpart #1 #2{\bigskip \noindent {\bf #1} {#2}}
\def\xbrpart #1 #2{\noindent {\bf #1} {#2}}
\def\irpart #1 #2{\bigskip \noindent {\it #1} {#2}}
\def\xirpart #1 #2{\noindent {\it #1} {#2}}

\def\o {\overline}

\def\mod{{\rm{\ mod\;}}}

\def\case #1{\bigskip\noindent{{\bf Case} {\em #1}:}}
\def\subcase #1{\bigskip\noindent{{\bf Subcase} {\em #1}:}}
\def\numcase #1 #2{\bigskip\noindent{{\bf Case #1} {\em #2}:}}

\def\claim #1 {\bigskip\noindent{{\bf Claim:} #1 \bigskip}} 
\def\nclaim #1 {\noindent{\bf Claim #1: }}

\def\obs #1 {\bigskip\noindent{\bf Observation #1: }} 

\def\mathy #1{\ifmmode {#1}\else{$#1$}\fi}
\def\o{\overline}

\newlength{\figwidth}
\newlength{\figindent}
\newlength{\vmargin}
\newlength{\Efigwidth}

\begin{document}

\title{A Weight-scaling Algorithm for $f$-factors of Multigraphs} 

\author{%
Harold N.~Gabow%
\thanks{Department of Computer Science, University of Colorado at Boulder,
Boulder, Colorado 80309-0430, USA. 
E-mail: {\tt hal@cs.colorado.edu} 
}
}

\maketitle

\def\today{\ifcase\month\or
January\or February\or March\or April\or May\or June\or
July\or August\or September\or October\or November\or December\fi
\ \number\day, \number\year}
\def\date#1.#2.{\ifcase#1\or
January\or February\or March\or April\or May\or June\or
July\or August\or September\or October\or November\or December\fi
\ #2, \number\year}
\def\ydate#1.#2.#3.{\ifcase#1\or
January\or February\or March\or April\or May\or June\or
July\or August\or September\or October\or November\or December\fi
\ #2, 199#3}
\def\nydate#1.#2.{\ifcase#1\or
January\or February\or March\or April\or May\or June\or
July\or August\or September\or October\or November\or December\fi
\ #2}
\def\doublespace{\multiply\baselineskip by3\divide\baselineskip by2%
                 \def\doublespace{}}
\def\bigdoublespace{\multiply\baselineskip by2%
                 \def\bigdoublespace{}}
\def\imp{\ifmmode {\ \Longrightarrow \ }\else{$\ \Longrightarrow \ $}\fi}
\def\rimp{\ifmmode {\ \Longleftarrow \ }\else{$\ \Longleftarrow \ $}\fi}
\def\ximp{\ifmmode {\Longrightarrow\ }\else{$\Longrightarrow\ $}\fi}
\def\xrimp{\ifmmode {\Longleftarrow\ }\else{$\Longleftarrow\ $}\fi}
\def\iff{\ifmmode {\ \Longleftrightarrow \ }\else{$\ \Longleftrightarrow \ $}\fi}
\def\xiff{\ifmmode {\Longleftrightarrow\ }\else{$\Longleftrightarrow\ $}\fi}
\def\tru{\ {\bf true}\ }
\def\fal{\ {\bf false}\ }
\def\wrt{\ {\it wrt}\ }
\def\endskip{\medskip}
\def\qed{$\Box$}
\def\qedn{\ \vrule width4pt depth-1pt height7pt }
\def\rqed{\hfill\hbox to 24 pt{\vrule width4pt depth-1pt
height7pt\hfil}\bigskip}
\def\rqedn{\hfill\hbox to 24 pt{\vrule width4pt depth-1pt height7pt\hfil}}
\def\log{\ifmmode \,{ \rm log}\,\else{\it log }\fi}
\def\con {\subseteq}
\def\pcon{\subset}
\def\firstnumstp#1 {\bigskip \noindent{\it Step} #1.\newquad}
\def\numstp#1 {\endskip\noindent{\it Step} #1.\newquad}
\def\newquad{\hskip1ex}
\def\stp#1.{\endskip
\noindent{\it #1 Step.}\newquad}
\def\firststp#1.{\bigskip
\penalty-1000
\noindent{\it #1 Step.}\newquad}
\def\cas#1 {\smallskip\noindent{\bf Case} #1.\ } 
%
%
%
\long\def\sec#1{\bigskip
\penalty-2000%
\noindent{\twelvebf #1}\par\ignorespaces\noindent\ignorespaces}
\def\aorbsec#1{\noindent{\twelvebf #1}}
\def\nsec#1{\penalty-2000%
\noindent{\bf #1\hfill\break}
\hbox to \parindent{\hfill}\ignorespaces}
\long\def\res #1. #2{\bigskip
\penalty-1000
\noindent {\bf #1.}\newquad%
#2 \bigskip}
\long\def\nres #1. #2{\bigskip
\noindent {\bf #1.}\newquad%
#2}
\def\pf{\noindent {\bf Proof.}\newquad}
\def\cont{\ifmmode\star\else$\star$\fi}
\def\+{\tabalign} 
\def\nskp{\def\bigskip{}}
\def\i{($i$) } \def\xi{($i$)}
\def\ii{($ii$) } \def\xii{($ii$)}
\def\iii{($iii$) } \def\xiii{($iii$)}
\def\iv{($iv$) } \def\xiv{($iv$)}
\def\pa{({\it a}) } \def\xpa{({\it a})} 
\def\pb{({\it b}) } \def\xpb{({\it b})}
\def\pc{({\it c}) } \def\xpc{({\it c})}
\def\hi{\hskip20pt\i} \def\hii{\hskip20pt\ii} \def\hiii{\hskip20pt\iii}
\def\ha{\hskip20pt\pa} \def\hb{\hskip20pt\pb} \def\hc{\hskip20pt\pc}
\def\tran{{\buildrel*\over\to}}
\def\n{\rlap{$\>/$}}
\def\({{\rm(}} \def\){{\rm)}}
\def\c#1{\lceil {#1} \rceil}
\def\f#1{\lfloor {#1} \rfloor}
\long\def\boxit#1{\vtop{\hrule
\hbox{\vrule\quad\vtop{\vskip5pt\hbox{#1}\vskip5pt}\quad\vrule}
\hrule}} 
\def\iboxit#1{\vtop{\hrule
\hbox{\vrule\quad\vtop{\vskip5pt\hbox{{\it #1}}\vskip5pt}\quad\vrule}
\hrule}} 
\def\x{\iffalse}
\def\b{\bigskip}
\def\set #1#2{\{ #1:#2 \}}
\def\pset #1#2{( #1:#2 )}
\def\h{\hskip20pt}
\def\hi{\advance\parindent by 20pt}

\def\o{\overline} 
\def\u{\underline}
\def\opn{\hangindent=40pt\hangafter=1}
\def\h{{\hskip 20pt}}
\def\v{\vfill}
\def\hi{\advance \parindent by 20pt}
\def\d{\cdot}
\def \il #1{\log^{(#1)} }
\def\al.{{\it add\_leaf}}
\def\alm{{\it add\_leaf}$\,$}
\def\O{o\hbox{-}smallest}
\def\os.{\ifmmode{ \o{\cal S} }\else{$\o {\cal S}$}\fi}
\def\oP.{\ifmmode{ \o{\cal P} }\else{$\o {\cal P}$}\fi}
\def\ot.{\mathy{ \o{\cal T} }}
\def\oG{\o G}
\def\oB{\o B}
\def\oE.{\mathy{\overline E}}
\def\p(#1,#2){\ifmmode p(#1,#2) \else{$p(#1,#2)$}\fi}
\def\op(#1,#2){\ifmmode \o{p}(#1,#2) \else{$\o{p}(#1,#2)$}\fi}
\def\lb{\ifmmode \,{ \rm log}_\beta \else{\it log XX }\fi}
\def\wh{\widehat}
\def\wx.{\ifmmode \wh x \else$\wh x$\fi}
\def\wy.{\ifmmode \wh y \else$\wh y$\fi}
\def\wz.{\ifmmode \wh z \else$\wh z$\fi}
\def\wv.{\ifmmode \wh v \else$\wh v$\fi}
\def\Px.{\ifmmode \wh x \else$\wh x$\fi}
\def\Py.{\ifmmode \wh y \else$\wh y$\fi}
\def\Pz.{\ifmmode \wh z \else$\wh z$\fi}
\def\Pv.{\ifmmode \wh v \else$\wh v$\fi}
\def\Pr.{\ifmmode \wh r \else$\wh r$\fi}
\def\Pr.{\ifmmode \wh r \else$\wh r$\fi}
\def\A.{\mathy{{\cal A}}}
\def\B.{\mathy{{\cal B}}}
\def\C.{\mathy{{\cal C}}}
\def\D.{\mathy{{\cal D}}}
\def\E.{\ifmmode {{\cal E}}\else{{$\cal E$}}\fi}
\def\F.{\mathy{\cal F}}
\def\H#1{\widehat{#1}} 

\def\L.{\mathy{\cal L}}
\def\M.{\mathy{\cal M}}
\def\P.{\mathy{\cal P}}
\def\Q.{\mathy{\cal Q}}
\def\R.{\mathy{\cal R}}
\def\S.{\ifmmode {{\cal S}}\else{{$\cal S$}}\fi}
\def\T.{\mathy{\cal T}}
\def\U.{\mathy{\cal U}}
\def\W.{\mathy{\cal W}}
\def\mathy #1{\ifmmode {#1}\else{$#1$}\fi}
\def\goin{\hspace{17pt}}
\def \algname #1.{{\sc #1}}
\def \algcall #1(#2){\algname #1.\mathy{(#2)}}
\def\DP{DismantlePath}
\def\SS{ShellSearch}

\def \shell (#1,#2){\mathy {(#1\backslash #2)}}
\def\mydef #1{\bigskip{\narrower#1}\bigskip}
\def\mmpr(#1){\mathy{\widehat{\,#1\,}}}
\def\mpr(#1){\mathy{\overline{#1}}}

\begin{abstract} 
The challenge for graph matching algorithms, and their generalizations to 
$f$-factors, is to extend known time bounds for bipartite graphs to 
general graphs. We discuss combinatorial algorithms for finding a maximum weight 
$f$-factor on an
arbitrary  multigraph, for given integral weights of magnitude at most $W$. 

For simple bipartite graphs 
the best-known time bound is
$O(n^{2/3}\, m\, \log nW)$ (\cite{GT89};
$n$ and $m$ are respectively  the number of vertices 
and edges).
A recent algorithm of Duan and He et al. \cite{DHZ} for $f$-factors
of simple graphs comes within logarithmic factors of this bound,
$\widetilde{O} (n^{2/3}\, m\, \log W)$. 
The best-known bound for bipartite 
multigraphs 
is $O(\sqrt {\Phi}\, m\, \log \Phi W)$ ($\Phi\le m$ is the 
size 
of the $f$-factor, $\Phi=\sum_{v\in V}f(v)/2$). This bound is 
more general than the restriction to simple graphs,
and is even superior on "small" simple graphs, i.e.,
$\Phi=o(n^{4/3})$. We 
present an algorithm that comes 
within a $\sqrt {\log \Phi}$ factor of this bound, i.e., $O(\sqrt {\Phi \log \Phi}\,m 
\,\log  \Phi W)$.

The algorithm is a direct generalization of the algorithm of Gabow and Tarjan 
\cite{GT} for the special case of ordinary matching ($f\equiv 1$). We present our 
algorithm first for ordinary matching, as the analysis 
is a simplified version of
\cite{GT}. 
Furthermore the algorithm and analysis both get incorporated
without modification into 
the multigraph algorithm.

To extend  these ideas to $f$-factors, the first step is
"expanding" edges (i.e., 
replacing an edge by a length 3 alternating path). \cite{DHZ} uses a one-time expansion of 
the entire graph. Our algorithm keeps the graph small by only expanding selected 
edges, 
and 
"compressing" them back to their original source when no longer needed.
Several other ideas are needed, including a relaxation of the notion of "blossom"
to e-blossom ("expanded blossom"). 
\end{abstract}

\ifcase 0
\section {Introduction} 
\label{IntroSec}
A guiding principle for developing matching algorithms is that
any asymptotic time bound achieved for bipartite graphs can be 
achieved for general graphs -- in spite of the complexity introduced by 
blossoms. 
(Successful examples include  the Hopcroft-Karp cardinality matching algorithm
\cite{HK} extended to general graphs by Micali and Vazirani\cite{MV};
the Hungarian algorithm for weighted matching
\cite{K} as implemented by Fredman and Tarjan \cite{FT},
extended to general graphs by Edmonds \cite{E} with
implementations of different components
in
\cite{GMG, Th, G18}.) 
We discuss the problem of finding a maximum weight $f$-factor on an
arbitrary  multigraph, for given integral weights of magnitude at most $W$.
The best approach for such edge weights is given by scaling algorithms.
To state their time bounds $n$ and $m$ denote the number of vertices and edges,
respectively, and 
$\Phi$ is the size
of the desired 
$f$-factor, i.e., $\Phi=\sum_{v\in V}f(v)/2$. (Clearly $n\le \Phi\le m $.)

The best-known time bounds for the problem on bipartite graphs,
given by Gabow and Tarjan \cite{GT89},
are
\begin{equation*}
\begin{cases}
  O(n^{2/3}\; m \log nW)&G \text{ a simple graph}\\
  O(\sqrt {\Phi} \; m \log \Phi W)&G \text{ a multigraph.}
  \end{cases}
\end{equation*}
These bounds are within a logarithmic factor of
the best bounds for the unweighted problem (due to Even and Tarjan \cite{ET}),
thus achieving the goal of scaling algorithms.
Note the multigraph bound
is more general than the restriction to simple graphs,
and is even superior on "small" simple graphs, 
$\Phi=o(n^{4/3})$.
(Note also these bounds can be improved for appropriately sparse bipartite graphs:
Cohen and Madry et al. \cite{CMSV} give a linear-programming based algorithm that
runs in 
$\widetilde O (m^{10/7} \log W)$ time.
It applies to simple graphs as well as small $b$-matchings.)

A recent algorithm of Duan and He et al. \cite{DHZ} for
simple (general) graphs comes within logarithmic factors of the bipartite bound,
$\widetilde{O} (n^{2/3}\, m\, \log W)$.
We
present an algorithm that comes
within a $\sqrt {\log \Phi}$ factor of the multigraph
bound, i.e., $O(\sqrt {\Phi \log \Phi}\,m
\,\log  \Phi W)$.

Both our algorithm and \cite{DHZ} are based on algorithms for the special
case of weighted matching ($f\equiv 1$).
 \cite{DHZ} extends the matching algorithm of
 Duan and Pettie et al. \cite{DPS}.
This matching algorithm has time bound 
$O(\sqrt n m\log(nW))$, the same as the best-known bound for bipartite graphs
(Gabow and Tarjan \cite{GT89}).
The algorithm is based on several distinct ideas (``large blossoms'',
the strong polynomial algorithm of \cite{G18}, and the heavy path algorithm of
\cite{G}) and
\cite{DPS}
poses the open problem of a simpler optimum algorithm.

Our $f$-factor algorithm is a direct generalization of the matching
algorithm ($f\equiv 1$) of Gabow and Tarjan \cite{GT}.
This matching algorithm runs in time $O(\sqrt{ n \log n}\, m \,\log(nW))$.
(The bound in 
\cite{GT} has an extra
$\sqrt{\alpha(m,n)}$ factor, which is automatically removed by using
the data structure of Thorup \cite{Th}
for list splitting.) Note the extra factor $\sqrt {\log n}$ compared to
\cite {DPS}, although the simpler approach offers potential regarding the open problem of
\cite{DPS}.

The current paper first presents the matching algorithm of \cite{GT}
(the ''GT'' algorithm) and then the extension to $f$-factors.
Our detailed presentation of the GT algorithm incorporates an analysis that
is fundamentally equivalent to \cite{GT} but has the following advantage.
\cite{GT} presents the basic ideas at an abstract level
(especally the ``crossing function''  $\gamma$, p.834).
We give  an explicit description of the graph structures
that 
measure progress of the algorithm
(``objective reducers'', Section \ref{dFSec}).
Continuing, we give a concrete ``credit system'' 
to derive the algorithm's time bound. The system gives
an explicit description of the involved recursive timing argument.
Our algorithm for $f$-factors uses exactly
the same objective reducers and exactly the same credit system.
Presenting these ideas first
in the context of matching helps  isolate
the difficulties introduced by  $f$-factors, and also offers a supplement to
\cite{GT} for readers interested only in matching.

We extend the GT algorithm to find a maximum weight
$f$-factor of a  multigraph. Several difficulties must be overcome.
Unlike ordinary matching, blossoms for $f$-factors have pendant edges
called ``$I$-edges'' (for ``incident'' in
\cite{G18}, see Appendix \ref{fAppendix}; these are called $F$-edges in
\cite[Ch.32-33]{S}).
$I$-edges invalidate 
the main mechanism of the GT algorithm
(the ``unit translation'',
for removing ``inherited blossoms'' -- see Section \ref{AlgSec}).
We solve this problem by replacing each $I$-edge with its ``expansion'', i.e.,
a length 3 path through 2 artificial vertices (Fig.\ref{ExpandedEdgeFig}).
This transformation is often used in matching
(e.g., \cite[Ch.32-33]{S}, \cite{DHZ}).
Duan et.al.
\cite{DHZ} use it to overcome problems of $I$-edges. Specifically \cite{DHZ}
expands every original edge (the ``blowup graph'').
We cannot use this graph, since
it increases the size of an $f$-factor (and its blossoms)
up to $\Theta(m)$
rather than $\Theta(\Phi)$.

Our edge expansions create several complications:

\bigskip

{\narrower

  {\parindent =0pt

Basic properties of blossoms change. On the plus side,
    $f$-factor blossoms do not have fixed parity. But they become odd sets,
    like ordinary matching blossoms, when edges
    are expanded. On the minus side, blossoms
lose their cyclic structure when edges get expanded.
    We call these expanded blossoms
    ``e-blossoms'' (defined after Lemma \ref{ExpVertLemma})
    and we derive and apply their fundamental properties.

\smallskip

$f$-factors have structures that do not occur in ordinary matching
and complicate the GT algorithm (edges ``undervalued'' by the dual variables,
and blossoms that are ``heavy'' rather than ``light'').
We handle these possibilities using  an efficient algorithm
(paragraph ``Tightening base edges'', Lemma \ref{EtaEligibleLemma}).

     \smallskip

     Expanded edges must be ``compressed'', i.e., replaced by their
     original source edge, to keep the graph small.
     Compression may turn an e-blossom into an
     ``ill-formed'' traditional blossom. We give a simple procedure
     to 
eliminate ill-formed blossoms (paragraph ``Compression: ill-formed blossoms''
after \eqref{yzCompressedEqn}).

  }
}

\bigskip

The paper is organized as follows.  
Section \ref{AlgSec}
presents the GT matching algorithm. Various low-level details are
omitted but nothing of significance for the analysis. 
Section
\ref{dFSec} derives the basic inequality
of scaling, \eqref{GoalCrossedEqn}.
It explicitly identifies the objective reducers.
Section \ref{CreditSec} uses \eqref{GoalCrossedEqn}
in a credit system. It proves the Gabow-Tarjan
time bound for ordinary matching,
Theorem \ref{MainTheorem}.
Section \ref{fFactorSec} presents our algorithm for $f$-factors.
Our main result is stated as
Theorem \ref{fMainTheorem},
which gives an outline of the proof. The proof is presented over
several sections as follows:
Section \ref{fdFOverallSec} derives the basic properties of the $f$-factor algorithm.
These properties support a derivation very similar
to Section \ref{dFSec}, which is given in Section \ref{fdFSec}.
The final result of that section is
inequality \eqref{fGoalCrossedEqn}, the analog of \eqref{GoalCrossedEqn}.
The credit system of Section \ref{CreditSec} is valid with \eqref{GoalCrossedEqn}
replaced by \eqref{fGoalCrossedEqn}, so it completes the proof for $f$-factors.

Appendix \ref{EdAppendix} gives a complete statement of Edmonds
matching algorithm, supporting Section \ref{AlgSec}.
Appendix \ref{Phase2Appendix} gives the data structures for efficient
implementation of both the GT matching algorithm and our $f$-factor algorithm.
Appendix \ref{fAppendix} reviews the $f$-factor algorithm of
\cite{G18}.

\paragraph*{Terminology}
We use a common summing notation: If $f$ is a function on elements
and $S$ is a set of elements then $f(S)$ denotes $\sum_{s\in S}f(s)$.
$\log n$ denotes logarithm to the base two.

We often omit set braces from singleton sets, denoting $\{v\}$ as
$v$. So $S-v$ denotes $S-\{v\}$. We abbreviate expressions $\{v\}\cup S$ to $v+S$.

In a graph $G=(V,E)$
for $S\con V$ and $M\con E$,
$\delta_M(S)$ ($\gamma_M(S)$) denotes the set of edges of $M$
 with exactly
one (respectively two) endpoints in $S$.
It is sometimes more convenent  to make  $M$ an argument rather 
than a subscript, e.g.,
$\delta(S,M_\omega)$.
We omit $M$ entirely (writing
$\delta(S)$ or $\gamma(S)$)
when $M=E$. A loop at $v\in S$ contributes
belongs to $\gamma(S)-\delta(S)$.

\section{The matching algorithm}
\label{AlgSec}
\def \algname #1.{{\sc #1}}
\def \algcall #1(#2){\algname #1.\mathy{(#2)}}
\def \shell (#1,#2){\mathy {(#1\backslash #2)}}
\def\D.{{\sc Dismantler}}

\setlength{\figwidth}{\textwidth}
\addtolength{\figwidth}{-1in}
\setlength{\figindent}{.5in}
\setlength{\vmargin}{.1in}

We presume familiarity with 
Edmonds' weighted matching algorithm and blossoms \cite{E} (see 
also \cite{CCPS,L,LP,S}) as well as
previous scaling versions of the algorithm \cite{GT} or 
\cite{DPS}. Here we give an overview that contains the details needed for the 
derivation of our result.

The various versions of weighted matching are essentially equivalent 
from an algorithmic viewpoint. Our discussion focuses on maximum weight 
perfect matching, which we shorten to {\em maximum weight matching}.

\subsection*{Edmonds' Algorithm and its Variants}
We  start with Edmonds' formulation of maximum weight 
matching as 
a linear program. The primal LP defines all possible perfect matchings.%
\footnote{Each variant of weighted matching, be it 
perfect
matching, maximum 
cardinality matching, etc., has a variant of the
primal LP.}
The dual LP has variables $y(v), v\in V$ and $z(B)$, $B$ a blossom. \B. 
denotes the family of nontrivial blossoms. The dual LP requires $z$ to 
be nonnegative, and every edge $e$ must satisfy
\begin{equation}
\label{yzInequalityEqn}
\H{yz}(e) \ge w(e),
\end{equation}
where $w$ is the given weight function and we define
\[\H{yz}(e) = y(e)  + z\set {B} {e \con B\in \B.}.\]
Here our summing convention
implies that for $e=vw$, $y(e)$ 
denotes $y(v)+y(w)$, and $z\set {B} {e \con B\in \B.}$
denotes $\sum_{e \con B\in \B.} z(B)$.
The unique complementary slackness condition is that equality holds in
\eqref{yzInequalityEqn} for every edge that is matched.

For convenience our algorithm  treats the vertex set
$V$ as a blossom, but it differs from all other
blossoms. $|V|$ is even (we assume $G$ has a perfect matching)
but every other blossom is odd. Our algorithm will assign
nonpositive values to $z(V)$, but  every other $z$ value must be
nonnegative. There is no harm in values $z(V)<0$, since they
are easily eliminated:
Add $z(V)/2$ to every $y$-value and change $z(V)$ to 0.
This preserves
the value $\H{yz}(e)$ for every edge $e$, so the modified duals are
valid.

Edmonds' algorithm finds a maximum weight matching by repeatedly 
searching the graph for a maximum weight augmenting path $P$. Wlog the 
input graph has a perfect matching, so such a $P$ always exists. The 
algorithm uses $P$ to augment the current matching. Eventually every 
vertex gets matched and the algorithm halts.

The algorithm is a primal-dual scheme. It maintains LP duals $y,z$
that are always feasible, i.e., \eqref{yzInequalityEqn} holds for every
$e$, with equality for every $e$ that is matched or in a blossom
subgraph. We say $e$ is {\em dominated} when \eqref{yzInequalityEqn}
holds, and {\em tight} when \eqref{yzInequalityEqn} holds with equality.

\begin{figure}[t]
\centering
\input{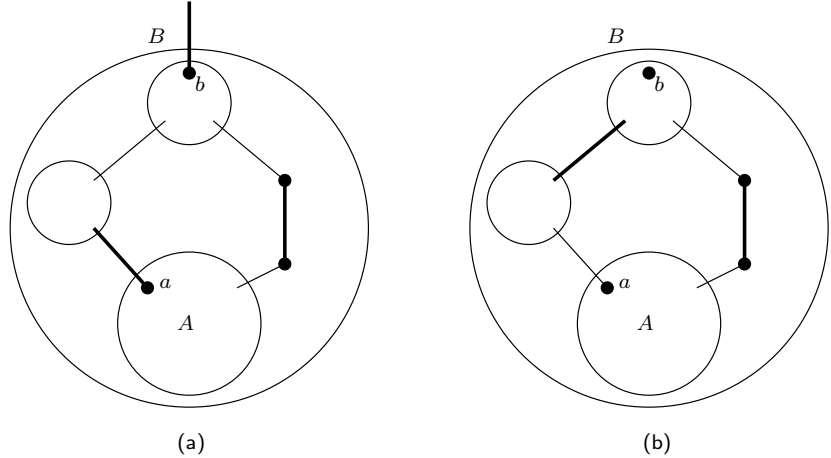}
\caption{Blossom $B$ and its maximal subblossoms.
Heavy edges are matched. $B$ has base vertex $b$, $A$ has base
$a$.
(b) $B$ is rematched along the alternating path from $a$ to $b$.}
 \label{BlossomFig}
\end{figure}

The algorithm assigns nonzero $z$ values to subsets called {\em 
blossoms}.
The blossoms form a laminar family with a 
corresponding blossom tree $T$. 
The root is the vertex set $V$.
Any interior node besides $V$ is a blossom, whose children are 
its 
constituent subblossoms. The leaves of $T$ are the vertices of $G$.
Each blossom has a corresponding subgraph of tight edges.

Edmonds' algorithm maintains
a {\em structured  matching} which consists of

\smallskip

$\bullet$ a matching, 

$\bullet$ LP
duals $y,z$, that dominate every edge, and are tight
on every edge that is matched or in a blossom subgraph,

$\bullet$ a blossom tree.

\smallskip
\noindent
When the matching is perfect
the structure is an {\em optimum structured matching}.
The corresponding matching 
has maximum weight, and 
$y,z$ are optimum LP duals.
So our ultimate goal is to find an optimum 
structured matching.

Edmonds' algorithm finds augmenting paths by building
a {\em search structure}. This structure is a forest in a graph
formed by contracting various blossoms. The edges of this forest
are all tight. Similarly the edges forming the contracted blossoms
are tight. These conditions are preserved as the algorithm modifies
dual variables and enlarges the search structure.
We next give a high-level sketch of Edmonds' original algorithm.
For completeness Appendix \ref{EdAppendix} gives detailed pseudocode for
the algorithm.

Edmonds' algorithm has five types of steps: {\em grow, blossom,
 expand, augment} and {\em dual adjustment}.  The search structure is
constructed by repeatedly executing grow, blossom and expand steps.
An augmenting path $P$ (of tight edges) may be discovered in a
blossom step.  In that case the algorithm proceeds to the augment
step.  It enlarges the matching by rematching $P$.  Then an entirely
new search structure is constructed and the procedure repeats.
The algorithm halts when an augment makes the matching perfect.
(We assume the input graph has a perfect matching.)

If the search structure becomes maximal -- i.e., no grow, blossom, or
expand step can be executed, and no augmenting path has been found --
then a dual adjustment step is executed. 
It
modifies the dual variables to
guarante
the search structure can change (via grow, blossom or augment steps).
Eventually (after some number of dual adjustments)
the desired augmenting path of tight edges is found.
(For further details of Edmonds' algorithm see e.g.,
\cite{E,CCPS, L, S, GT, DPS}.)

\paragraph*{Near optimality}
Edmonds' original algorithm is easily modified to use
a
variant of LP dual variables:
Define {\em near-optimum} duals by
\begin{equation}
  \label{NearOptimumDualsEqn}
\H{yz}(e)
\begin{cases}
\ge w(e)-2&e\in E\hskip 91pt\mbox{near domination}\\
\le w(e)&e\in M\cup \bigcup_{B\in \B.}E(B)\hskip 20pt\mbox{near
tightness.}
\end{cases}
\end{equation}
Here $M$ denotes the current matching and $E(B)$ denotes the set of
edges in the odd cycle that defines blossom $B$.

Consider a graph 
that satisfies the conditions for an optimum structured matching
except that the duals are near-optimum rather than optimum.
The perfect matching $M$
weighs $\ge W^*-n$, for $W^*$ the maximum weight
of a perfect matching of $G$
(\cite[Lemma 2.1]{GT}, \cite[Lemma 2.2]{DPS}).
Thus if the given weight function $w$ is replaced by
an integer multiple $aw$, $a > n$, the matching
given by the blossom tree has maximum weight.
The duals $y,z$ need not be optimum LP duals.
Such duals are required in some applications. They
can be easily derived from the near-optimum duals, see
\cite[Theorem 10.4]{GT}. Thus we can take our goal to be finding
an optimum structured matching with near-optimum duals.

The advantage of
near-optimum duals is that augmenting paths can be found in 
batches rather than individually. Specifically, when Edmonds' algorithm uses LP 
duals a search is guaranteed to find one augmenting path, if such exists.
Near-optimum duals provide a stronger guarantee:
After a given search, rematching
a maximal
collection of disjoint augmenting paths guarantees that 
duals must change
to get another augmenting path. 
Details of a
depth-first search that finds the maximal collection
are
given in \cite{GT} and also  \cite{G17}.
Figure \ref{EdAlgFig} gives  high level pseudocode
for this batching version of Edmonds' algorithm.

\begin{figure}
\begin{center}
  \fbox{
    \setlength{\Efigwidth}{\textwidth}
    \addtolength{\Efigwidth}{-.3in}
\begin{minipage}{\Efigwidth}
\setlength{\parindent}{.2in}

\narrower{
\setlength{\parindent}{0pt}
\vspace{\vmargin}
\setlength{\parindent}{20pt}

\noindent
Repeat the following procedure until the matching becomes
perfect (in the augmenting step).

\smallskip

rematch a maximal set of disjoint augmenting paths of eligible edges

construct a maximal search structure of eligible edges in $S$

adjust dual variables in the search structure 

}

\vspace{\vmargin}

\end{minipage}
}
\caption{Pseudocode for Edmonds' algorithm with near-optimum duals.}
\label{EdAlgFig}
\end{center}
\end{figure}

The advantage of batching augmenting paths dissipates in the later 
searches of the algorithm: As time goes on the "batches" 
have smaller size, eventually
just a few augmenting paths.
In keeping with this our algorithm operates in ``phases''.
In Phase 1 batching is advantageous.
In Phase 2  
batches have shrunk to small size. Phase 3
actually has no augmenting paths.

In order to modify
Edmonds' algorithm to use near-optimum
duals, we no longer use tightness as the  
criterion for being in the 
search structure. Call an edge  {\em eligible} if it can be in
the search structure. 
The condition determining eligiblity depends on the phase.
In Phase 1 an edge
$e$ is eligible if
it
essentially satisfies \eqref{NearOptimumDualsEqn}
with equality, specifically
\begin{equation}
  \label{EligibleEqn}
\H{yz}(e)=
\begin{cases}
w(e)&\text{$e$ is matched}\\
w(e)-2&\text{$e$ is unmatched}.
\end{cases}
\end{equation}
Edges that form contracted blossoms of the search structure
satisfy
\begin{equation}
  \label{EligibleTwoEqn}
\H{yz}(e)\in \{w(e),w(e)-2\}.
\end{equation}
This modification is needed since blossom edges enter the search 
structure
as eligible edges (satisfying
  \eqref{EligibleEqn}) but can get rematched in augments.

In Phases 2 and 3 the definition is less restrictive:
$e$ is eligible if it satisfies \eqref{EligibleTwoEqn}.
Again the reason for the change is due to
blossoms.

Modulo these changes, the search structure is constructed
as in Edmonds' original algorithm.
Consider now Figure \ref{EdAlgFig}.
Note the rematch step is given first.
This handles the
possibility that augmenting paths
exist for the starting dual variables.
This will be the case in our algorithm.

For succinctness
 Figure \ref{EdAlgFig} omits obvious logic to increase efficiency: When no augmenting path is found,
 the current search structure can be extended rather than constructed starting from scratch. Figures \ref{ScaleAlgFig}--\ref{Phase3Fig} present
 our algorithm
 following the same philosophy,
 omitting obvious optimizations in favor of conceptual simplicity.

\subsection*{Overall Algorithm: The Scaling Loop}
Our overall algorithm is a scaling loop. Here we follow the 
description in \cite{DPS}.
Take $W$ so that the given weight function,
denoted $\widehat w$, takes on integral values in $[0,W]$.
(Negative weights are easily eliminated since every perfect matching of
$G-v$ has cardinality $\f{n/2}$.)
Define a  derived weight function by
$\overline w =     
(n+1)\widehat w$.
The algorithm operates in 
$s=\f{\log (n+1) W}$ scales.
For $i=1,\ldots, s$, the $i$th scale
finds a near-optimum structured matching
for the  weight function 
\[w(e)=2\times(\text{the leading $i$ bits of }\overline w(e)).\]

To justify this outline note that any value $\overline w(e)$ 
is at most $(n+1) W$, 
so it is an integer of
$\le \f{\log (n+1)W}+1$   
bits. The last bit is 0 ($n+1$ is even).
So the $s$th scale uses the weight function  $(n+1) \widehat w$,
and it gives the desired near-optimum structured matching.

The $i$th scale starts by scaling up:
\begin{equation*}
\begin{array}{lll}
w(e)&\gets 2(w(e)+ \text{ the $i$th bit of } \overline w(e))&\forall e\in E\\
y(v) &\gets 2y(v)+2&\forall v\in V\\
z(B)&\gets 2z(B)&\forall \text{ blossom } B.
\end{array}
\end{equation*}
The matching of the previous scale is discarded,
the new matching is empty.

The new duals nearly dominate the new weights:
For any edge $e$, the previous scale ends with
$\H{yz}(e) \ge w(e)-2$,
so
$2\H{yz}(e) \ge 2w(e)-4$.
Increasing both sides by 4
makes the right-hand side an upper-bound
on the new scale's quantity $w(e)-2$. 
(The first scale is a special case:
Each $w(e)$ is 0 or 2, every $y$ value is 0, and there are no 
blossoms.)

\subsection*{Algorithm for a Scale}

The \D. algorithm finds a near optimum matching for the
new edge weights. The algorithm starts with
a collection of {\em inherited blossoms} from the previous scale.
Since the matching is initially empty these blossoms are no longer valid.
A main goal of \D. is to discard them.
An inherited blossom  {\em dissolves}
when its $z$ value has decreased to 0, and it is no longer
relevant. 

We consider $V$ to be an inherited blossom.
It never dissolves: The initialization makes $z(V)=0$, after
which $z(V)$ becomes negative and never increases in value.

As \D. progresses it  creates new, current blossoms. These are valid
for the current matching.
As in \cite{DPS} $\Omega$ denotes the set of all current blossoms,
$\Omega^-$ denotes the set of all inherited blossoms. ($V\in \Omega^-$.)

Let $T$ be the blossom tree of the inherited blossoms.
($T$ corresponds to the laminar structure of blossoms.
Its interior nodes are the blossoms of $\Omega^-$, including
the root $V$; its leaves
are the vertices of $V$.) Our strategy for dissolving blossoms
is to use the heavy path decomposition of $T$ \cite{T}, defined as 
follows.

\begin{figure}[t]
\centering
\input{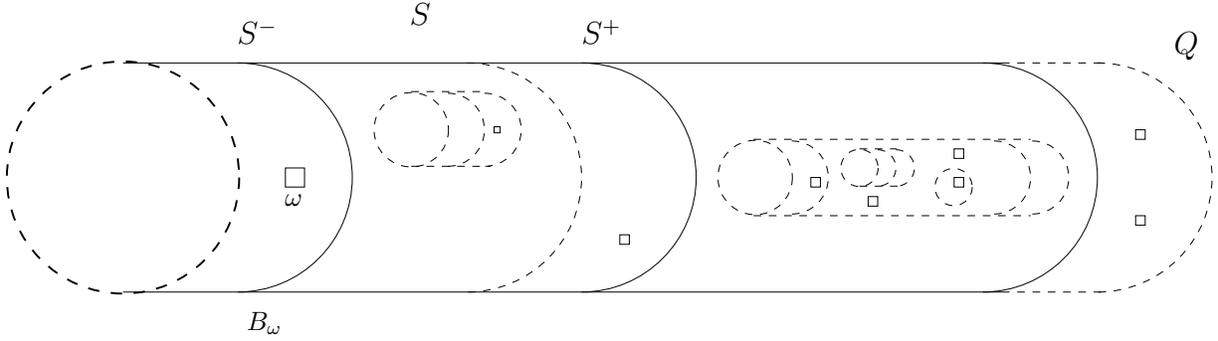}
\caption{\algcall DismantlePath(Q) searching major path $P(Q)$.
  Blossoms of $P(Q)$ are circular arcs or circles.
  Dissolved blossoms are dashed. Shell $S$ is atomic.
  Free vertices are squares.
  $Q$ contains 4 smaller major paths.
}
 \label{MajorPathFig}
\end{figure}

Let a blossom $B\in \Omega^--V$ have height $>1$ in $T$, i.e.,
$B$ has at least one subblossom that is inherited.
The {\em major child} of $B$ is the subblossom 
$C$ of $B$ with the greatest size $|V(C)|$.
A subblossom with
$|V(C)| > |V(B)|/2$ is obviously the unique major child.
If such $C$ does not exist there may be a tie for major child, in which
case
the major child is chosen arbitrarily
from all candidates. 
Since $|V(B)|$ is odd a tie can occur only when
every child $C$ has size
$|V(C)| < |V(B)|/2$.
A blossom $Q\in \Omega^-$ that is not a major child has a {\em major path}
$P(Q)$,
formed by starting at $Q$ and repeatedly descending to the major child
(see Fig.\ref{MajorPathFig}).
$Q$ is a major path root; we abbreviate major path root to {\em mpr}.

In addition we take $V$ to be a major path root, with major path simply $V$.
The major paths form a partition of $\Omega^-$.

\begin{figure}
\begin{center}
\fbox{
\begin{minipage}{\figwidth}
\setlength{\parindent}{.2in}

\narrower{
\setlength{\parindent}{0pt}
\vspace{\vmargin}

decompose the blossom tree $T$ into major paths $P(Q)$

traverse the mpr's $Q$ in  a bottom-up fashion, 
executing \algcall DismantlePath(Q)

}
\vspace{\vmargin}
\end{minipage}
}
\caption{\D. algorithm for a scale, pseudocode.}
\label{ScaleAlgFig}
\end{center}
\end{figure}

The \D.  dissolves blossoms using  a routine
\algcall DismantlePath(Q), where $Q$ is an mpr.
This routine works on the blossoms of 
$P(Q)$, eventually eliminating them all.
\algcall DismantlePath(V) is a special case: It eliminates all free vertices,
eventually constructing a perfect matching of $G$.
Figure \ref{ScaleAlgFig} gives a high-level statement of
\D..


On entry to 
\algcall DismantlePath(Q) every blossom $B\in P(Q)$
has $z(B)$ equal to its scaled up value, and all other
blossoms descending from $Q$ are now valid, i.e.,
either $z(B)=0$ or $B$ is a current blossom.
Clearly \algcall DismantlePath(V) is executed with every
inherited blossom besides $V$ dissolved.

\subsubsection*{Algorithm {\algcall DismantlePath(Q)}}

The algorithm works on shells of $P(Q)$:
A {\em shell} is a subgraph of $G$ induced by
a vertex set of the form $C-D$, where $C$ is a blossom of $T$ and
$D\pcon C$ is either a
descendant of $C$ or $\emptyset$.
(In the first case $D$ may be either a blossom or a vertex.)
The shell has
{\em outer boundary} $C$ and 
{\em inner boundary} 
$D$.
The shell is {\em even} or {\em odd} depending on the parity of $|C-D|$.
For $Q\ne V$ a shell is  odd iff $D=\emptyset$ (since any blossom is odd,
as is a single vertex).
$P(V)$ has the unique shell $V-\emptyset$, an even shell.
Shells that   have $D$ a vertex are only used in the analysis of the algorithm,
not the algorithm itself.

We denote the above shell as $(C,D)$. Alternatively
for a shell $S$ we may use the notation
$S^-,S^+$ to denote the inner and outer boundaries
of $S$, respectively. Thus $S^-\pcon S^+$
and $S=S^+ - S^-$. 
A blossom
$B$
with $S^-\pcon B\pcon S^+$ 
is an {\em interior} blossom
of $S$. ($B$ is interior to the $T$-path joining $S^-$ and $S^+$.) 
\algcall DismantlePath(Q) works on shells $S$ of $P(Q)$, specifically,
$S^+$ is a blossom of $P(Q)$, $S^-$ is either a blossom of  $P(Q)$ or
$\emptyset$.

The main operation to dismantle blossoms is a {\em unit translation}.
To {\em translate} a blossom $B$ by 1 means to decrease $z(B)$ by 2
and increase every $y(v)$, $v\in B$ by 1.
Clearly this preserves the value of $\H{yz}(e)$ for 
every edge with both ends in  $B$.

As discussed previously the details of
\algcall DismantlePath(Q) depend on the phase.
Phase 1 is the workhorse of the algorithm, and we start by discussing it
in detail. Phases 2--3 require less elaboration,
being   variations of Phase 1.

\begin{figure}
\begin{center}
\fbox{
  \begin{minipage}{\figwidth}
\setlength{\parindent}{.2in}
\narrower{
\setlength{\parindent}{20pt}
\vspace{\vmargin}
\noindent{\bf Phase 1:}
Repeat the following procedure until it ends Phase 1.

\smallskip

rematch a maximal set of disjoint augmenting paths of eligible edges

{\hi in every atomic shell of $P(Q)$}

$/*$ each pass starts here $*/$

if the atomic shells contain $\le \pi$ free vertices then end Phase 1

$/*$ $\pi$ is set 
to $\Theta(\sqrt{|Q|\log |Q|})$ in \eqref{piDefnEqn},  
balancing Phases 1 and 2 $*/$

let \A. be the set of atomic shells containing a free vertex

for shell $S\in \A.$ with size $|S|$ nonincreasing 

{\hi
if neither boundary of $S$ has dissolved then \algcall ShellSearch(S)
}
\vspace{\vmargin}
}
\end{minipage}
}

\vspace{.3in}

\fbox{
\begin{minipage}{\figwidth}
\setlength{\parindent}{.2in}
\narrower{
\setlength{\parindent}{20pt}
\vspace{\vmargin}
\noindent{\bf Algorithm} \algcall ShellSearch(S):

construct a maximal search structure of eligible edges in $S$

translate the boundaries of $S$ by 1

adjust dual variables in the search structure by 1
\vspace{\vmargin}
}
\end{minipage}
}

\caption{Pseudocode for Phase 1 of \algcall DismantlePath(Q).}
\label{Phase1Fig}
\end{center}
\end{figure}

\paragraph*{Phase 1: algorithm}
Phase 1 is presented in Figure \ref{Phase1Fig}.
It starts with a rematch step that processes augmenting paths
that may exist from previous executions of \algname DismantlePath.
or the previous scale.
The rest of the 
execution of Phase 1 is divided into
``passes'', where each pass processes every relevant shell
of $P(Q)$.
In Fig.\ref{Phase1Fig}
a comment line marks the demarcation point between passes.
Thus a {\em pass} begins by  constructing \A., continues by processing
the shells of \A., and ends with the global rematching step (in the next iteration of Pass 1).

At the start of a pass
a shell of $P(Q)$ is {\em atomic} if 
every interior blossom has dissolved but neither boundary has dissolved.
(A boundary $S^-=\emptyset$ never dissolves.)
The atomic shells partition the vertices of the maximal undissolved 
blossom of $P(Q)$.

At any point in \algname DismantlePath.
define
\[B_\omega = \text{  the currently minimal
undissolved blossom of $P(Q)$.}\]
\A. always contains the shell $S=(B_\omega,\emptyset)$.
$S$ is odd (even) for $Q\ne V$ ($Q=V$), respectively.
As mentioned
$S^-=\emptyset$ never dissolves
and it never gets translated (in \algcall
ShellSearch(S)).

During the pass 
\algname ShellSearch. is executed for every atomic shell 
$S\in  \A.$
whose boundaries are still intact when $S$ is examined. 
The dual adjustment made in 
\algcall ShellSearch(S) is valid because of the
rematch step in the preceding pass.
(In detail, the duals can be adjusted in a search structure 
that contains no augmenting path. This is guaranteed by the
rematch step. This also explains why 
\algcall ShellSearch(S) does not do the rematch:
A boundary of $S$ may dissolve in later executions of ShellSearch
during the current pass. The new shell containing $S$ may contain
augmenting paths that leave $S$.)

The dual adjustment in \algcall ShellSearch(S) 
may allow the search structure to be modified by 
new grow, blossom, and
expand steps. These steps are not executed in 
\algcall ShellSearch(S) for two reasons.
The first is given above, i.e., the search structure needs to be 
calculated for the atomic shell that contains $S$ at the end of the 
pass. Second, the  current search 
structure in $S$ may become invalid due to expand steps:
As mentioned above, the edges that replace an expanded blossom $B$ may 
not 
be eligible for Phase 1, due to rematching of $B$. 
Thus as part of the rematching step, the begins by computing the
entire search structure for each shell.
and then finds the maximal set of augmenting paths.

Each pass ends with the rematch step. Note that an atomic shell $S$ in that step
may consist of a number of shells that
were atomic at the start of the pass and got merged because of blossoms
dissolving. ($S$ will be atomic at the start of the next pass.)

At the start of a pass let $S$ be the atomic shell having $S^+$ as the 
maximal undissolved blossom of $P(Q)$. ($S$ exists since
$\A. \ne \emptyset$.) Suppose $S\in \A.$, \algcall ShellSearch(S) is executed, 
and the unit translation dissolves $S^+$. (Each of these conditions may 
fail to occur.)
After \algcall ShellSearch(S) returns, the vertices of 
$S$ are ``inactive'' -- they will never be examined again in the current 
execution of \algcall 
DismantlePath(Q). (They are processed when the mpr containing $Q$ gets 
dismantled.)
The pass may go on to deactivate more shells in this manner.
Furthermore if $S^-$ dissolves along with $S^+$ then other vertices
of $P(Q)$ are  deactivated, specifically the vertices of atomic shell
$T$ with $T^+=S^-$.
If all remaining vertices of $P(Q)$ are deactivated then
\algname DismantlePath. 
terminates.
(We shall see there is no Phase 2 or 3.)
We say a shell $T$ becomes {\em inactive} whenever a unit translation
deactivates the vertices of $T$ in any of the
scenarios described above.
(The rightmost shell of Fig.\ref{MajorPathFig} is inactive.)

We say a shell $S\in \A.$ is {\em preempted} if \algcall 
ShellSearch(S) is not executed because one of its boundaries has 
dissolved. A deactivated shell
may or may not be preempted.
Other scenarios are possible for
a blossom $B$ to dissolve 
without causing a preemption: $B$ may be the boundary of a shell of \A. 
that was already processed in the pass, or the boundary of a shell not 
added to \A.  because it had no free vertex. A given pass may end with a 
(new) atomic shell that contains any number of such dissolved blossoms 
as well as preemption blossoms.

\bigskip

To illustrate the algorithm recall 
$B_\omega$ defined above,
the innermost undissolved blossom of $P(Q)$. 
Suppose $Q\ne V$, so  $B_\omega$ contains an odd number of free vertices.
If there is exactly one such free vertex
we call it $\omega$.
 $\omega$ evolves in Phase 1 as follows.  A vertex $v$ may at some point 
become
$\omega$ (when augments in the innermost shell leave only one free
vertex).  Later $v$ may cease being $\omega$ (when blossom $B_\omega$
dissolves, creating a new $B_\omega$ with more free vertices).
Later still $v$ or
a different vertex may become the next $\omega$ (due to more
augments).

\paragraph*{Phase 1: correctness}
The correctness of \algname ShellSearch. hinges on the fact that it
is essentially simulating Edmonds' algorithm (modified
for near optimality) on shell $S$. An iteration of
this simulation begins with the first step of Phase 1
(enlarging the matching on $S$),
the subsequent construction of the search structure,
and following dual variable
adjustment.
The search structure does not contain
an augmenting path.
This follows from the preceding enlargement of the matching.
If we were to continue Edmonds' algorithm in $S$,
it would adjust duals by some positive integral quantity $\delta$.
\algname ShellSearch. adjusts the duals by $1 \le \delta$.
We will show this is safe, i.e.,
the adjustment by 1 preserves near-optimality
on  every edge $e$ of $G$.

In proof, 
Edmonds' algorithm
is guaranteed to preserve near-optimality 
if duals are adjusted by any value $\le \delta$.
So our adjustment is safe if $e$ has both ends in $S$.
Suppose only one end $v$ is in $S$. $e$ is not matched, so
we need to show near-domination,
$\H{yz}(e)\ge w(e)-2$, after the dual adjustment. 
The unit translations of \algname ShellSearch. increase $y(v)$ by
1 and the following
dual adjustment decreases
it by $\le 1$. (In detail, $v$ is either outer, inner, or not
in the search structure, and $y(v)$ decreases by 1, $-1$, or 0, respectively.)
So  near-domination continues to hold.

A last detail of the simulation of Edmonds' algorithm is that
it requires all free vertices to have $y$-values of the same parity.
(This is needed to preserve integrality in blossom steps.)
A scale starts out with every $y$-value being even.
\algname DismantlePath.
maintains the property that every free vertex $f$
has $y(f)$ even. In proof $y$ values are only changed in
\algcall ShellSearch(S). 
For $f\in S$ the unit translation of $S^+$
increases $y(f)$ by 1 
and the subsequent dual adjustment decreases it by 1 ($f$ is necessarily outer).
For $f\in S^-$ the two unit translations increase $y(f)$ by 2.

Other details of correctness are exactly the same as \cite{GT} or \cite{DPS}.
We conclude that the simulation of Edmonds' algorithm is correct.

\paragraph*{Phase 1: efficiency}
Let $F$ denote the set of vertices in $Q$ that are currently free and
still active (i.e., still in an atomic shell of $P(Q)$).  Any vertex
$v\in F$ is processed in \algname ShellSearch.  unless its shell gets
preempted by a shell with at least as many vertices.  So a preemption
doubles the size of its new atomic shell.  Thus a free vertex $v$ can
be preempted $\le \log |Q|$ times.  This will lead to the crucial
inequality $|F|(\#\text{passes }-\log n)\le n\log n$.  (This
inequality is restated as \eqref{ProductInequalityEqn}.)
Each pass runs in linear time $O(m)$ -- data structure details
used to achieve this time bound are given in Appendix \ref{Phase2Appendix}.

\begin{figure}[t]
\begin{center}
\addtolength{\figwidth}{.5in}
\fbox{
\begin{minipage}{\figwidth}
\setlength{\parindent}{.2in}
\narrower{
\setlength{\parindent}{20pt}
\vspace{\vmargin}
\noindent {\bf Phase 2:}
Repeat the following procedure until
the atomic shells of $P(Q)$ contain $\le 1$ free vertex.

\smallskip

let $S$ be the atomic shell of $P(Q)$ that
contains a free vertex and has $S^+$ maximal

\algcall ShellSearch(S)

while $\exists$ augmenting path $P$ of eligible edges
in an  atomic shell

{\hi

  $/*$ the shell is $S$ or the current atomic shell containing $S$ $*/$
  
  augment $P$
  }

\vspace{\vmargin}
}
\end{minipage}
  }

  \caption{Pseudocode for Phase 2 of \algcall DismantlePath(Q).}
\label{Phase2Fig}
\end{center}
\end{figure}

\paragraph*{Phase 2: algorithm}
The algorithm is presented in Figure \ref{Phase2Fig}.
The major change from Phase 1 is switching to
the less restrictive eligibility condition
of \eqref{EligibleTwoEqn}. As previously mentioned this is done
since batching is no longer effective.
The switch makes the algorithm behave
like Edmonds original algorithm, in that
rematching an augmenting path may create a new augmenting path.
The augment step in Phase 2
allows for this by repeatedly augmenting
as many times as possible.

The second change from Phase 1 is
to work on just one shell $S$ rather than pass over all the shells.
This simplfies low-level details of the algorithm, as discussed
below.

The algorithm works on the chosen shell $S$
 until 
it gets
completely matched
or a boundary of $S$ dissolves. (If the latter,
$S$ either gets merged with an adjacent shell,
or $S$ is deactivated 
in \algcall DismantlePath(Q).)
Then the algorithm  repeats
the process on the next shell, etc.
If $Q=V$ Phase 2 ends with a perfect matching (the scale is now complete).
If $Q\ne V$
Phase 2 ends when
either every shell of
$P(Q)$ dissolves (\algname DismantlePath. is done) or
there is a unique free vertex $\omega$ (Phase 3 is executed).

\paragraph*{Phase 2: efficiency}
Figure \ref{Phase2Fig} is a high-level version of
the Phase 2 algorithm. Adjusting duals by only 1 (as done in
\algname ShellSearch.) is inefficient,
since larger adjustments are needed to make progress.
An efficient implementation of Phase 2 adjusts duals
by the same quantity $\delta$ computed in Edmonds' algorithm.
It is important to use
\eqref{EligibleTwoEqn} as the criterion for elibiblity,
so that inner blossoms can be expanded without invalidating the search forest.

The low-level implementation of Phase 2 
tracks events in Edmonds' search algorithm using
a bucket-based priority queue.
In addition to Edmonds events the queue
tracks the $z$-values of inherited blossoms
so they get dissolved at the appropriate time.
The simple details of the priority queue are given in Appendix
\ref{Phase2Appendix}.

The high-level analysis of Phase 2 must
ensure that the number of buckets in the priority queue
is acceptable --
this is done in Lemma \ref{PQEntriesLemma}. Assuming that result
it is clear that Phase 2 uses 
$O(m)$ time for each augment.

The algorithm of Fig.\ref{Phase2Fig}
works on one shell $S$ at a time
in order to simplify the priority queue.
It is possible to organize Phase 2 like Phase 1, where
each pass searches all relevant shells. But this
complicates the details of the bucket-based queue, because
preemptions of shell searches
necessitate the rescheduling of many events.

Choosing $S^+$ maximal also simplifies the low-level algorithm.
This choice guarantees that
once $S^+$ is chosen as the blossom $B_\omega$
it remains as $B_\omega$ for the duration of Phase 2.

\paragraph*{Phase 3: algorithm} 
Phase 3 is only executed when $Q\ne V$, i.e., $Q$ is a blossom.
It is essentially one unsuccessful Edmonds search.
Every blossom eventually dissolves, in preparation for
the execution of \algname DismantlePath. on the mpr containing $Q$.
Phase 3 continues to use the eligibility criteria
\eqref{EligibleTwoEqn}.
Figure \ref{Phase3Fig} gives the detailed statement.

It is worth explaining how blossoms dissolve in this phase.
Consider  an arbitrary point in Phase 3 when $B_\omega$
is the blossom $B\in \Omega^-$ . It is possible that some
execution of \algcall ShellSearch(B) makes $B$ a blossom
(of $\Omega$). Once that occurs, each subsequent execution of \algcall ShellSearch(B)
decreases $z_0(B)$ by 2 (in a unit translation) and then adjusts duals by increasing
$z(B)$ (the dual for $B$ as an $\Omega$ blossom) by 2. In other words the value of $z_0(B)$ gets transferred to $z(B)$, at which point $B$ dissolves.

\paragraph*{Phase 3: efficiency}
Phase 3 uses a simple implementation of Edmonds' algorithm,
e.g.\ the $O(m\log n)$ algorithm of \cite{GMG}.
The search runs on the current minimal shell $(B_\omega,\emptyset)$.
If $B_\omega$ becomes a blossom a dual adjustment of
$\delta=z(B_\omega)/2$ dissolves $B_\omega$.
The time for Phase 3 is strictly dominated by the
rest of our algorithm, so we ignore it. A potential issue in the analysis
is the fact that Phase 3 can perform many unit translations, far more than
Phases 1 and 2. We shall see this does not present a problem.

\begin{figure}[t]
\begin{center}
\addtolength{\figwidth}{.5in}
\fbox{
\begin{minipage}{\figwidth}
\setlength{\parindent}{.2in}
\narrower{
\setlength{\parindent}{20pt}
\vspace{\vmargin}
\noindent {\bf Phase 3:}
Repeat the following procedure until
every blossom of $P(Q)$ has dissolved.

\smallskip

$B_\omega \gets$ the minimal undissolved blossom

ShellSearch($(B_\omega,\emptyset)$)

\vspace{\vmargin}
}
\end{minipage}
  }

  \caption{Pseudocode for Phase 3 of \algcall DismantlePath(Q).}
\label{Phase3Fig}
\end{center}
\end{figure}

\paragraph*{The overall algorithm}
The input graph is assumed to be perfectly matchable.
The algorithm is initialized by setting every weight $w(e)$ to 0,
every $y$ value to $-1$,
and taking $z$ identically 0 with no blossoms.
These are near optimum duals for any  perfect matching $M$.
The algorithm starts by scaling
the initial $y$ and $z$  values up for the first scale, as described above.
Note that the matching $M$ is not needed to start the first scale, and there are no inherited blossoms.

The output of the algorithm is an optimum structured matching
(with near-optimum duals).
This is accomplished using an outer loop that does scaling.
Each scale 
executes the \D. algorithm of Figure \ref{ScaleAlgFig}. 
Eventually \algcall DismantlePath(V) ends
in Phase 2 with a perfect matching.
The last scale ends with a perfect matching of maximum weight
(for the given weights).
Although the duals at the end of \D. are near-optimum, they are easily 
converted to optimum duals in linear time \cite{GT}.
We  ignore the time for this conversion in the rest of the analysis.

\section{Objective reducers} 
\label{dFSec} 
We turn to 
the analysis of the algorithm.  Similar to the analysis of cardinality 
matching algorithms, the goal is to prove a tradeoff between the number 
of passes in \algname DismantlePath. and the number of free vertices.  
This is ultimately achieved as equation \eqref{ProductInequalityEqn}, 
but this section derives the main precursors, \eqref{GoalEqn} and 
\eqref{GoalCrossedEqn}.

\begin{figure}[t]
\centering
\input{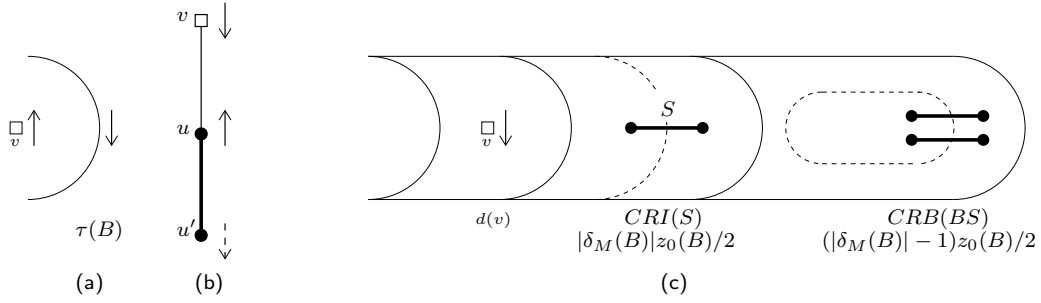}
\caption{Difficulties and basic notions. (a) Unit translation
  of a blossom increases the dual objective function by 1. (b)
  Dual adjustment with near optimum duals need not
  decrease the dual objective. (c) Dual objective reducers:
  $d(v)$, dual adjustments of free vertices;
  $CRI(S)$, crossings of interior blossoms;
  $CRB(BS)$, double crossings of contained blossoms.
}
 \label{NotionsFig}
\end{figure}

We begin by providing motivation, specifically
discussing the difficulties of the analysis and the basic
notions that we  use.  Fig.\ref{NotionsFig} illustrates these.
It also introduces our notation for these fundamental concepts.
(The precise definitions are below,
\eqref{TauDDefEqn} and \eqref{NotionsDefnEqn}.)

Previous scaling algorithms make progress by 
decreasing the dual objective
function or the dual objective $y$ values (e.g., \cite{DHZ}).
The fundamental difficulty in nonbartite matching
is that translating a blossom increases the dual objective
by 1, Fig.\ref{NotionsFig}(a). As in that figure 
$\tau(B)$ denotes
the number of unit translations of an inherited blossom $B$. A
further difficulty introduced by near optimum duals
is illustrated in  Fig.\ref{NotionsFig}(b) which illustrates
a dual adjustment of 1. When optimum duals are used the dual objective
for the two edges decreases by 1, but for near optimum duals
the objective can stay the same.
Specifically in both cases
the free vertex $v$ decreases its $y$ value $y(v)$ by 1.
The inner vertex $u$ increases $y(u)$ by 1.
For optimum
duals  matched edges are  tight,
$\H{yz}(e)=w(e)$. So
$y(u')$ decreases by 1, and dual objective decreases by
1. But
near optimum duals may have
$\H{yz}(uu')<w(uu')$. In that case $y(u')$ does not change
and the dual objective remains the same. If there are $s$ similar edges
$uu'$ the dual objective increases by $s-1$.

Our analysis uses 3 structures that contribute to decreases
in the dual objective. As illustrated in Fig.\ref{NotionsFig}(c),
the first is the aforementioned decrease in $y$ values of
free vertices in a dual adjustment.
We track this using the quantity $d(v)$, the number of
dual adjustments made for free vertex $v$. The two other structures
involve inherited blossoms that are crossed
by the current matching $M$ of the algorithm.
(see Fig.\ref{NotionsFig}(c) and also Fig.\ref{bcrFig}).
$CRI(S)$ (``crossed interior'')
counts the number of crossings by $M$
of interior blossoms $B$. Since the shell $(B,S^-)$ is even,
each such crossing decreases the objective of the previous scale
by $z_0(B)/2$, for $z_0$ the $z$ values of the previous scale.
(We prove this below. As an example,
in Fig.\ref{bcrFig}
the two crossings of $B_1$ replace the possibility
of one matched edge in $B_1$, thus decreasing the objective by
$z_0(B_1)$.)
$CRB(BS)$ (``crossed blossoms of a blossom set $BS$'')
counts the number of crossings by $M$, beyond the first,
of inherited blossoms $B$ contained in a shell.
Since a blossom $B$ is odd,
each crossing beyond the first
decreases the objective of the previous scale
by $z_0(B)/2$. (Again see Fig.\ref{bcrFig}.)

The difficulty of working with these dual objective function reducers
is their ephemeral nature: As the matching gets augmented, free vertices as well as blossom crossings may disappear, in essentially arbitrary
fashion.
\eqref{GoalCrossedEqn} gives the bound on the reducers
that we eventually deduce.

\subsection*{Preparation}
Consider
\algcall DismantlePath(Q) for any mpr $Q$ (including  $Q=V$).
We will analyze an arbitrary 
point in the execution of Phase 1 or Phase 2.
Recall the time for Phase 3 (one search of Edmonds' algorithm)
is not an issue. But the analysis must track
how Phase 3 
influences 
Phases 1--2, in terms of number of unit translations as well as possible
objective reducers.

Let $S=(C,D)$ be an even shell. (For generality 
$S$ need not be a shell of $P(Q)$, although this paper  does not use that case.)
Let $|S|=n$, so $n$ is even.
When $Q=V$ $S$ is the unique shell $(V,\emptyset)$.

$y_0$ and $z_0$ denote the duals immediately after scaling up,
i.e., when \D. begins.
$y$ and $z$ denote the algorithm's current duals, $M$ denotes the 
current matching on $S$
($M\con \gamma(S)$) and $F$ denotes its set of free vertices
($F\con S$). 
Our main assumption is that $M$ does not cross a boundary of $S$. Note 
this does not prevent an $\Omega$-blossom from crossing a boundary of 
$S$ -- it can cross arbitrarily many times on unmatched edges,
assuming $Q\ne V$.

Note the set $F$ may be empty -- our analysis is still valid 
(see the $CRI$ and $CRB$ terms in \eqref{GoalCrossedEqn}) but we will
not use the case of empty $F$.
The notion of ``core'' defined in Section \ref{CreditSec}
(e.g., Fig.\ref{creFig}) introduces the possibilty of
blossoms crossing $S$.

When $Q\ne V$
$B_\omega$
denotes the smallest undissolved blossom of $P(Q)$,
and $\omega$ denotes an arbitrarily chosen vertex of
$B_\omega -V(M)$. 
$\omega$ exists since $|B_\omega|$ is odd.
The analysis of this section applies to any shell $(B,\omega)$, $B\in P(Q)$,
although we only use it for $(B_\omega,\omega)$
with $\omega$ free.
When $|B_\omega-V(M)|=1$, $(B_\omega,\omega)$
has $F=\emptyset$.

\bigskip

We begin the analysis with three preliminary  properties,
\eqref{yzPreliminariesEqn}--\eqref{BcapSEqn}.
Start by defining two quantities, both
of which ignore Phase 3.
Let $B\con Q$ be an arbitrary
inherited  blossom ($B$ may be contained in an mpr $Q'\pcon Q$).
Let $v$ be a vertex that is free at the chosen point of
\algcall DismantlePath(Q).
\begin{eqnarray}
\tau(B)&=&\text{the number of Phase 1 or 2 unit translations
  made on $B$,}\notag\\
&&\text{up to and including the chosen point.}\label{TauDDefEqn}\\
d(v)&=&\text{the number of Phase 1 or 2 dual adjustments 
  made for $v$,}\notag\\
&&\text{up to and including the chosen point.}\notag
\end{eqnarray}
Note that a blossom that dissolves before Phase 3 has
$\tau(B)=z_0(B)/2$. A blossom that dissolves during
Phase 3 has $\tau(B)<z_0/2$.
(And 
$B=V$ never dissolves.)

Every free vertex $v$ satisfies
\begin{equation}
\label{yzPreliminariesEqn}
y(v)=y_0(v)-d(v)+ \sum_{v\in B}\tau(B).
\end{equation}
In proof note that whenever $v$ is the free vertex of Phase 3,
$y(v)$ does not change, since
every unit translation that increments $y(v)$ is offset by the 
following
dual adjustment that decrements it.

Next we show that
every $B\in \Omega^-$ with $B\cap S\ne \emptyset$
satisfies
\begin{equation}
  \label{RelevantBlossomsEqn}
B\pcon S \text{ or } D\pcon B.
\end{equation}
First assume $S\ne V$ and consider the blossom tree $T$:
$B$ is either \i a proper ancestor of $C$,
or 
a descendant of $C$ that is either
\ii a proper ancestor of $D$, or \iii  a nondescendant of $D$.
\i and \ii imply 
$D \pcon B$. \iii implies $B\pcon S$.
(Note this argument holds even when 
$D=\omega$.)
When $S=V$ both alternatives trivially hold
although we will only use
the first one.

\bigskip

Preparing for  \eqref{BcapSEqn},
recall the dual objective function
for shell $S$ is
\[\H{yz}(S)=y(S) + \sum_{B} z(B)\f{|S\cap B|/2}.\]
When $y,z$ are standard LP duals $\H{yz}(S)$
is an upper bound on the weight of a perfect matching on $S$. 
Our analysis will use $\H{yz}(S)$ (even though we use near-optimum duals).
We now show for the matching $M$ of the algorithm, 
 the above summation term has
\begin{equation}
  \label{BcapSEqn}
\f{|S\cap B|/2} =|\gamma_M(B)|+
  \begin{cases}
    0&B\in \Omega\\
    |F\cap B|/2&B\in \Omega^-\text{ is undissolved}.
  \end{cases}
\end{equation}
\noindent Note that the case
``$B\in \Omega^-\text{ dissolved}$'' is irrelevant:
$B$ dissolves when $z(B)$ decreases to 0. At that point
$B$ does not contribute to the dual objective and so is irrelevant.
Later an $\Omega$-blossom $B'$ with the same vertex set as $B$ may be formed,
and its dual $z(B')$ may become positive. $B'$ is treated
in the first case of
\eqref{BcapSEqn}.

To prove \eqref{BcapSEqn},
since $B$ consists of vertices that are either matched or free,
and $F\con S$,
\begin{equation}
\label{aBcapSEqn}
|S\cap B|= 2|\gamma_M(B)|+|\delta_M(B)|+
    |F\cap B|.
\end{equation}
Here we also use $\gamma_M(S\cap B) =\gamma_M(B)$ and
$\delta_M(S\cap B)=\delta_M(B)$, since $M\con \gamma(S)$.

First suppose ${B\in \Omega}$.
Since $M$ does not cross a boundary of $S$,
$S \cap B$ consists of the two vertices of various matched edges
and possibly one other vertex, the base vertex $b$ of $B$.
The base is either free or matched to a vertex
of $S-B$.
So \eqref{aBcapSEqn} has
$|\delta_M(B)|+
  |F\cap B|\le 1$.
  (Note this sum is 0 when $C=B_\omega$ and $\omega\in B$.)
    This implies  \eqref{BcapSEqn}.
    
    Now suppose $B\in \Omega^-$ and
        $B$ is undissolved at the chosen instant of time.
    The latter implies no matched edge crosses $B$,
        $|\delta_M(B)|=0$.
      Also $B$ undissolved implies either $B$ is interior to 
$S$
      or $S\con B$. In both cases $|S\cap B|$ is even.
      This implies  \eqref{BcapSEqn}.
      (Note that $|F\cap B|$ may be arbitrarily large.)

\subsection*{The analysis}
Let $M_\omega$ be the perfect matching of $S$ from the previous
scale. In detail, if $Q=V$ then $M_\omega$ is the entire
matching from the previous scale.
If $Q\ne V$ then $Q$ is a blossom.
This blossom gives a perfect matching of $Q-v$ for every
vertex $v\in Q$.
Since $\omega \in S^-$, $S$
is perfectly matched in the
perfect matching of $Q-\omega$.  $M_\omega$ is that perfect matching
($M_\omega\con\gamma(S)$).

The argument consists of 5 steps. The first four apply the
two dual functions, successively using
near tightness of $\H{y_0z_0}$, near domination of $\H{yz}$,
near tightness of $\H{yz}$, and near domination of $\H{y_0z_0}$.
The 5th step analyzes 
how $M$ crosses the blossoms of $\Omega^-$.

\bigskip

Any edge $e\in M_\omega$ was nearly tight in the previous scale,
since it was a blossom edge.
Thus in the new scale $w(e)\ge \H{y_0z_0}(e)-4$ (recall 
scaling up operation $y(v)\gets 2y(v)+2$). Summing these inequalities
gives
\begin{alignat*}{2}
    \hspace{80pt}
  w(M_\omega)&\ge \H{y_0z_0} (S)-4(n/2)&\hskip100pt\mbox{\bf scaled near tightness of
    \boldmath {$\H{y_0z_0}$}.}
\end{alignat*}
Here we use the fact that a blossom $B\in \Omega^-$ contains
exactly $\f{|S\cap B|/2}$ edges of $M_\omega$. In detail,
\eqref{RelevantBlossomsEqn}
implies
when $S\cap B\ne \emptyset$
\begin{equation}
  \label{mDetailOmegaMEqn}
|\gamma_{M_\omega}(S\cap B)|=
\begin{cases}
  \f{|B|/2}&B\pcon S\\
  |S\cap B|/2&\text{$B$ interior to $S$ or
    $\omega\in B\pcon B_\omega$ if $S=(B_\omega,\omega)$}\\
|S|/2&C\con B.
\end{cases}
\end{equation}
All three quantities equal
$\f{|S\cap B|/2}$.

\eqref{BcapSEqn} shows we can rewrite the current dual objective function as
\[\H{yz}(S)=\H{yz}(M)+y(F)+\sum_{D\pcon B\in \Omega^-} z(B)|F\cap B|/2.\]
The range of the summation is justified by
\eqref{RelevantBlossomsEqn} and the fact that $\Omega^-$-blossoms $B\pcon S$ 
have dissolved.
(Similarly we can assume $D\ne \omega$ in the summation
but we don't,
 just to simplify notation.)
Using this and the above bound on $w(M_\omega)$ gives
\begin{xalignat}{2}
\label{LHSEqn}
\H{y_0z_0} (S)-3n &\le w(M_\omega)-2(n/2)\\
&\le \H{yz}(S)&\hskip25pt \mbox{\bf near domination of \boldmath {$\H{yz}$}}\notag\\
&\le y(F) + w(M) +
\sum_{D\pcon B\in \Omega^-} z(B)|F\cap B|/2&\mbox{\bf near tightness of \boldmath {$\H{yz}$}.}\notag
\end{xalignat}

To upper bound the last quantity first sum \eqref{yzPreliminariesEqn}
for every $v\in F$:
\begin{equation}
y(F)=y_0(F)-d(F)+\sum_{B} |F\cap B|\tau(B),
\notag
\end{equation}
and then bound the matched edges by
\begin{eqnarray*}
  \hspace{80pt}
w(M)\le y_0(V(M))+ \sum_{B}  z_0(B)|\gamma_M(B)|+2(n/2)
&\text{\hspace{10pt}
  {\bf near domination of \boldmath{$\H{y_0z_0}$}.}}
\end{eqnarray*}
Combining the last two inequalities gives
\begin{align}
\label{yz1Toyz0Eqn}
y(F)+w(M)\le         
y_0(S) -d(F)+
\underbrace{\sum_{B} \Big(  |F\cap B|\tau(B) +z_0(B)|\gamma_M(B)|\Big)}_{SUM}
+n.
\end{align}

\subsubsection*{Analysis of blossom crossings}
\begin{figure}[t]
\centering
\input{bcr.pstex_t}
\caption{Crossings of inherited blossoms by matched edges.
    Two $\Omega^-$-blossoms form $SUB(S)=\{B_3,B_4\}$, and
  one $\Omega$-blossom crosses $S$.
    $CRI(S)=z_0(B_1)+z_0(B_2)$, 
  $CRB(SUB(S))=z_0(B_4)/2$.
}
 \label{bcrFig}
\end{figure}

We will show
\begin{equation}
\label{UpperboundedByEqn}
SUM\le \sum_{B} z_0(B)\f{|B\cap S|/2}
-\sum_{D\pcon B\in \Omega^-}|F\cap B|z(B)/2
+\Delta
\end{equation}
where
$\Delta$ is a quantity that we will derive.
$\Delta$ consists of the $\tau, CRI$ and $CRB$ terms
mentioned above.
Fig.\ref{bcrFig} illustrates
the latter two.
The blossoms $B$ contributing to $SUM$
belong to $\Omega^-$
with $B\cap S\ne \emptyset$, so they satisfy
\eqref{RelevantBlossomsEqn}.
We  consider the two  corresponding possibilities,
choosing the possibility $B\pcon S$ when $S=Q=V$.

\case{$D\pcon B$}

\subcase{$D\ne \omega$}
With $S\ne V$, i.e., $Q$ a blossom, this implies
either $B\in P(Q)$ or $Q\pcon B$.
In the first case the 
unit translations
of \algcall DismantlePath(Q)
maintain the invariant $\tau(B)+z(B)/2 = z_0(B)/2$.
This holds trivially in the second case, since $\tau(B)=0$,
$z(B) = z_0(B)$. Rearranging the invariant to
$\tau(B)= z_0(B)/2-z(B)/2$ shows
$B$'s term in $SUM$ is
\begin{eqnarray*}
|F\cap B|\tau(B) +z_0(B)|\gamma_M(B)|
&=&(z_0(B)/2)\big(|F\cap B|+2|\gamma_M(B)|\big)-(z(B)/2)|F\cap B|\\
&=& z_0(B)\big(|S\cap B|-|\delta_M(B)|\big)/2 -(z(B)/2)|F\cap B|.
\end{eqnarray*}
A blossom $B$ of this case has   $|S\cap B|$ even
(recall the last two cases of \eqref{mDetailOmegaMEqn}).
So $|S\cap B|/2=\f{|S\cap B|/2}$. Using this  and rearranging terms changes the last line to
\[z_0(B)\f{|S\cap B|/2}-(z(B)/2)|F\cap B|
 -|\delta_M(B)|z_0(B)/2.\]
The first
two terms match the
terms for $B$ in the two summations of
\eqref{UpperboundedByEqn}. We include the third term
\[ -|\delta_M(B)|z_0(B)/2\]
in $\Delta$. 
It is a nonpositive quantity,
nonzero only on interior blossoms of $S$.

\subcase{$D= \omega$}
Using $\tau(B)\le z_0(B)/2$ the term for $B$ 
in $SUM$ is at most
\begin{equation*}
  z_0(B)(| F\cap B|+2|\gamma_M(B)|)/2=
  z_0(B)(|S\cap B|-|\delta_M(B)|)/2
  \le z_0(B)|S\cap B|/2.
\end{equation*}
Since $S=(B_\omega,\omega)$, $S\cap B=B-\omega$ has even cardinality.
So
the above right-hand side is
$z_0(B)\f{S\cap B|/2}$.
This  matches the
term for $B$ in the first summation of
\eqref{UpperboundedByEqn}.
As mentioned above there is no contribution to the
second summation ($z(B)=0$).

\case{$B\pcon S$}

\subcase{$B$ crossed}
Again 
using $\tau(B)\le z_0(B)/2$, the term for $B$ 
in $SUM$ is at most
\begin{eqnarray*}
z_0(B)(|F\cap B|+2|\gamma_M(B)|)/2&=& z_0(B)(|B|-|\delta_M(B)|)/2\\
&=&z_0(B)\f{|B\cap S|/2} +z_0(B)( 1-|\delta_M(B)|)/2.
\end{eqnarray*}
The first term 
of the last line
 matches the
term for $B$ in the first summation of
\eqref{UpperboundedByEqn}.  Recalling the assumption
$|\delta_M(B)|\ge 1$, we include the second term
\[ z_0(B)( 1-|\delta_M(B)|)/2\]
in $\Delta$ when $|\delta_M(B)|>1$.
It is a nonpositive quantity.

\subcase{$B$ uncrossed}
Since $B$ contains a free vertex, $B$'s term in $SUM$ is
\[|F\cap B|\tau(B) +z_0(B)|\gamma_M(B)|\le
\tau(B) +z_0(B)(|B|-1)/2=
\tau(B) +z_0(B)\f{|B\cap S|/2}.\]
In the rightmost bound the second term matches
term for $B$ in the first summation of
\eqref{UpperboundedByEqn}.
We include the nonnegative term
\[\tau(B)\]
in $\Delta$.

\bigskip

We conclude \eqref{yz1Toyz0Eqn} gives
\begin{xalignat*}{2}
y(F)+w(M)\le
&y_0(S)-d(F) 
+\sum_{B} z_0(B)\f{|B\cap S|/2}
-\sum_{D\pcon B\in \Omega^-}|F\cap B|z(B)/2+\Delta+n
\notag\\
=&\H{y_0z_0}(S) -d(F)
-\sum_{D\pcon B\in \Omega^-}|F\cap B|z(B)/2+\Delta+n. 
\end{xalignat*}
Combining this with
\eqref{LHSEqn}
gives
\[\H{y_0z_0} (S)-3n 
\le \H{y_0z_0}(S) -d(F)+\Delta+n.
\] 
Hence
\begin{equation}
\label{GoalEqn}
d(F)\le 4n+\Delta.
\end{equation}

We introduce notation to make the terms of $\Delta$ explicit.
Let $S$ denote a shell, $M$ a matching on $S$,
$BS$ an arbitrary collection of
inherited blossoms. $B$ always denotes an inherited blossom.
\begin{eqnarray}
{\U.}    &=& 
\set {B} {\text{$B\pcon S$ an inherited blossom not crossed by $M$}},
\notag\\
  INT(S) &=& \set {B} {\text{$B$ interior to $S$}},\notag\\
  SUB(S) &=& \set {B} {\text{$B\pcon S$}},\label{NotionsDefnEqn}\\
  CRI(S)&=& \sum_{B\in INT(S)}|\delta_M(B)| z_0(B)/2,\notag\\
  CRB(BS)&=&\sum_{B\in BS,\; |\delta_M(B)|>1} (|\delta_M(B)|-1)z_0(B)/2.\notag 
\end{eqnarray}
\U.
and the crossed blossom functions $CRI,CRB$ depend on the matching $M$.
This matching will usually be clear from context, but if not
 we will identify it in a comment.
Finally define the constant $c=4$. Now \eqref{GoalEqn} becomes
\begin{equation}
\label{GoalCrossedEqn}
d(F)+ CRI(S)+ CRB(SUB(S)) \le cn+\tau(\U.).
\end{equation}
For convenience we
reiterate the setting for \eqref{GoalCrossedEqn}.
\eqref{GoalCrossedEqn} applies 
at any chosen point of Phase 1 or 2 of \algcall DismantlePath(Q)
where the matching is $M$,
$S$ is an
even shell of $P(Q)$  uncrossed by $M$, 
$F$ is the set of vertices of $S$ that are free in $M$,
and $M$ is used to define $CRI$ and $CRB$.

We also extend the $d$ function, as follows.
For any mpr $Q$ and any vertex $v$, $d(v,Q)$ denotes the number of
dual adjustments made on $v$ during \algcall
DismantlePath(Q) while $v$ is free.
We allow the possibility $v\notin Q$, in which case $d(v,Q)=0$. 
For $v\in Q$
the quantity depends on the chosen
moment in the execution of \algcall DismantlePath(Q): When we are
discussing a point during the execution of \algcall DismantlePath(Q)
$d(v,Q)$ counts all adjustments up to that moment. When discussing a
point when \algcall DismantlePath(Q) has already returned $d(v,Q)$
counts all adjustments made during Phases 1 and 2. So for 
example $d(v)=\sum \set
{d(v,Q)} {v\in Q} $.

For any mpr $Q$ 
define
\[{\mmpr(Q)}   =\set  {Q'} {Q'\text{ an mpr properly contained in $Q$} }. 
\]
 Our summing conventions imply
$d(F, Q+\mmpr(Q))=\sum
\set{d(v,Q')}{v\in F, \text{ mpr }  Q'\con Q}$.
Thus
\begin{equation*}
d(F,Q+\mmpr(Q))=d(F,Q)+d(F,\mmpr(Q)).
\end{equation*}

\bigskip

The bound on $d(F)$ provided by
\eqref{GoalCrossedEqn} will lead to our desired inequality
\eqref{ProductInequalityEqn}.
To use \eqref{GoalCrossedEqn} we need to
upper bound the quantity 
$\tau(\U.)$.
This is done in Section
\ref{CreditSec}, giving our bound Lemma \ref{CreditSystemLemma}.
The desired bound on $d(F)$ is then proven in
Lemma \ref{MainLemma}, which again uses inequality
\eqref{GoalCrossedEqn}. This culminates in the time bound for the entire algorithm,
Theorem \ref{MainTheorem}.



\section{The credit system}
\label{CreditSec}

\def\BM{\mathy{B_{\rm max}}}
\def\c4{\mathy{c\,}}

We bound $\tau(\U.)$ using a system of credits,
based on \eqref{GoalCrossedEqn}.
This inequality also holds for $f$-factors, \eqref{fGoalCrossedEqn}.
So the credit system applies to both ordinary matching and $f$-factors.
We  introduce an oracle to account for slight differences
for $f$-factors. Readers iterested only in ordinary matching
can skip the description of the oracle below, and assume
the oracle simply executes an Edmonds search for ordinary matching.

\paragraph*{The oracle for $f$-factors}
In general the \D. algorithm invokes 
an oracle to return the results of each Edmonds search.
The oracle provides arbitrary results subject only to
the following constraints.

\bigskip

{\narrower

  {\parindent=0pt
    
The search decreases each value $y(v)$, $v$ free, by 1.

The search does an arbitrary number of augments, possibly 0,
along arbitrary
augmenting paths  (within the current shell).

Inequality \eqref{GoalCrossedEqn} holds after every search.
\eqref{fGoalCrossedEqn} has the term $d(\o F)$ instead of
  $d(F)$, and that term is interpreted in the natural way:
  If a vertex $v\in V$ is free, i.e., it is on $|\delta(v,M)|<
  f(v)$ matched edges, then it contributes $d(v)\times
(f(v)-|\delta(v,M)|)$ to the quantity $d(F)$.

}}

\bigskip

Duals of nonfree vertices may change arbitrarily.
So the only numerical information available to the credit system
are the $\tau$ values and the quantities $d(v), v$ free. The oracle
does not report any details of blossoms in the search graph.
So the credit system cannot use any information about $\Omega$
blossoms.

The section can be interpreted for $f$-factors in two equally
valid ways. In
the first we view the $f$-factor as a matching on the graph
where each vertex $v$ has been expanded to $f(v)$ copies,
each either free or matched to the appropriate other vertex.
For example the vertex $\omega$ is one copy of some given $f$-factor
vertex.
An expression $|A|$, for $A$ a vertex set like
a shell or an mpr, etc., is intepreted as $\sum_{v\in A} f(v)$.
An inherited  blossom is called an e-blossom in Section
\ref{fFactorSec}, and
Lemma \ref{lBlossomLemma} shows an e-blossom $B$ has odd size
$f(B)$. Thus a blossom in this section has odd size $|B|$,
a shell $S$ with  two blossom boundaries has even size
$|S|$, etc.

The second way to view this section, for readers familiar with
Sections \ref{fFactorSec}--\ref{fdFSec}, is to make the obvious
changes in notation. For instance
 ``matching'' is interpreted as ``$f$-factor'',
the size $|S|$ of a shell $S$
is interpreted as $f(S)$, etc.

\bigskip

\paragraph*{Basis of the credit system}
The system is based on a simple structural property:

\begin{proposition}
\label{StructuralProp}
Consider a shell $S$ with two blossom boundaries. Let $M$ be
a matching  that crosses one of the boundaries, say $B\in \{S^+,S^-\}$, but 
not the other.  Either $S$
contains a free vertex of $M$ or $|\delta_M(B)|\ge 2$.
\end{proposition}

\begin{proof}
  For any shell $S$, its complement $\overline S$ is the disjoint union of
  $S^-$ and $\overline{S^+}$.
Thus for any set of edges 
$\delta (S)$ is the set of edges crossing exactly one of the boundaries
$S^+,S^-$. Applied to  the proposition
we get $\delta_M(S)=\delta_M(B).$
Suppose $S$ contains no free vertex.
Since $S$ is an even shell, $|\delta_M(S)|$
is even.
$B$ crossed implies $|\delta_M(B)|\ge 1$. Thus
$|\delta_M(B)|=|\delta_M(S)|\ge 2$.
\end{proof}

\noindent
This proposition will allow us to charge a translation to 
either some $d(f)$, $f$ a free vertex, or some doubly crossed blossom.

\subsection*{The credit system}
Consider an execution of  \algcall DismantlePath(X) (``X'' for ``executing'').
Choose any point  in  Phase 1 or Phase 2.
The goal is to prove at the chosen point that 
\[d(F,X) \le \c4 |X|\log |X|\]
(achieved in Lemma \ref{MainLemma}).
We use the notation of Section \ref{dFSec}
for the mpr $X$.
So 
$S=(C,D)$ is  an even shell,
$M$ is the current matching on $S$
($M\con \gamma(S)$) and $F$ is its set of free vertices
($F\con S$). 
We will show 
every minimal uncrossed even shell $S$ of $P(X)$
satisfies
\[d(F\cap S,X) \le \c4 |S|\log |X|.\] 
Summing these inequalities implies the goal.
(If $X=V$  there is only one shell,
$S=(V,\emptyset)$. If $X\ne V$
the  relevant shells $S$ partition
the set $X' -\omega$, where $X'$ is the largest undissolved blossom in $X$,
and $\omega$ is the chosen vertex in the minimal undissolved blossom
$B_\omega$.)

To accomplish this apply \eqref{GoalCrossedEqn} to $S$ and $M$:
\[d(F\cap S)+ CRB(SUB(S)) \le \c4 |S| +\tau(\U.).\]
(We ignore the term $CRI(S)$ for 
crossed blossoms interior to  $S$.) 
This shows it suffices to prove
\[\tau(\U.)\le d(F\cap S,\widehat X)+ CRB(SUB(S))+ \c4 |S|\log(|X|/2).\] 
The proof is based on a system of credits.
One credit can pay for one unit translation of an inherited blossom.
We wish to pay for the $\tau(\U.)$ translations.

Distribute the right-hand side credits as follows:
For each free vertex $f\in S$ give
$d(f,Q)$ credits to $f$'s occurrence in $Q$, for each mpr
$Q$ where
$f\in Q\pcon X$.
Each crossed blossom $B\in SUB(S)$  gets
$(|\delta_M(B)|-1)z_0(B)/2$ credits.
(If $B$ is crossed at least twice, this quantity
is included in $CRB(SUB(S))$. If $B$ is crossed once
no credits are given.
The credits will naturally be used in the mpr $Q$ where $B\in P(Q)$.)
Finally give each
mpr $Q\pcon S$ a total of $\c4 |Q|$ credits. 
The total of all such credits is $\le \c4 |S|\log (|X|/2)$,
since the mpr's containing a given vertex double in size
(i.e., for any mpr $Q$, any maximal mpr $Q'\pcon Q$
has $|Q'|<|Q|/2$.)

Consider a fixed  major path $P(Q)$, $Q\con S$. 
We will show how to pay for every
Phase 1 or 2 translation  of every \U.-blossom in $P(Q)$.
The analysis does not depend on
the behavior of
\algcall DismantlePath(X). Instead we hypothesize
some simple properties of the matching on $Q$.
It will easily be seen that the properties hold for the above matching
$M$ of \algcall DismantlePath(X).

Let $M$ be a matching with at least one uncrossed blossom in $P(Q)$.
(References below to ``crossing'' and ``uncrossing'' refer to
$M$. If every blossom of $P(Q)$ is crossed then $Q$ does not contribute to
$\tau(\U.)$.)
Let $F$ be the set of free vertices of $M$.
Assume every vertex of $F$ is free at the end of
\algcall DismantlePath(Q). Beyond that
$M$ is arbitrary -- it may contain edges with both ends in $Q$,
edges crossing $Q$, and vertices of $F$.
Let $B_\omega$ and $\BM$
be the minimal and maximal uncrossed blossoms of $P(Q)$, 
respectively.
(Possibly $B_\omega=\BM$.)

Partition the vertices of $Q$ into the following shells:

\bigskip

$(B_\omega,\emptyset)$;

\smallskip

{\narrower

{\parindent = 0pt

minimal shells $S$ with two blossom boundaries $S^-,S^+$ that are both uncrossed (i.e., 
every interior blossom of $S$ is crossed);

}
}

\smallskip

$(Q,\BM)$ if $Q$ is crossed.

\bigskip

Say that a shell $S$ with two uncrossed blossom boundaries
is {\em void} if it is maximal for the condition
$S\cap F=\emptyset$.
Clearly a void shell is perfectly matched.
Void shells will be treated later.
Note that $(Q,\BM)$ is not void, even though it may not contain a
free vertex. Also $(B_\omega,\emptyset)$ is not void, since it has only one blossom boundary. More importantly it necessarily contains a free vertex, $\omega$.

\begin{figure}[t]
\centering
\input{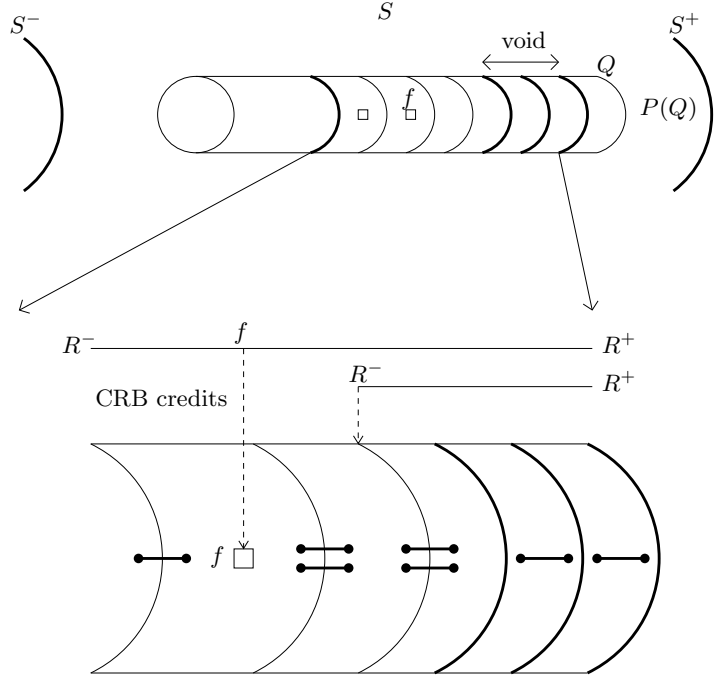}
\caption{Uncrossed-blossom Payment.
  Inequality \eqref{GoalCrossedEqn} applied to shell $S$
  gives $CRB$ credits used to pay for {\sc ShellSearches} in
  \algcall DismantlePath(Q).
  Translations of \algcall ShellSearch(R)
  are paid for by $z_0(R^-)/2$ or $d(f)$.
  $\cal U$ blossoms are drawn heavy. $Q\ne \BM.$ is crossed.
  The void shell is void even for matchings that
  cross its interior blossom.
  }
 \label{uncrFig}
\end{figure}

The following lemma shows how to pay for all {\sc ShellSearches}
except those contained in a void shell. It is illustrated in
Fig.\ref{uncrFig}.

\begin{lemma} [Uncrossed-Blossom Payment]
\label{NonVoidLemma}
Assume $Q$ has these credits:

$\bullet$ Every crossed blossom $B\in P(Q)$ not interior to a void shell
has $(|\delta_M(B)|-1)z_0(B)/2$ credits.

$\bullet$ Every 
free vertex $f\in Q$ has $d(f,Q)$ credits, where this quantity
is evaluated at the end of

\hskip 10pt \algcall DismantlePath(Q). 

\noindent
These credits can pay for every Phase 1-2 translation of an uncrossed blossom
of $P(Q)$, except  for the translations executed in \algcall ShellSearch(R)
where $R$ is contained in a void shell.
\end{lemma}

\remark{The hypothesis is satisfied by the distribution of credits
  given above. In particular vertex $\omega\in Q$ has $d(\omega,Q)$ credits.}

\begin{proof}
Consider an execution of \algcall ShellSearch(R) in \algcall DismantlePath(Q).
It is irrelevant if both boundaries $R^-,R^+$  are crossed.
So the following three cases exhaust all possibilities:

\case {$R$ has exactly one uncrossed blossom boundary, and
$R\cap (F+\omega)\ne \emptyset$}
For any  free vertex $f\in R$, the translation
(of the uncrossed blossom boundary) is counted in $d(f,Q)$.
This follows since $f$ was free when \algcall ShellSearch(R)
was executed, so its dual value $y(f)$ was decremented.

\smallskip

Note this case allows  
$R$ to have two blossom boundaries or only one. The second possibility,
i.e.,
$R$ an odd shell $(R^+,\emptyset)$, is always covered by this case.
In proof $R^+$ uncrossed
and the minimality of $B_\omega$ 
implies $B_\omega \con R^+$. Thus $\omega\in R^+$.

\case {$R$ has exactly one uncrossed blossom boundary, and
$R\cap (F+\omega)=\emptyset$}
Let $R^+$ be uncrossed. (The case $R^-$ uncrossed is symmetric.)
$R$ is an even shell and $R^-$ is crossed.
$R$ has no free vertex, so $|\delta_M(R^-)|$
is even and $\ge 2$ (Proposition \ref{StructuralProp}).
If $R^-$ is  interior to a void shell $S$ then 
$R\con S$ (by the maximality of $S$).
If $R^-$ is not interior it
has $z_0(R^-)/2$ credits.
One such credit can pay for the translation of $R^+$
($R^-$ was translated in \algcall ShellSearch(R), and $\tau(R^-)\le z_0(R^-)/2$).


\smallskip

An example of this case is the shell $(Q,\BM)$
when it contains no free vertex.

\case {$R^-$ and $R^+$ are both uncrossed blossoms}
If $R$ contains a free vertex, it has at least two, say $f,f'\in F$,
since $R$ is even.
The two unit translations are counted in $d(\{f,f'\},Q)$.
If $R$ has no free vertex then it is a subset of a void shell.
\end{proof}

\paragraph*{Comment on \boldmath {$z_0$} charges}
Blossoms that survive into Phase 3 can have very large $z_0$ values.
But the above argument, and the next lemma, only charges a term $z_0(B)/2$ for translations made
in Phases 1-2.

\bigskip

As noted
we have already distributed credits so the lemma's hypothesis is satisfied.
So to complete the analysis
we need only pay for translations in void shells.

Consider such a void shell $S$. The lemma shows we must pay for
 the translations executed in \algcall ShellSearch(R)
 where $R\con S$.
We can assume the boundaries of $R$
are among $S^-,S^+$, and the interior blossoms of $S$.
This follows since any blossom $B\pcon S$ 
is crossed, since $S$ is perfectly matched.
By definition $S^-$ and $S^+$ are uncrossed
(recall $S$ is maximal) and they must be paid for.
There seems no way to distinguish
the interior blossoms of $S$ that are
uncrossed from those that are crossed.
(As extreme examples, every interior blossom might be uncrossed,
or they all might be crossed.)
So we will pay for every translation of such blossoms.
To summarize
the task for void shells $S$ is to pay for {\em every} unit translation
made in an invocation of \algcall ShellSearch(R)
from \algcall DismantlePath(Q), for any $R\con S$.
We start by presenting a lemma similar to the previous one, showing
how hypothesized credits can pay for translations
made in subshells of $S$.

To set the stage let $S$ be an arbitrary shell of $P(Q)$.
Consider
the last invocation of \algname ShellSearch.
in Phase 1 or 2 whose argument is 
a shell contained in $S$.
Let \algcall ShellSearch(R) be that invocation.
Let $M$ be the matching on $R$ in that invocation.
Define the {\em core} $C$ of $S$ to be
a minimal uncrossed shell contained in $R$
and containing a free vertex $f$ of $M$.
$C$ exists since $R$ is a candidate. $C$ may be $R$ itself.
At the other extreme there may be various choices for $C$,
in which case the choice is arbitrary.
Note also that an $\Omega$-blossom
may cross a core, e.g., in Fig.\ref{bcrFig}
take $S$ to be the core.

In the following lemma,
$d(f,R)$ denotes this quantity
evaluated at the end of \algcall ShellSearch(R)
(i.e., the number of dual adjustments made on $f$
from the start of \algcall DismantlePath(Q) up to and including
the adjustment made in  \algcall ShellSearch(R)).
This may be less than $d(f,Q)$.
(For example, $f$ may get matched before the chosen instant in
\algcall DismantlePath(X) but after  \algcall ShellSearch(R) returns.)
The lemma is illustrated in Fig.\ref{creFig}.

\begin{figure}[t]
\centering
\input{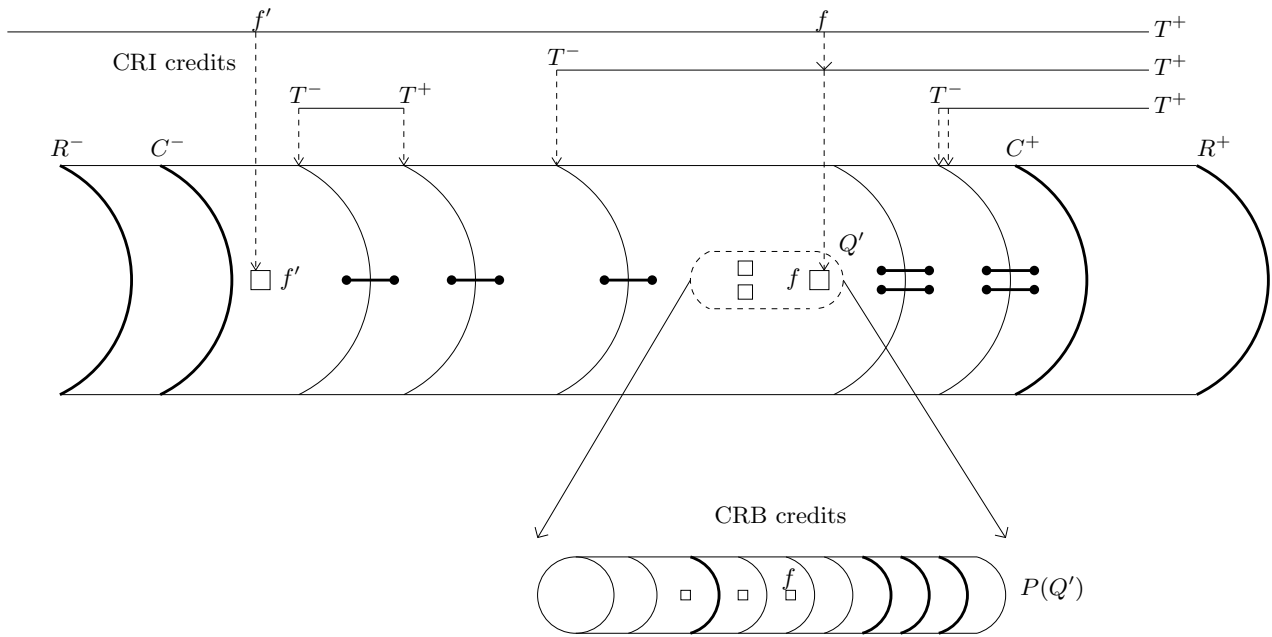}
\caption{All-blossom Payment.
  Inequality \eqref{GoalCrossedEqn} applied to core $C$ of atomic
  shell $R$
  gives credits used to pay for {\sc ShellSearches} in
  \algcall DismantlePath(Q).
  Translations of \algcall ShellSearch(T)
  are paid for by $z_0(T^-)$, $z_0(T^+)$,
  $d(f)$, $d(f')$. New debts for uncrossed blossoms of $P(Q')$,
  $f\in Q'\pcon Q$,
  are settled by Uncrossed-blossom Payment (Fig.\ref{uncrFig}).
  }
 \label{creFig}
\end{figure}

\begin{lemma}[All-Blossom Payment]
\label{CoreLemma}
Assume $C$ has these credits:

$\bullet$ Every blossom $B$ interior to $C$
has $|\delta_M(B)|z_0(B)/2$ credits.

$\bullet$ Every free vertex $f\in C$ has $d(f,R)$ credits.

\noindent
Then $C$ can pay for every Phase 1-2 translation
in an invocation of \algcall ShellSearch(T)
from \algcall DismantlePath(Q)
with $T\con S$ and $T\cap C\ne \emptyset$.
\end{lemma}

\begin{proof}
We consider two cases.
  The main argument is for
a shell $S$ whose boundaries are
both blossoms, i.e., $S^-\ne \emptyset$. Recall this holds for
a void shell, and hence for any of its subshells.
This assumption allows a minor simplification of the lemma's proof.
The case
$S^-=\emptyset$ is not needed for the credit system.
But it
will be 
needed to prove Lemma \ref{PQEntriesLemma}.
We treat that case at the end.

  Every blossom $B$ interior to $C$ is crossed
and so has $\ge z_0(B)/2$ credits. $B$ uses these credits to
pay for every translation of itself.
In particular when both boundaries of $T$ are interior
to $C$, the translations of $T$ are paid for (Fig.\ref{creFig}).

In the remaining  possibility
one or both boundaries of $T$
are not interior to $C$ yet $T$ intersects $C$. This gives
two symmetric
cases: $T^+$ not interior, implying
\[T^-\pcon C^+ \con T^+,\]
and $T^-$ not interior, implying
\[T^-\con C^- \pcon T^+.\]
Assume $S^-\ne \emptyset$. The opposite case is
degenerate and treated after the main argument.
Since $C$ is an uncrossed even shell (it has two blossom boundaries)
it contains $\ge 2$ free vertices, say $f,f'$.
Choose $f$ ``closest to'' $C^+$, i.e.,
for any blossom $B\con C^+$
interior to $Q$, $f\in B$ implies
the shell $(C^+,B)$ has no free vertex.
Associate $f$ with the first case above.
Treat the second case symmetrically, i.e.,
associate it with a free vertex $f'\in C$ that is closest to
$C^-$. Clearly this $f'$ can be chosen distinct from $f$.
We now analyze the first case.
Here the goal is to pay for every translation of $T^+$.
The analysis of the second case is
symmetric.
For a shell $T$ in both cases (i.e.,
$T^-\con C^- \pcon  C^+ \con T^+$)
we need only account for $T^+$, leaving $T^-$ to the second case.

\case {$f\notin (C^+,T^-)$}
This case, with $f\in C^+$,
implies $f\in T^-\pcon C^+$.
This makes $T^-$ interior to $C$, and
the choice of $f$ implies $(C^+,T^-)$ has no free vertex.
Since $C^+$ is uncrossed and $T^-$ is crossed,
$|\delta_M(T^-)|\ge 2$ (Proposition \ref{StructuralProp}).
Thus $T^-$ has $\ge 2\times z_0(T^-)/2$ credits.
One such credit pays for itself, the other for $T^+$.

\case {$f\in (C^+,T^-)$}
 Our assumption $C^+\con T^+$
implies $f$ is in the shell $T$. $f$
is free when \algcall ShellSearch(T) is executed.
(In proof observe the definition  of $R$ along with $f\in T\cap R$
makes $T\con R$.
Furthermore $f$ is free in  \algcall ShellSearch(R).)
So
the translation of $T^+$
is counted in $d(f,R)$. We use the corresponding credit.

\bigskip

This completes the argument
for $S^-$ a blossom so we turn to the case 
$S^-=\emptyset$.
If $C^-\ne \emptyset$
the argument is as before. (A further degenerate possibility
is $T^-=\emptyset$, which has no charge for $T^-$.)
If $C^-=\emptyset$ then $C$ may have only one free vertex.
It is chosen as $f$ and the argument for the first case
$T^-\pcon C^+ \con T^+$ is unchanged.
The other possibility is $T^+$ interior to $C$. $T^+$ pays
for itself, 
as before.
\end{proof}

Any shell
$T\con S$ either intersects $C$ (as in the lemma)
or is disjoint from $C$, i.e.,
$T$ is contained in one of the subshells
$(S^+,C^+)$ or $(S^-,C^-)$.
We can process these shells
recursively. This leads to the following procedure.

\bigskip

\noindent
{\bf Void-Shell Payment:}
Let $S$ be a void shell, and
$C$ its core. Apply Lemma \ref{CoreLemma}
to pay for every translation made in an atomic shell $T\con S$
where $T\cap C\ne \emptyset$.
Process the shells
$(S^+,C^+)$ and $(S^-,C^-)$ recursively.

\bigskip

It is clear that Void-Shell Payment
correctly pays for every unit translation
made in \algcall DismantlePath(Q)
for an atomic
shell contained in $S$.
We note two subtleties.
The collection of cores making  payments
need not form a partition of $S$ (e.g., if \algname ShellSearch.
is never executed for a shell with inner boundary
$S^-$).
Also a core $C$ need not have been an atomic shell
in \algcall DismantlePath(Q)
(since $C$ depends on the matching at the time of \algcall ShellSearch(R),
and we may have $C\pcon R$).

To complete the credit system we need only
show where the credits in Lemma \ref{CoreLemma}
come from.
Recall  $M$ is the matching on $R$
in the invocation  \algcall ShellSearch(R).
Let $F$ be the set of vertices
in $C$ that are free in $M$.
Apply \eqref{GoalCrossedEqn} to $C$
right after the dual adjustment
made in  \algcall ShellSearch(R):
\begin{equation}
\label{CoreEqn}
d(F,R+\widehat Q)+ CRI(C)+ CRB(SUB(C)) \le \c4 |C| +\tau(\U._C).
\end{equation}
Here $\U._C$ is the set of inherited blossoms properly contained in $C$
and not crossed by $M$.

We will use \eqref{CoreEqn} to change  right-hand side credits
into
left-hand side credits. To do this the $\tau(\U._C)$ term represents a
debt that must be paid.
So we must now  make two types of payments:
The original task, applying Lemma \ref{CoreLemma},
we call ``All-Blossom Payment''.
The new debt, to be paid by applying
Lemma \ref{NonVoidLemma}, we call
``Uncrossed-Blossom Payment''.
Note that the blossoms of $\U._C$ are
contained in $C$. As such they have not been reached in the payment process,
i.e., there are no credits or debits on them or their major paths,
aside from the initial distribution of $\c4 |X|\log |X|$ credits.
So we need not worry about duplicate credits or debts.
We allocate credits for the terms of inequality \eqref{CoreEqn} as follows.

On the right, the first term $\c4 |C|$ is paid 
from our initial distribution of $\c4 |X|\log |X|$ credits.
The $\tau$ term will be paid for as mentioned, as a new debt.

We convert the left-hand side to credits term by term.
For the first term consider each vertex $f\in F$.
Give
$d(f,R)$ credits to $f$'s occurrence in $C$. This
is the number of credits required for $f$
in Lemma \ref{CoreLemma} for All-Blossom Payment. 
In addition  give
$d(f,Q')$ credits to $f$'s occurrence in $Q'$
for each mpr
$Q'\pcon C$.
 This
is the number of credits required for $f$
in Lemma \ref{NonVoidLemma} for Uncrossed-Blossom Payment
of the new debt  $\tau(\U._C)$.

Next consider the term $CRI(C)$.
Recalling
$CRI(C)= \sum_{B\in INT(C)} |\delta_M(B)| z_0(B)/2$,
give $|\delta_M(B)| z_0(B)/2$ credits to each interior blossom
of $C$. This
is the number of credits required for $B$
in Lemma \ref{CoreLemma}  for All-Blossom Payment.
The conditions for All-Blossom Payment of $C$ are now satisfied.

Finally consider the third term $CRB(SUB(C))$.
Each crossed blossom $B$ properly contained in $C$ gets
$(|\delta_M(B)|-1)z_0(B)/2$ credits. This
is the number of credits required for $B$
in Lemma \ref{NonVoidLemma} for Uncrossed-Blossom Payment
of $\tau(\U._C)$.
The conditions for Uncrossed-Blossom Payment of $\tau(\U._C)$
in each mpr $Q'\pcon Q$ are now satisfied.

\bigskip

This completes the definition of the credit system:
Void-Shell Payment will apply the above procedure
to each core defined by the recursion.
Each such application spawns recursive applications of
Uncrossed-Blossom Payment within that core.
(These in turn spawn further Void-Shell and Uncrossed-Blossom Payments.)
No two of all these recursive processes charge the same shell
for credits.

\paragraph*{Comment on the recursion}
Lemma \ref{CoreLemma} can be applied to pay for all
translations in \algcall DismantlePath(Q). In fact this will be done in
Lemma \ref{PQEntriesLemma}. But this does not allow us to eliminate
Uncrossed-Blossom Payment. The following example illustrates why
Uncrossed-Blossom Payment is needed.

In the execution of \algcall DismantlePath(X), consider
a sequence of mpr's $Q_i, i=1,\ldots$, with each $Q_i$ properly contained in $Q_{i-1}$
and each major path $P(Q_i)$ containing various blossoms of $\U.$.
Our credit system pays for all these uncrossed blossoms using credits
obtained by
applying \eqref{GoalCrossedEqn}
to
$X$. If we use All-Blossom Payment to pay for each $Q_i$,
we use credits obtained by applying \eqref{GoalCrossedEqn}
to
each $Q_i$. This incurs a second charge for each $Q_i$, $i\ge 2$ (from the $\tau$ term
of \eqref{GoalCrossedEqn}). The two payments for $Q_2$ mean two more payments
for each $Q_i$, $i\ge  3$, etc.

\paragraph*{Bound for {\boldmath $\tau(\U.)$}}

To state the result formally
we enlarge the domain of the function $SUB(S)$ from Section \ref{dFSec}:
For an mpr $Q$,
\[SUB(Q) = \set {B} {\text{$B\con Q$}},\]
where again $B$ denotes an inherited blossom.
Note that unlike Section \ref{dFSec}, $P(Q)\con SUB(Q)$.

Consider an mpr $Q$. Let $M$ be the algorithm's matching
at some time 
after \algcall DismantlePath(Q) returns.
$M$ may cross blossoms of $Q$ arbitrarily.
Define
\[\U.= \set{B}{B\con  Q \text{ is inherited and uncrossed by $M$}}.\]
Let $F$ be the set of free vertices of $M$.
Assume $F\cap Q\ne \emptyset$ since otherwise
there is nothing to prove (every blossom contained in $Q$ is crossed
and so does not belong to any \U. set).

Let $Q$ have height $h$ in $Q+ \mmpr(Q)$, i.e.,
the longest chain of mprs
contained in $Q$ has $h$ mprs.
So $h=1$ means $Q$ is a minimal mpr, i.e., every
inherited blossom $B\con Q$ belongs to $P(Q)$.

\begin{lemma}
\label{CreditSystemLemma}
\begin{equation*}
   \tau(\U.)\le  d(F, Q+ \mmpr(Q)) +   CRB(SUB(Q))  +\c4 |Q|h.
  \end{equation*}
  Here the matching
 $M$ on $X$ is used to define $\U.$,
 $F$, and $CRB$.
\end{lemma}

\begin{proof}
The proof is by induction on $h$.
Let $Q'$ range over the maximal mprs properly contained in $Q$. 
A
blossom of \U. belongs to either $P(Q)$ or some family $SUB(Q')$.

View the upper bound  of the lemma as a collection of credits.
Using the identity $d(F,Q+ \mmpr(Q) )= d(F,Q)+ \sum_{f,Q'} d(f,Q'+\mmpr(Q'))$,
distribute the credits for each free vertex $f$ accordingly,
i.e., each
$f$ in $Q$ ($Q'$) gets
$d(f,Q)$ ($d(f,Q'+\mmpr(Q'))$) credits, respectively.
For each blossom $B$ contributing to  $CRB(SUB(Q))$
distribute the corresponding credits  $(|\delta_M(B)|-1)z_0(B)/2$
to either $Q$ if $B\in P(Q)$ or the mpr $Q'$ where $B\con Q'$.

Each mpr $Q'$ has height $\le h-1$.
The inductive hypothesis shows
\[\tau(\U.\cap SUB(Q'))\le  d(F\cap Q', Q'+ \mmpr(Q')) +   CRB(SUB(Q'))  +\c4 |Q'|(h-1).\]
We pay for the right-hand side using the credits on $Q'$
plus an additional charge of $\c4 |Q'|(h-1)$.
The total additional charge for all mpr's $Q'$ is at most
\begin{equation}
  \label{UNonemptyEqn}
  \sum_{Q'\cap F\ne \emptyset}\c4 |Q'|(h-1).
  \end{equation}
The restriction to $Q'$ sets containing a free vertex follows since
the opposite imples
$\U.\cap SUB(Q')=\emptyset$ so $\tau(\U.\cap SUB(Q')=0$.

It remains only to pay for the uncrossed blossoms of $P(Q)$.
Lemma \ref{NonVoidLemma}
Uncrossed-Blossom Payment
shows the distributed credits
pay for every translation of a blossom of $\U.\cap P(Q)$
except for the translations made in \algcall ShellSearch(T)
where $T$ ranges over 
$T\con S$, $S$ a void shell  of $P(Q)$.

We complete the payment for $\U.\cap P(Q)$ 
using Void-Shell Payment.
For each void shell $S$
this Payment forms a  collection of disjoint cores $C\con S$.
Consider the application of \eqref{CoreEqn} to such a core $C$.
We start by distributing credits of the left-hand side
to pay for
Lemma \ref{CoreLemma}  All-Blossom Payment
for $C$. (As discussed after \eqref{CoreEqn}
these credits come from terms $d(F,R)$ and $CRI(C)$.)
It remains to pay for the uncrossed blossoms in
each set  $SUB(C)$.

If $h=1$
$C$ contains no such blossoms ($Q$ is a minimal mpr.)
 The right-hand side of \eqref{CoreEqn} shows we need $\c4 |C|$ credits.
 Summing this over all cores (in every void shell $S$ of $Q$) gives a total charge of
 $\c4 |Q|$. This matches the term $\c4 |Q|h$ in the lemma.
This establishes the base case of the induction.
(There are no charges from \eqref{UNonemptyEqn}.)

Suppose $h>1$.
The uncrossed blossoms in $SUB(C)$
are contained in various maximal mprs $Q^c\pcon C$.
Consider such a $Q^c$.
The left-hand side
of  \eqref{CoreEqn} gives
$Q^c$
$d(F,Q^c+\mmpr(Q^c))$ credits from free vertices
and
$(|\delta_M(B)|-1)z_0(B)/2$ 
credits for blossoms $B\con Q^c$.
By induction any mpr $Q^c$ satisfies
\[\tau(\U._C\cap SUB(Q^c))\le
  d(F, Q^c+ \mmpr(Q^c)) +   CRB(SUB(Q^c))  +\c4 |Q^c|(h-1).\]
So we need an additional credits numbering
$\c4 |Q^c|(h-1)$ to pay for
$\U._C\cap Q^c$.
The right-hand side of \eqref{CoreEqn} shows we also need
$\c4 |C|$ more credits. The charge for
$C$ thus amounts to
\[\c4 |C|+ \sum_{Q^c\con C} \c4 |Q^c|(h-1)\le \c4 |C|h.\]
The total charge
for void shells $S$ of $P(Q)$ is at most
\begin{equation}
  \label{VoidEqn}
\sum_S \c4 |S|h,
\end{equation}
since the cores of each void $S$ are disjoint, and the void shells $S$
are themselves disjoint.

The total charge for $Q$ is the sum of
quantities \eqref{UNonemptyEqn} and \eqref{VoidEqn}.
The sets $Q'$ are 
nonvoid and so disjoint from the sets $S$ of \eqref{VoidEqn}.
So the total charge is at most
$\c4 |Q|h$ as required by the lemma.
\end{proof}

\vspace{-10pt}

\paragraph*{The {\sc Dismantler}'s time bound}
The lemma implies our final bound as follows.

Consider the execution of \algcall DismantlePath(X)
at any point in
Phase 1 or 2. Here $X$ is an arbitrary major path root,
possibly $X=V$.
Let
$M_X$ be the current matching on $X$. Let
$F_X$ be its set of free vertices where, if $X\ne V$, we
exclude $\omega$, a free vertex in the smallest possible 
blossom $B_\omega \in P(X)$. 

\begin{lemma}
\label{MainLemma}
$d(F_X,X)\le c|X|\log |X|$.
\end{lemma}

\begin{proof}
Let  $X$ have height $h$ in the set of major path roots.
  Let $S$ be a minimal even  shell containing a vertex of $F_X$.
  $S$ may even be a dissolved shell of $P(X)$.
Also $S$ may be $(B_\omega,\omega)$ if it contains a vertex of $F_X$.
  Let $F$ denote $F_X\cap S$.
\eqref{GoalCrossedEqn}
 gives
\begin{equation}
\label{goalAppliedEqn}
  d(F)+ CRI(S)+CRB(SUB(S)) \le c|S| + \tau(\U.).
\end{equation}
(We ignore interior blossoms of $S$.)
Apply Lemma \ref{CreditSystemLemma} to each
maximal mpr $Q'\pcon S $  and sum the inequalities to get
\begin{equation}
\label{FinalTauBBoundEqn}
\tau(\U.) \le
d(F,\mmpr(X))+CRB(SUB(S))
 + c\,|S|\,(h-1).
\end{equation}
Combining
\eqref{goalAppliedEqn} with
\eqref{FinalTauBBoundEqn}
(and recalling $d(F)= d(F,X)+d(F,\mmpr(X))$)
gives
\begin{equation}
\label{FandCRIBoundEqn}
d(F,X)+ CRI(S)\le c|S|+c|S|(h-1)\le c|S|h.
\end{equation}
Adding \eqref{FandCRIBoundEqn} for all
shells $S$ gives
$d(F_X,X)\le c|X|h$ proving the lemma.
\end{proof}

Let $F_p$ denote the subset of free vertices $F_X$
that are still active at the end of the $p$th pass of Phase 1.
The $p$ passes collectively perform $\ge p - \log |X|$
dual adjustments on every vertex of $F_p$.
So the lemma implies
\begin{equation}
\label{ProductInequalityEqn}  
  |F_p|(p-\log |X|) \le c|X|\log |X|.
\end{equation}

Define the parameter $\pi$ of Phase 1 (Fig.\ref{Phase1Fig}),
\begin{equation}
\label{piDefnEqn}
\pi=c\sqrt{|X|\log |X|} +1.
\end{equation}
Suppose pass $p$ is not the last pass of Phase 1.
Recalling vertex $\omega$ implies
$|F_p|>\pi-1= c\sqrt{|X|\log |X|}$. 
Thus
\eqref{ProductInequalityEqn}  
implies
$p< \sqrt{|X|\log |X|} +\log |X|\le 2\sqrt{|X|\log |X|}$.

So Phase 1 uses total time $O(\sqrt{|X|\log |X|} \,m(X))$.

Phase 2  begins with $\le \pi$ free vertices
in active shells. So Phase 2 finds $O(\sqrt{|X|\log |X|})$ augmenting paths.
As mentioned we implement Edmonds' algorithm using 
a straightforward bucket-based priority queue, say $PQ$, for
the search step events.
The algorithm steps through each time unit of $PQ$, invoking \algcall ShellSearch(S)
to
execute the Edmonds steps (if any) scheduled for that time.
The time for all Edmonds steps to find one augmenting path is
$O(m(X))$.
So these steps use total time $O(\sqrt{|X|\log |X|} \,m(X))$,
as desired. It remains to bound the number of time slots in
$PQ$. Since the \algname ShellSearch. of each time unit performs 2 unit translations,
this amounts to bounding the total number of unit translations
in Phase 2. The next lemma shows the number of time slots
is $O(|X|\lg |X|)$, so the overhead for $PQ$ is irrelevant.

\def\tb{\mathy{\o {\tau}}}

Let $S$ be a shell of $P(X)$.
For instance $S$ can be $(X,\emptyset)$. Define
\begin{eqnarray*}
\o\tau (S) 
&=&\text{the total number of unit translations made in Phase 1-2 searches}\\
  &&\text{of $P(X)$-shells $R\con S$.}
\end{eqnarray*}

\begin{lemma}
\label{PQEntriesLemma}
$\tb(X,\emptyset)\le |X|\log |X|$.
\end{lemma}

\begin{proof}
Let $S$ by an arbitrary shell of $P(X)$.
Let $R$ the last subshell of $S$ that is 
searched in Phase 1 or 2, and $C$ its core.
Let $F$ be the set of free vertices in $C$ when
$R$ is searched for the last time.
Lemma \ref{CoreLemma} translates to 
\begin{equation}
\tb(S)\le CRI(C) + d(F,R)+\tb(C^-,S^-)+\tb(S^+,C^+).
\end{equation}

Let $h$ be the height of $X$.
 \eqref{FandCRIBoundEqn}
applied to the core $C$ gives
$d(F,R)+ CRI(C)\le c|C|h$.
Hence we have the recurrence
$\tb(S)\le c|C|h +\tb(C^-,S^-)+\tb(S^+,C^+)$.
It
easily solves to $\tb(S)\le c||h$
by induction on $|S|$.
\end{proof}

We conclude
\algcall DismantlePath(X) uses
total time $O(\sqrt{|X|\log |X|}\, m(X))$.

\bigskip

The blossoms $X$ for mpr's with $2^{i-1}<|X|\le 2^{i}$
are vertex disjoint. So \algname DismantlePath. for
all of these blossoms
uses total time $O(\sqrt{2^ii}\; m)$. Summing over all $i\le \log n$
gives total time $O(\sqrt{n\log n }\;m)$ for each scale, as desired.

\begin{theorem}
\label{MainTheorem}
The scaling algorithm (using the \D. of Fig.\ref{ScaleAlgFig} for each scale)
finds a maximum weight perfect matching
in time $O(\sqrt{n\log n }\;m\, \log nW)$.\hfill$\Box$
\end{theorem}
\section{\boldmath {$f$}-factors}
\label{fFactorSec}
\def\fD.{$f$-{\sc Dismantler}}
\def\New{\bigskip \noindent *********************************}
\def\Ol.{\mathy{\Omega^{\rm e}}}
\def\Oe.{\mathy{\Omega^{\rm e}}}
\def\oe.{e-}

\begin{figure}[t]
\centering
\input{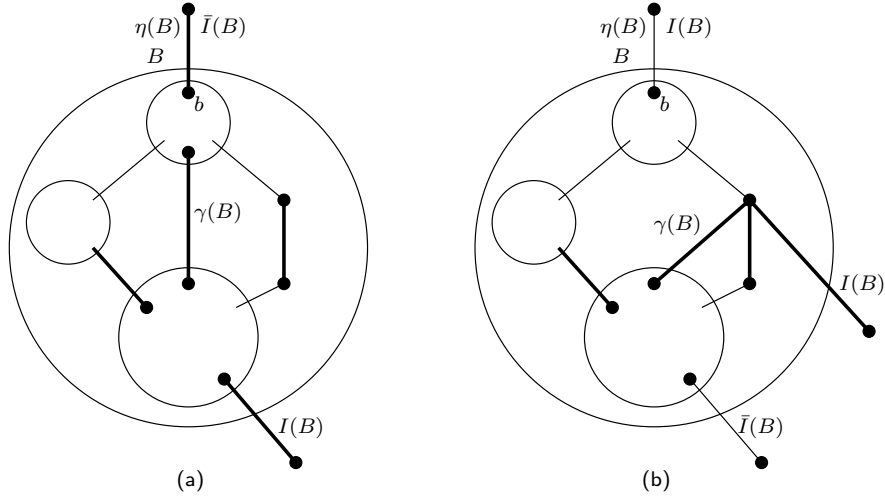}
\caption{An $f$-factor blossom $B$ with base vertex $b$. 
Edges of $I(B)\cup \gamma(B)$ include $z(B)$ in $\H{yz}(B)$.
Edges of $\delta(B)$
are marked as belonging to $I(B)$ or its complement $\bar I(B)$.
Edges of $\gamma(B)$ but not $B$'s blossom cycle are shown.
}
\label{fBlossomFig}
\end{figure}

The reader is assumed familiar with the $f$-factor algorithm
of \cite{G18}.
The cogent points are reviewed in Appendix \ref{fAppendix}.
Many details of the scaling algorithm are
identical to or obvious analogs of ordinary matching.
We postpone discussing those details until
the very end of Section \ref{ECSection}
(paragraph ``Remaining details $\ldots$'').
Instead this section is organized as follows.

Section \ref{DSection} presents the
difficulties that arise in adapting  the matching algorithm
to $f$-factors. It introduces  two mechanisms we use to treat
these difficulties.
Section \ref{ECSection}
gives the $f$-factor algorithm (Fig.\ref{fFactorAlgFig}) and the details of all its steps.

\subsection{Difficulties introduced by \boldmath {$f$}-factors}
\label{DSection}
To present the new issues we first give the important
definition that generalizes ordinary matching:

\paragraph*{Near optimality}
$f$-factors introduce the possibility
of matched edges that cannot be nearly dominated.
Let $M$ be the current matching.
The duals are {\em near-optimum} if edges $e$ not in
blossom subgraphs  satisfy
\begin{equation}
\H{yz}(e)
\begin{cases}
\ge w(e)-2&e\notin M\hskip 91pt\mbox{near domination}\\
\le w(e)&e\in M \hskip 91pt\mbox{near
tightness,}
\end{cases}
\end{equation}
and
edges $e$ in blossom subgraphs 
($e\in E(B)$, $B$ a blossom) satisfy
\[\H{yz}(e)\in \{w(e)-2, w(e)\}.\]
$e$ is called {\em undervalued}%
\footnote{The term ``undervalued''
  is the analog of ``strictly underrated''
  when optimum duals are used. See Appendix \ref{fAppendix}.}
if
\[\H{yz}(e)<w(e)-2.\]
Near domination implies an
undervalued edge must be matched.

The phase-dependent definition of eligibility
(\eqref{EligibleEqn}, etc.)
 is unchanged in the $f$-factor algorithm.


\paragraph*{The difficulties} 
In ordinary matching
unit translations maintain validity of
the dual variables. This fails for $f$-factors,
for a variety of reasons.
As an example
consider a blossom $B$ with
edge $uv\in I(B)$,
$u\in B$. A unit translation of $B$
decreases $\H{yz}(uv)$ by 1
($y(u)$ increases by 1 and $z(B)$ decreases by 2).
$\H{yz}(uv)$
may decrease even more: In an execution of
\algcall ShellSearch(S) where $S^+=B_u$,
$u$ may be an outer vertex, so $y(u)$ decreases by 1.
Such decreases may eventually destroy near dominance of $uv$.

Continuing, suppose  $uv$ does become undervalued, so it
must be matched. But $u$ or $v$ may already be completely matched, prohibiting
$uv$ getting matched. Continuing further, suppose $uv\notin I(B_v)$.
So unit translations of blossoms $B, v\in B\not \ni u$ increase 
$\H{yz}(uv)$ (as usual).
ShellSearches making $v$ inner give further increase.
The effect is that  $uv$ may oscillate between being
undervalued ($<w(uv)-2$, must be matched) and not even nearly tight
($>w(e)$, cannot be matched).

We avoid all these problems in several ways.

\paragraph*{Edge expansion}
We ``expand'' an edge $uv$ by replacing it with
the length 3 path shown in Fig. \ref{ExpandedEdgeFig}.
This leads to a modified notion of blossoms,
``e-blossoms'' (e.g., Fig. \ref{IEdgeExpansionFig}).
Expansion   breaks the linkage between change-of-dual-variable and
change-of-edge-status (dominance, undervalued, etc.).
We expand {\em $I$-edges}, i.e., edges $uv\in I(B_u)\cup I(B_v)$.
To prevent the graph from growing too big, we ``compress'' expanded edges
back to their original source  $uv\in E(G)$, when expansion is no longer needed.

Expansion by length 3 paths is used in various reductions of
general matching to ordinary matching \cite[Ch.32]{S}.
This replacement is used by Duan et al.
\cite{DHZ} to overcome problems of $I$-edges.
In more detail they use the ``blowup graph'', wherein every
original edge gets expanded. We cannot use this graph, since
it increases the size of an $f$-factor (and its blossoms)
up to $\Theta(m)$
rather than $\Theta(\Phi)$.
Also as mentioned, the replacements and associated e-blossoms
change from scale to scale.

Edge compression can create two new difficulties:
violations of standard blossom structure
(``ill-formed blossoms'') and
problematic structures (ineligible base edges).
Fig.\ref{ExceptionalSearchFig} of Appendix \ref{fAppendix}
illustrates how both these configurations can be introduced
by compression. We resolve the problem of ineligible base edges
in a general fashion, that may find other applications.
For that reason we begin by presenting our resolution, before proceeding to
our $f$-factor algorithm (Section \ref{ECSection}).

\paragraph*{Tightening base edges}
This difficulty is caused by 
base edges of blossoms that are ineligible.
(An ineligible  base edge that is matched is undervalued,
causing the problems illustrated above. An ineligble
base edge that is unmatched can invalidate the analysis of the algorithm.)
The base edge of a maximal blossom may be ineligible.
(This is the only possibility -- other base edges belong
to a blossom subgraph, and hence are eligible.)

\begin{figure}
\begin{center}
  \fbox{
    \setlength{\Efigwidth}{\textwidth}
    \addtolength{\Efigwidth}{-.3in}
\begin{minipage}{\Efigwidth}
\setlength{\parindent}{.2in}

\narrower{
\setlength{\parindent}{0pt}
\vspace{\vmargin}
\setlength{\parindent}{20pt}

\noindent

while $\exists$ a cycle $C$ of $\eta$-edges

{\hi
  
break $C$

}

while $\exists$ an ineligible $\eta$-edge  

{\hi
  let $e$ be a root of the forest of
  ineligible $\eta$-edges

tighten $e$

}

}

\vspace{\vmargin}

\end{minipage}
}
\caption{Algorithm to eliminate ineligible base edges.}
\label{TightenFig}
\end{center}
\end{figure}

We use the high-level procedure of Fig.\ref{TightenFig}
to eliminate base edges that are under- or or over-rated.
The procedure proves the following structural fact.
An {\em $\eta$-edge} is the base edge of some blossom
(recall $\eta(B)$ denotes
the base edge of  blossom $B$).

\begin{lemma}
  \label{EtaEligibleLemma}
  A maximum weight $f$-factor
  always has (optimum or near optimum)
  dual variables wherein every $\eta$-edge is eligible.
  \end{lemma}

The rest of this section gives the details of the procedure,
thereby proving the lemma. We also show the procedure uses $O(n)$ time.
For completeness we present the algorithm for both
optimal and near optimal duals, even though we only use the latter.
To do this define $t(e)$, the ``target'' value for eligibility,
to be $w(e)-2$ for an unmatched edge $e$ with near optimum duals,
and $w(e)$ otherwise (optimum duals, or $e$ matched with
near optimum duals).

For any $f$-factor with arbitrary 
duals and blossoms,  use the $\eta$-edges to
define a directed pseudoforest $P$ as follows.
Contract every maximal blossom.
Retain only their 
$\eta$-edges, each directed away from
its base vertex.
(All other edges are discarded.
An edge is bidirected if it is the base edge of both 
its end blossoms.)

Every contracted blossom of $P$ has outdegree 1
and every atom has outdegree 0.
So $P$ is a directed
pseudoforest, i.e., a collection of
connected components,
each consisting of

{\hi
{\narrower

  {\parindent=0pt
    
a subgraph $R$ plus zero or more out-trees,
each rooted at a vertex of $R$,

where $R$ is either
an atom, or a bidirected edge,
or a directed  cycle.

}}}

The algorithm uses unit translations to reduce $z$-values
and possibly dissolve blossoms. The first step modifies the
pseudoforest to have no cycles.
The second step does a top-down traversal of each remaining (acyclic)
component.
Each 
ineligible $\eta$-edge is either made eligible, or
its tail becomes atomic (so it is no longer an $\eta$-edge).

The first step executes the following procedure
on each connected component $C$ of $P$.

\bigskip

{\hi

while $C$ contains a directed cycle of contracted blossoms $R$

{\hi

  $\delta \gets \min \set {z(B)/2}{B \text{ a maximal blossom in $R$}}$

translate every maximal blossom of $R$ by $\delta$

}}

\bigskip

First observe that the cycle $R$ may change
from iteration to iteration. This is because
a blossom on $R$ may dissolve and get replaced by
a path of subblossoms and atoms.
The  processing of $C$ is complete the first time
$R$ contains an atom.
Note also that as vertices leave $R$, 
some
trees of $C$ may no longer be rooted in $R$.
This causes no problem.

For correctness we must show
each unit translation preserves validity of the duals.
There are two cases.
First consider an $\eta$-edge $uv$ on $R$.
It suffices to show $\H{yz}(uv)$ does not change.
Clearly $y(uv)$ increases by $1+1=2$.
So it suffices to show the $z$ terms contributing to  
$\H{yz}(uv)$ decrease by 2.
Let $A_u$ and $A_v$ be the maximal blossoms containing $u$ and $v$ respectively. It suffices to show $uv$ belongs to exactly one of the
sets $I(A_u), I(A_v)$.
Observe $uv = \eta(A_u)\ne\eta(A_v)$
(this follows since $R$ is a cycle, not a bidirected edge).
If $uv$ is matched then $uv \in I(A_v)-I(A_u)$, as desired.
If $uv$ is unmatched then $uv \in I(A_u)-I(A_v)$, as desired.
We conclude the duals are valid on the edges of $R$.

It remains to consider edges  $e\in \delta(R)$.
Suppose $e\in \delta(B)$, $B$ a maximal blossom of $R$.
If $e$ is matched
then $e\in I(B)$ (since $e\ne \eta(B)$). So
$\H{yz}(e)$ decreases by 
1, and the requirement $\H{yz}\le t(e)$ is preserved.
If $e$ is unmatched then $e\not\in I(B)$ (since $e\ne \eta(B)$). So
$\H{yz}(e)$ increases by 
1, and the requirement $\H{yz}\ge t(e)$
is preserved.

Since we 
use this algorithm (in the compression step, below), we note that
the total time for step one is $O(n)$. The idea is as follows.
We  compute $t$, the total of all $\delta$ quantities needed
to make $R$ contain an atom.
We do this
vertex by vertex rather than blossom by blossom. Specifically
let $uv$ be an $\eta$-edge initially on $R$.
We compute $t_u$, the total number of translations needed to make $u$
atomic. We also do this for every $\eta$-edge that enters
$R$ in a later iteration.
The smallest of all values $t_u$ is the desired quantity $t$.
Over all connected components, this algorithm
spends $O(1)$ time on each blossom $B$ of the graph.
So the total time is $O(n)$.

We turn to step two. Let $e=uv$ be the edge of step two. 
We tighten $e$ using
the following procedure.

\bigskip

{\hi

  let $e=uv=\eta(B)$ with $u=\beta(B)$
  
  while $u$ is not atomic and $uv$ is ineligible
  
{\hi

  $\delta \gets \min \set {z(B)/2, |\H{yz}(e)-t(e)|}
  {B \text{ the maximal blossom containing $u$}}$

translate $B$ by $\delta$

}}

\bigskip

Every unit translation decreases the distance from eligibility,
$|\H{yz}(e)-t(e)|$. This follows since
$e$ matched makes $\H{yz}(e)\le t(e)$ and $e\notin I(B)$
implies $\H{yz}$ increases by 1.
Symmetrically
$e$ unmatched makes $\H{yz}(e)\ge t(e)$ and $e\in I(B)$
implies $\H{yz}$ decreases by 1.
As in step one every unit translation maintains feasiblity
of the duals for edges  $f\in \delta(B)-\eta(B)$.

It is possible that the translation changes $f$ from eligible to
ineligible.
So we must show that $f$ has not been previously
tightened by step two.
To do this
let $P$ denote the pseudoforest immediately after step one.
It is easy to see that step two chooses $e$ according to a top-down
traversal of $P$. So the above $f$ has not been previously tightened.

We conclude that step two  is correct.
The time for step two in its entirety is $O(n)$ as before.
This concludes the proof of Lemma \ref{EtaEligibleLemma}.

\subsection{The algorithm: expansion and compression}
\label{ECSection}

\begin{figure}
\begin{center}
  \fbox{
    \setlength{\Efigwidth}{\textwidth}
    \addtolength{\Efigwidth}{-.3in}
\begin{minipage}{\Efigwidth}
\setlength{\parindent}{.2in}

\narrower{
\setlength{\parindent}{0pt}
\vspace{\vmargin}
\setlength{\parindent}{20pt}

\noindent

compress all expanded edges

{\hi dissolving ill-formed blossoms and tightening $\eta$-edges}

scale up the dual values

expand all $I(B)$ edges

{\hi matching every undervalued edge}

use the \fD. to dismantle all inherited blossoms 

\hi and find a nearly optimum $f$-factor  on $\o G$

}

\vspace{\vmargin}

\end{minipage}
}
\caption{Pseudocode for a scale of the $f$-factor algorithm.}
\label{fFactorAlgFig}
\end{center}
\end{figure}

The high-level algorithm for $f$-factors
is presented in Fig.\ref{fFactorAlgFig}.
The algorithm applies to any given multigraph $G$, loops allowed.
We call the generalization of the algorithm of Section
\ref{AlgSec} (Fig.\ref{ScaleAlgFig}) the \fD.. It is essentially identical
to the \D. (see paragraph ``Remaining details $\ldots$'').
As before we treat $V$ as a blossom (obviously $I(V)=\emptyset$).

Our result on $f$-factors is the following theorem.
The theorem is proved in this section and the next.
To give a broad overview we outline the proof below.

\begin{theorem}
\label{fMainTheorem}
The scaling algorithm (using 
the algorithm of Fig.\ref{fFactorAlgFig} for each scale)
finds a maximum weight $f$-factor
in time $O(\sqrt{\Phi\log \Phi}\;m\, \log \Phi W)$.
\end{theorem}

\noindent {\bf Proof Outline:}
Section \ref{fFactorSec} gives the details of the
steps preceding the \fD. in Fig.\ref{fFactorAlgFig}.
The total time  for
these steps is $O(m)$.

Section \ref{fdFOverallSec}
analyzes the \fD.:
Section \ref{PrepSec} 
establishes basic properties of the expanded graph $\o G$.
Then Section \ref{fdFSec}
follows the analysis of the \D. 
in Section \ref{dFSec}. It derives inequality \eqref{fGoalCrossedEqn},
the analog of \eqref{fGoalCrossedEqn}. This inequality is the basis of the credit system
of Section \ref{CreditSec}. That system applies to $f$-factors so it completes
the proof of the theorem.
\hfill$\Box$.

\bigskip

We start by giving the details of edge expansion.
In a graph with a matching the {\em M-type} of an edge is M or U,
depending on whether the edge is matched or unmatched
respectively.
For any edge $e$ we write $\H{yz}(e)= w(e) + \Delta(e)$.  So
$\Delta(e)=-u(e)$ when $e$ is a matched edge with
$\H{yz}(e)<w(e)$.

\begin{figure}[t]
\begin{center}
\mbox{\input{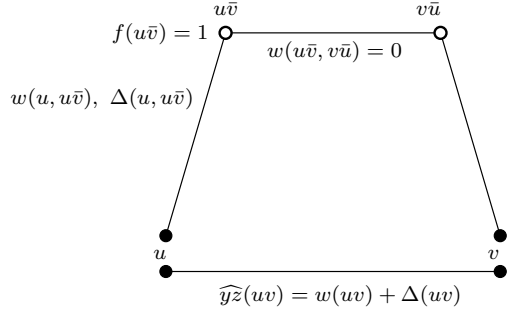}\hspace {70pt}}
\caption{Function values for the expansion of an edge $uv$.
e-vertices  are drawn hollow.
}
\label{ExpandedEdgeFig}
\end{center}
\end{figure}

An edge $uv$ that belongs to at least one set $I(B)$ gets expanded.
The expansion, illustrated in 
Figs.\ref{ExpandedEdgeFig}--\ref{IEdgeExpansionFig},
is defined in the following discussion.
We explicitly give 
the details for the new vertex $u\bar v$, and we rely on symmetry 
to infer the details for $v\bar u$.
These two vertices are called the {\em e-vertices}. So an expanded edge
has two $G$-vertices (its ends) and two e-vertices.
$\o G$ denotes the graph with all $I(B)$ edges expanded.

\paragraph*{Expansion: functional values}
Functional values are illustrated in
Fig.\ref{ExpandedEdgeFig}.
The two e-vertices have $f$-value 1. 
Where 
$f$-factors are concerned we use the natural correspondence
between an edge and its expansion:
$uv$ has the same M-type  as the two 
end edges $(u, u\bar v)$ and $(v\bar u,v)$
opposite M-type from  $(u\bar v, v\bar u)$.

Starting with the known values $w(uv)$ and
$\Delta(uv)$ 
we distribute them to the two end expansion edges via the definitions 
\begin{eqnarray*}
w(uv)&=&w(u, u\bar v)+w(v,v\bar u)\\
\Delta(uv)&=&\Delta(u, u\bar v)+\Delta(v,v\bar u).
\end{eqnarray*}
(Note $\Delta(uv)$ is defined in the figure. A value
$\Delta(uv)<0$ is equal to the linear programming dual
value $-u(uv)$.)
The exact values of the expansion edge quantities
(i.e., $w(u, u\bar v)$, etc.)
are flexible but required to be
even. 
This is always possible since
$w(uv)$ and $\Delta(uv)$ are both even,
and an even value $4a+b$, $b\in \{0,2\}$
can be expressed as $4a+b= 2a + (2a+b)$.
The only other constraints on $\Delta(\cdot)$ values
come from 
undervalued edges, which must be matched.
The definition of $\Delta(\cdot)$ for these edges will be given in
Fig.\ref{NearOptFig}.

The $f$-factors  on $G$ and $\o G$
correspond 1-to-1 with edge weights preserved.
Our algorithm will halt with a maximum weight
$f$-factor on the $\o G$ graph of the last scale. The
correspondence gives the desired $f$-factor on $G$. The dual
variables are near optimum on $\o G$. If optimum duals
on $G$ are desired they can be found using the techniques
from ordinary $f\equiv 1$ matching \cite{GT}.

The e-vertices have $y$ values defined so the end expansion 
edges have dual values analogous to $uv$. In other words  we want
\begin{equation}
\label{HExpansionEdgeEqn}
\H{yz}(u,u\bar v)=w(u, u\bar v)+\Delta(u,u\bar v). 
\end{equation}
This is easily achieved by defining $y(u\bar v)=w(u,u\bar v)
+\Delta(u,u\bar v)-(y(u)+ \sum \set{z(B)}{(u,u\bar v)\in B})$.
As required $y(u\bar v)$ is even.
$y(v\bar u)$ is treated symmetrically.
We show below that this makes
$(u\bar v, v\bar u)$ tight, i.e., 
$\H{yz}(u\bar v, v\bar u)=0$.

\paragraph*{Expansion: vertex placement}
Recall $B_u$ is the smallest blossom containing vertex $u$.
It always exists since we define $V$ as a blossom.
$B_{uv}$ is the smallest blossom containing both $u$ and $v$.
Again it aways exists.
Note that $uv$ gets expanded exactly when $uv\in I(B_u)\cup I(B_v)$. 

\begin{figure}[t]
\centering
\input{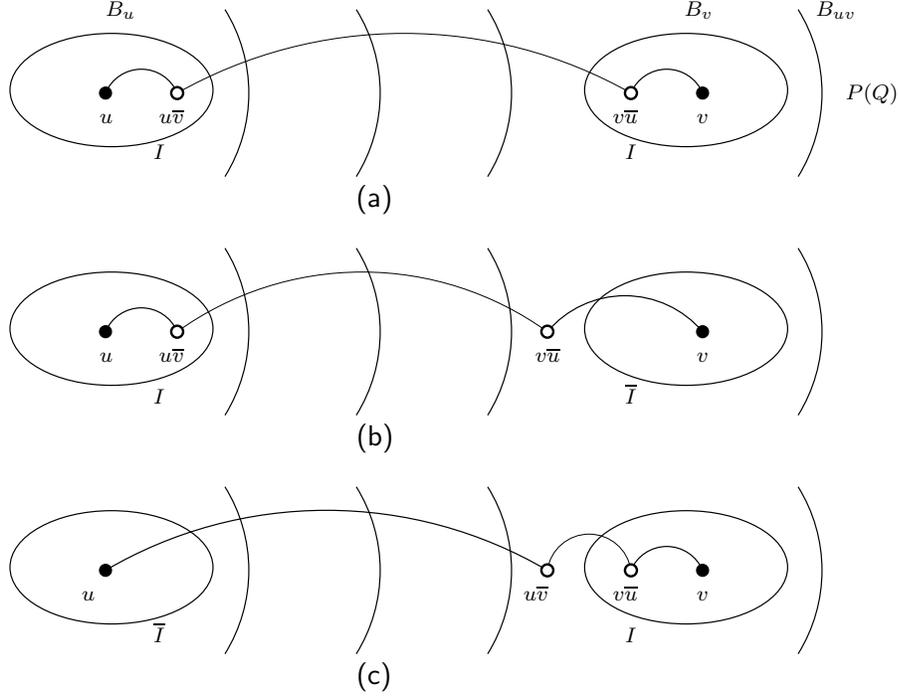}
\caption{Expansion of an edge $uv\in I(B_u)\cup I(B_v)$.
  $uv$ may be matched or unmatched.
  Blossoms of
$P(Q)$, where $B_{uv} \in P(Q)$,
are shown as circular arcs; the minimal blossoms
$B_u$ and $B_v$ are shown as ovals.
(a) $uv\in I(B_u)\cap I(B_v)$.
(b) $uv\in I(B_u)-I(B_v)$.
(c) $uv\in I(B_v)-I(B_u)$.
}
 \label{IEdgeExpansionFig}
\end{figure}

We define the location of  e-vertices in 
the laminar family of blossoms
by specifing $B_{u\bar v}$, the minimal blossom
containing $u\bar v$.
(There is no danger of interpreting this as the minimal blossom
containing $u$ and $\bar v$, since no vertex
$\bar v$ exists.) The following definition is
illustrated in Fig.\ref{IEdgeExpansionFig}:
\begin{equation*}
B_{u\bar v}=
\begin{cases}
B_u&uv\in I(B_u)\\
B_{uv}&uv\notin I(B_u)
\end{cases}
\end{equation*}
The same definition holds {\em mutatis mutandis}  for
$B_{v\bar u}$, i.e., it equals $B_v$ when $uv\in I(B_v)$.

Some special cases are worth noting.
When no blossoms of $P(Q)$ separate $u$ and $v$ (three circular arcs
disappear)
and $uv\in I(B_u)\oplus I(B_v)$,
Fig.\ref{IEdgeExpansionFig}(b) and  (c) are identical up to renaming.
(This case includes the possibility $B_{uv}=V$.)
Also throughout 
Fig.\ref{IEdgeExpansionFig}
$B_u$ or $B_v$ may
be a blossom of $P(Q)$ (the corresponding oval disappears).

The figure illustrates the following characterization of
e-vertices.

\begin{lemma}
  \label{ExpVertLemma}
  Consider a blossom $B$ in the expanded graph,  and an expanded edge $uv$.
  \begin{eqnarray*}
    u,v\notin B &\Longrightarrow& u \bar v, v\bar u \notin B\\
    u\in B \not\ni v &\Longrightarrow& v\bar u \notin B,\ 
    u \bar v \in B \mbox{\rm \ iff\ }    u v \in I(B)\\ 
        u,v\in B &\Longrightarrow& u \bar v, v\bar u \in B.
\end{eqnarray*}
  \end{lemma}

\begin{proof}
The definition of $B_{u\bar v}$ immediately gives
  \begin{eqnarray*}
\label{uBarvEqni}
\text{(a)}\hskip 10pt  u\bar v\in B_{uv}&&\\
\label{uBarvEqnii}
\text{(b)}\hskip 10pt u\bar v\in B\hskip 10pt & \Longrightarrow & u \in B
\end{eqnarray*}
  Consider the three cases of the lemma.

  $u,v\notin B$: Property (b) shows
  $u \bar v, v\bar u \notin B$.

  \bigskip
  
   $u\in B \not\ni v$: Property (b) again shows $v\bar u \notin B$.
Since $B_u\con B\pcon B_{uv}$ the definition of $B_{u\bar v}$ shows
$u \bar v \in B$ if     $u v \in I(B_u)$
and $u \bar v \not\in B$ if  $u v \not\in I(B_u)$.
So $u \bar v \in B$ iff     $u v \in I(B_u)$.

Since $uv$ leaves $B$ it leaves $B_u$.
The definition of $I(B)$ shows
$u v \in I(B_u)$
iff
$u v \in I(B)$.
(We use the fact that
$uv=\eta(B_u)$ iff $uv=\eta(B)$.)

\bigskip

$u,v\in B$: $B_{uv}\con B$ and property (a)
gives
$u \bar v, v\bar u \in B$.
  \end{proof}

\begin{figure}[t]
\centering
\input{EBlossom.pstex_t}
\caption{A heavy e-blossom $B$, and shell $S$ with outer boundary $B$.
  Edges in $C(B)$ and their expansions are solid. Other
  expanded edges are dashed.}
 \label{EBlossomFig}
\end{figure}
 
Define
\[\Oe. = \text{ the family of all inherited blossoms, as modified by edge 
expansion.}
\]
We use the term {\em \oe.blossom} to refer to a member of \Oe..
When necessary we use the term {\em $\Omega$-blossom} to refer
to the $f$-factor blossoms defined in \cite{G18}.
So the \fD. algorithm dismantles e-blossoms and constructs $\Omega$-blossoms.
As illustrated in Fig.\ref{EBlossomFig}
e-blossoms have slightly different structure than $\Omega$-blossoms:
There may be e-vertices in the closed path
defining the blossom and as well as e-vertices not on that path.
Like  blossoms for the case $f\equiv 1$,
\oe.blossoms do not have 
$I(B)$ sets. 
Like all blossoms \Oe. is
a laminar family.
Its blossom tree is used to determine mprs and hence the 
invocations \algcall DismantlePath(Q).

\paragraph*{Analysis of expansion duals}
The goal of this section is to show the expanded edge
$(u \bar v, v\bar u)$ is tight.

Our approach to analyzing dual variables in expansion and compression
is based on a "compression quantity" $C$ and "expansion quantity" $E$,
\begin{eqnarray*}
C&=&\H{yz}(uv)\\
E&=&\H{yz}(u,u\bar v)
+\H{yz}(v,v\bar u)
-\H{yz}(u\bar v,v\bar u)
\end{eqnarray*}
When $uv$ gets expanded we start with the known quantity $C$
and analyze the new quantity $E$; vice versa for compression.
$C$ and $E$ are closely related. For instance any
{\em arbitrary} values of the four $y$ terms make equal contributions to $C$ 
and $E$,
\[y(u)+y(v)=
(y(u)+y(u\bar v))+
(y(v)+y(v\bar u))
-(y(u\bar v)+y(v\bar u)).\]
We now show 
the $z$ terms also make identical contributions on expansion.
(For compression the contributions needn't be identical 
but 
they relate in the proper fashion, see below.)
\begin{lemma}
An expanded edge $e$ has $C=E$.
\end{lemma}

\begin{proof}
An e-blossom $B$  contributes $z(B)$ to $\H{yz}(e)$
iff $e\in\gamma(B)$. We apply this to the three expanded edges as follows:

\[(u,u\bar v)\in \gamma(B)
\Longleftrightarrow
uv \in \gamma(B) \text { or $u\in B$ and $uv \in I(B)$.}\]
This follows by observing the equivalence holds
in each of the three cases of Lemma \ref{ExpVertLemma}.
Obviously the similar characterization holds for 
the contribution of $(v,v\bar u)$.
Finally
\begin{equation*}
(u\bar v, v\bar u)\in \gamma(B) \iff uv \in \gamma(B).
\end{equation*}
Again this follows by observing the equivalence holds
in each case of Lemma \ref{ExpVertLemma}.

We have shown the net contribution of $B$ 
to $E$ when $uv\in \gamma(B)$ is 
$z(B)+z(B)-z(B)=z(B)$. 

So the three expanded edges contribute exactly $z(B)$ to $E$ when
$uv\in \gamma(B)\cup I(B)$. This exactly matches $B$'s contribution to
$C=\H{yz}(uv)$.
We conclude $C=E$.
\end{proof}

The lemma implies
$(u\bar v, v\bar u)$ is tight:
Substituting for $\H{yz}$ in the equation $C=E$ gives
\begin{eqnarray*}
w(uv)+\Delta(uv)&=&\H{yz}(u,u\bar v)+ \H{yz}(v,u\bar u)- \H{yz}(u\bar v, v\bar u)
= w(uv)+\Delta(uv)- \H{yz}(u\bar v, v\bar u)\\
0&=&\H{yz}(u\bar v, v\bar u).
\end{eqnarray*}

\paragraph*{Edge compression}
Compression begins by
replacing every expansion by its source.
We present that step, first
stating it and then analyzing it, similar to the previous sections.
After that we give the two post-processing steps listed in Fig.\ref{fFactorAlgFig}.

We set the stage by reviewing the structure of the $f$-factor
on $\o G$ returned by the \fD..
Let $B$ be a blossom ($B \in \Omega$) 
and let $uv$ be an edge whose expansion is in $\o G$. Then
\begin{equation}
  \label{oGBlossomStructureEqn}
  u\bar v \in B \imp u, u\bar v, v \bar u, v \in B.
\end{equation}
This follows since in general,
every vertex in a blossom $B$ 
is on at least two edges of the subgraph of $B$.
(In detail, $B$'s subgraph contains 
the edges of the cycle $C(B)$ that forms $B$, plus
the edges in the subgraphs of the
contracted blossoms  on $C(B)$.)

The first  compression step replaces
each edge expansion $(u, u\bar v,  v\bar u,v)$ by $uv$.
The end  edges $(u, u\bar v), (v\bar u,v)$ 
have the same M-type as $uv$, and $(u\bar v,  v\bar u)$
has opposite M-type.
In addition consider a blossom $B$ where
\[u\in B \not\ni u\bar v, v\bar u.\]
({\em Mutatis mutandis} for $v$ but not $v\bar u$.)
We define
\begin{equation}
  \label{CompressStepOneEqn}
\begin{cases}
uv\in I(B)& v\notin B,\; (u,u\bar v)\in I(B)\\ 
uv=\eta(B)& v\notin B,\; (u, u\bar v)=\eta(B)\\
\eta(B)=\emptyset & v\in B,\; (u, u\bar v)=\eta(B).
\end{cases}
\end{equation}
Note the first two cases are consistent with each other:
$uv$ and $(u,u\bar v)$ have the same M-type,
so the defining property $\eta(B) \in I(B)$ iff $\eta(B)\notin M$
is maintained.
The last case makes $B$ (and possibly other blossoms)  {\em ill-formed}.
The post-processing  remedies this. But first we analyze how compression
affects the
dual variables.

\begin{lemma}
Compressing the expansion of edge $uv$ makes
\begin{equation*}
\begin{cases}
C\le E \text{ if $uv$ gets matched}\\ 
C\ge E \text{ if $uv$ gets unmatched.} 
\end{cases}
\end{equation*}
In addition $C=E$ if $uv$ becomes a blossom subgraph edge
(i.e., the expanded edges are contained in $E(B_{uv})$).
\end{lemma}

\begin{proof} \eqref{oGBlossomStructureEqn}
implies that
$\{u,u\bar v ,v \bar u,v\}\cap B$
is either
\[\{u,u\bar v ,v \bar u,v\}, \{u,v\}, \text{ or }\{u\}\]
where the last case holds modulo a renaming.
We consider the corresponding three cases. We will see
the first two
have equal contributions to $C$ and $E$.

\case  {$\{u,u\bar v ,v \bar u,v\}\con B$}
$B$ contributes to all three expansion edges, so the net contribution to
$E$ is $z(B)+z(B)-z(B)=z(B)$. So $B$ makes the same contribution to
$C$ and $E$. 

If the compression makes  $uv$ a blossom subgraph edge,
this is the only case that applies (although many blossoms $B$ are possible).
Thus $C=E$ when $uv$ becomes a blossom edge.  

\case{$\{u,u\bar v ,v \bar u,v\}\cap B=\{u\}$}
$B$ contributes to $E$ iff 
$(u, u\bar v) \in I(B)$.
We define $uv$ to be $\eta(B)$ iff
$(u, u\bar v) =\eta(B)$.
This is valid since $uv$ is matched iff
$(u, u\bar v)$ is matched.
(Also $uv\in \delta(B)$.)
Now
$(u, u\bar v) \in I(B)$ iff 
$uv\in I(B)$.
So $B$ contributes to $E$ iff it contributes to $C$,
and in both cases the contribution is $z(B)$.
(Note this argument covers 4 cases,
$(u, u\bar v)$ can be matched or unmatched,
as well as equal or unequal to
$\eta(B)$.)

\case {$\{u,u\bar v ,v \bar u,v\}\cap B=\{u,v\}$}
$uv\in \gamma(B)$ so $B$ contributes $z(B)$ to $C$.

Suppose $(u,u\bar v)$ and  $(v,v \bar u)$ are both matched. 
One of these edges
must belong to $I(B)$,
the other 
belongs to $I(B)$ iff it is not $\eta(B)$.
$B$ contributes $2z(B)$ to $E$ in the first case,
$z(B)$ in the second.
In both cases the contribution to $E$ is $\ge $ the contribution to
$C$, as desired.

Suppose $(u,u\bar v)$ and  $(v,v \bar u)$ are both unmatched.
If one of these edges is $\eta(B)$, 
$B$ contributes $z(B)$ to $E$.
If neither is $\eta(B)$ the contribution is 0.
So the contribution to $E$ is $\le $ the contribution to
$C$, as desired.
\end{proof}

At the end of the edge compression step 
let $M$ be the $f$-factor (i.e., edges $uv$ that got compressed 
are matched or unmatched according to their expansions).
Every edge $e$ now belongs to $G$ and satisfies
\begin{equation}
  \label{yzCompressedEqn}
\H{yz}(e) = w(e)+\Delta(e) \text{ where } \Delta(e) \, 
\begin{cases}
\ge -4 & e \notin M\\
\le 2& e \in M\\
\in [-4,2]&  e \in \cup \set{E(B)}{B \text{ a blossom}}.
\end{cases}
\end{equation}
Note in the third condition
$B$ is
an $\Omega$-blossom, not an e-blossom (there are no e-blossoms
at the end of the edge compression step).

In proof, these inequalities are obviously weak if $e$ 
does not result from a compression.
In the opposite case let
$e=uv$ with corresponding e-vertices
$u \bar v$, $v \bar u$.
Recall $w(uv) =w(u \bar v)+w(v \bar u)$
and $w(u \bar v,v \bar u)=0$.
An unmatched edge 
$e$ has $C \ge E$ and  $(u \bar v,v \bar u)$ matched, so
\begin{equation*}
\H{yz}(uv) \ge (w(u \bar v)-2)+(w(v \bar u)-2)-0 =w(uv)-4
\end{equation*}
(using near domination and near tightness).
Similarly a matched edge $e$ has $C \le E$ and
$(u \bar v,v \bar u)$ unmatched, so
\begin{equation*}
\H{yz}(uv) \le w(u \bar v)+w(v \bar u) -(-2)=w(uv)+2.
\end{equation*}
Finally if $e$ is a blossom subgraph edge,
the above inequalities  all hold, since
before compression every edge $e'$ in a blossom subgraph has
$\Delta(e')\in [-2,0]$. 


\paragraph*{Compression: ill-formed blossoms}
We will eliminate blossoms with $\eta(B)=\emptyset$ after compression.
First note this possibility can actually occur in the algorithm:
Fig. \ref{ExceptionalSearchFig} illustrates how such a blossom $B$
can be created 
(in an invocation \algcall DismantlePath(Q) for arbitrary $Q$): In part (a),
assume the $u\bar v$ and $v\bar u$ are the two expansion vertices.
In parts (b) and (c)
$z(B)>0$ and the compression operation makes $\eta(B)=\emptyset$.

As shown in the lemma's proof, $\eta(B)=\emptyset$ when
\[u,v\in B \not\ni u\bar v, v\bar u
\text{ and } \eta(B)=(u, u\bar v).\]
(The fact that 
$v\bar u \not \in B$ follows since we are in the lemma's third case.)

The preceding scale
ended with
$B$ a maximal blossom in $\o G$. In proof, suppose for contradiction that
$B$ is a maximal subblossom of some blossom $A\ne V$. 
The cycle defining $A$ contains $B$, so it must 
consist of $B$ and the three expanded edges. $\eta(A)$ is an edge leaving $A$, 
but there is no such edge, contradiction.

The algorithm 
dissolves $B$, i.e., 
\begin{equation*}
\begin{array}{lll}
y(v) &\gets y(v)+z(B)/2&\forall v\in B\\
z(B)&\gets 0.
\end{array}
\end{equation*}
This change preserves inequalities
\eqref{yzCompressedEqn}.
In proof,
consider an edge with one or both vertices in $B$.
There are three possibilities:

$e\in \gamma(B)$: $\H{yz}(e)$ is preserved.

$e\in \delta(B)\cap M$:
$e\ne \eta(B)$ since $\eta(B)=\emptyset$.
So $\H{yz}(e)$ decreases. $e$ 
may become an undervalued matched edge.

$e\in \delta(B)-M$:
$e\ne\eta(B)$ since $\eta(B)=\emptyset$.
$\H{yz}(e)$ increases, so $e$ remains dominated.

\bigskip

It remains to verify the correctness of
the other change to  $\eta(B)$ in compression, i.e., the middle case
of  \eqref{CompressStepOneEqn} which assigns
$\eta(B)=uv$ when
$(u, u\bar v)=\eta(B)$ and
$ v\notin B$.

Let $B$ be the maximal blossom with base $u\bar v$. Let $B'$ be
the minimal blossom containing $B$. So $u \bar v\in B'-B$.
\eqref{oGBlossomStructureEqn} implies
$v \bar u, v\in B'-B$ (we are also using the assumption  $v\notin B$).
The cycle $C(B')$ forming blossom $B'$
originally 
contained the subpath $u, u\bar v,  v \bar u, v$, which gets transformed
to the subpath $uv=\eta(B)$. So $C(B')$ remains a valid closed path forming $B'$.

We conclude $B'$ and 
all other blossoms on $G$ are valid after edge compression.
(Recall there are no \oe.blossoms,
they all dissolve in the \fD..) So the final blossom structure is valid.
Thus
the compression step ends with a valid structured
matching on $G$ (having started with one on $\o G$ of the previous scale).

\paragraph*{Compression: tightening $\eta$-edges}
We make all $\eta$-edges eligible by executing the
procedure of Lemma \ref{EtaEligibleLemma}.
(Regarding step one of that procedure, 
the example of Fig. \ref{ExceptionalSearchFig} 
easily extends to show the 
compression step can create cycles of $\eta$-edges.)
Let $ET$ be the set of all $\eta$-edges.
After this step we can extend the third case of
\eqref{yzCompressedEqn} to
\begin{equation*}
\Delta(e) 
\in [-4,2]\hskip20pt  e \in \cup \set{E(B)}{B \text{ a blossom}}
\cup ET.
\end{equation*}

\paragraph*{Scaling up}
The $i$th scale starts by scaling up:
\begin{equation*}
\begin{array}{lll}
w(e)&\gets 2(w(e)+ \text{ the $i$th leading bit of } \overline w(e))&\forall e\in E\\
y(v) &\gets 2y(v)+4&\forall v\in V\\
z(B)&\gets 2z(B)&\forall \text{ blossom } B.
\end{array}
\end{equation*}
As before the updated dual functions are denoted $y_0$ and $z_0$.
Note that as previously claimed $\Delta(e)=\H{y_0z_0}(e)-w(e)$ is even.

Let $M_0$ be the matching after edge compression. ($M_0=\emptyset$
when we are starting the first scale.)
Let $e$ be any edge in the compressed graph.
Comparing its  new weight $w(e)$ with
its weight in the previous scale, denoted $w^-(e)$,
\begin{equation}
\label{ScaledWeightsEqn}
2w^-(e)\le w(e)\le 2w^-(e)+2.
\end{equation}
Combining \eqref{ScaledWeightsEqn} with
\eqref{yzCompressedEqn} and its extension to $ET$
gives
\begin{equation}
  \label{yzScaledEqn}
\H{y_0z_0}(e)=w(e)+ \Delta_0(e),
\text{ where }\Delta_0(e)
\begin{cases}
 \ge -2&e\notin M_0\\
 \le 12&e\in M_0\\
 \in [-2,12]& e \in \cup \set{E(B)}{B \text{ a blossom}} \cup ET.
\end{cases}
\end{equation}

\paragraph*{Expansion: final details}
\eqref{yzScaledEqn} continues to hold when edges are expanded.
To be precise we must
translate
various notions from $G$ to the expanded graph $\o G$. In \eqref{yzScaledEqn}
  ``$M_0$'' denotes the expanded matching. 
A ``blossom'' is an e-blossom $B$.
Its blossom subgraph $E(B)$ is the $\o G$-image of $E(A)$,
for $A$ 
the $\Omega$-blossom that expands to $B$. (For example in Fig.\ref{EBlossomFig}
$E(B)$ excludes the dashed edges.)
Similarly $ET$ is the $\o G$-image of that set in $G$.
(So for an expanded $\eta$-edge, all three of its expansion edges
belong to $ET$.)

\begin{proposition}
  \label{yzScaledProp}
  The bound of \eqref{yzScaledEqn}, interpreted for $\o G$ as above,
  holds after edge expansion, for any distribution of $\Delta(uv)$
  into $\Delta(u,\bar v)$ and $\Delta(v,\bar u)$. 
\end{proposition}

\begin{proof}
Consider an edge expansion.
$(u \bar v, v \bar u)$ is always tight. Edges
$(u, u \bar v)$ and $(v, v \bar u)$ have the same M-type as
$uv$, and the quantities
$\Delta(u, u \bar v)$ and $\Delta(v, v \bar u)$ are
of smaller magnitude
than $\Delta(uv)$.
\end{proof}

The upper bound on $M_0$ edges will be used in the analysis of the algorithm.
We put it aside until Section \ref{fdFSec}.
We turn to the task of establishing near optimality of $y_0,z_o$.
Unlike ordinary  $f\equiv 1$ matching, the Scaling Up step does not
guarantee near optimality, because 
undervalued edges must be matched.
We use the procedure of Fig.\ref{NearOptFig}.

\begin{figure}[t]
\begin{center}
\fbox{
\begin{minipage}{\figwidth}
\setlength{\parindent}{.2in}
\narrower{
\setlength{\parindent}{20pt}
\vspace{\vmargin}
\noindent
For every undervalued edge $uv$

{\hi

  if 
    $uv$ joins two atoms of a shell, i.e.,
    $B_u=B_v=B_{uv}$

  {\hi
    match $uv$}

  else

  {\hi
    wlog assume $u$ is in a blossom $B$

    $/$ $uv\in I(B)$ has been expanded $*/$

    $\Delta(u,u\bar v) \gets  \Delta(uv)$,
    $\Delta(v,v\bar u) \gets 0$

    match $(u,u\bar v)$

  }
}

\vspace{\vmargin}
}
\end{minipage}
  }

\caption{Edge expansion procedure to establish
  near optimality.}
\label{NearOptFig}
\end{center}
\end{figure}

For correctness consider
an undervalued edge $uv$.
Clearly in blossom $B_{uv}$, either $u$ and $v$ are both atomic
or at least one of them, say $u$, is in a blossom $B\pcon B_{uv}$.
Since $uv$ is matched in $M_0$ it belongs to $I(B)$. Thus
$u \bar v$ is a vertex in $B_u$.
The algorithm makes it undervalued and matches it.
We conclude the \fD. begins with feasible duals.

In addition
no edge matched by the algorithm crosses a shell boundary.
So we avoid the problem of undervalued edges detailed in the Difficulties
section. The problem of unmatched $\eta$-edges that are
ineligible has already been avoided by Lemma \ref{EtaEligibleLemma}.

\paragraph*{Remaining details of scaling and the \fD.}
The algorithm operates in $\Theta(\log \Phi\wh W)$ scales.
We give the justification, which  is entirely 
analogous to ordinary matching.

Let $\wh w$ be the given weight function,
$\wh W= \max \; \wh w$, and $\wh W_f$ the maximum given
weight of an $f$-factor.
Let $a$ be a parameter to be determined.
The algorithm replaces $\wh w$ by weight function $\o w=aw$.
There are $s=\f{\log a \wh W}$ scales.
For $i=1,\ldots, s$, the $i$th scale
finds a near-optimum structured $f$-factor
for the  weight function 
\[w(e)=2\times(\text{the leading $i$ bits of }\overline w(e)).\]
It is easy to see the last scale uses weights $\ge a\wh w$.

In each scale
let $\o f$ be the degree constraint function of the expanded graph
$\o G$, and
let $W_f$ be the maximum weight of an $\o f$-factor
(using the scale's weights $w$).
A near optimum 
$\o f$-factor 
weighs $\ge W_f-\o f(V)$.
(The proof is the same as \cite[Lemma 2.1]{GT}, \cite[Lemma 2.2]{DPS},
as well as \cite{DHZ}.)
Let
the $f$-factor of the last scale weigh $a W$,
so $aW\ge a \wh W_f-\o f(V)$.
If $a>\o f(V)$ we get $aW> a\wh W_f -a$. This implies
$W=\wh W_f$, so the last scale gives a maximum weight
$f$-factor on the expanded graph $\o G$, as desired.
We will prove the expanded graph has $\o f(V)\le 3f(V)$
(see \eqref{oGSizeEqn}).
So we take
$a=3f(V)+1$.
Thus the number of scales is $\Theta(\log \Phi\wh W)$ as desired.

\bigskip

Next consider the \fD.. It is essentially the same as
the \D.. It uses the $f$-factor algorithm of \cite{G18}.
The only other change is to interpret 
quantities appropriately for the new context.
This occurs in two places, as follows.
Note that in light of \eqref{oGSizeEqn}, for convenience we
omit overlines, e.g., writing $f$ instead of $\o f$.

The first change is in the definition of major path.
We use the blossom tree of
\oe.blossoms to define the major paths and associated notions.
To choose the major child, each node $B$ has size $f(B)$ rather than
$|B|$.
(Recall from the start of Section \ref{CreditSec}
that the size of a vertex set like the mpr $X$
always gets replaced by $f(X)$.)

The second change is interpretation of the termination test in
Phases 1 and 2.
In both tests
each free vertex is counted according to
its deficiency, i.e., a vertex $v$ in an atomic shell contributes
$f(v)-deg(v)$ to the count, for $deg(v)$ the degree of $v$ in the current matching.
In Phase 1
the definition of $\pi$, 
\eqref{piDefnEqn}, becomes
$\pi=c\sqrt{f(X)\log f(X)} +1$.
In Phase 2, the termination condition requires
that
the matching on the atomic shells has
total deficiency $\le 1$.

\bigskip

With these changes Lemma \ref{MainLemma} becomes
$d(F_X,X)\le cf(X)\log f(X)$ for $f$-factors.
We conclude
\algcall DismantlePath(X) uses
total time $O(\sqrt{f(X)\log f(X)}\, m(X))$.

As in ordinary matching, using
major paths gives the desired time bound for the \fD..
We reiterate the timing argument, which is essentially that 
given at the very end of
Section \ref{CreditSec}.
As in that section
the time for \algcall DismantlePath(Q)
is $O(\sqrt {f(Q)\log f(Q)}\;  m(Q))$.
The mpr's $Q$ with $2^{i-1}<\f(Q)\le 2^{i}$
are vertex disjoint. So the total time
for invocations 
of \algname  DismantlePath. on these mpr's is
$O(\sqrt{2^ii}\;  m)$.
Summing over all $i\le \log  f(V)$
gives total time
$O(\sqrt { f(V)\log  f(V)}\; m)$.
\eqref{oGSizeEqn} shows $f$ and $m$ in
$\o G$ and $G$ are
the same 
to within constant factors.
Using $\Phi=\sum_v f(v)/2$
from Section \ref{IntroSec} the time
for the \fD. is
$O(\sqrt {\Phi\log \Phi}\;m)$.

\section{Objective reducers for $f$-factors}
\def\oM.{\mathy{\o M}}
\label{fdFOverallSec}
The fundamental inequality is essentially unchanged from ordinary matching
(\eqref{fGoalEqn} and \eqref{fGoalCrossedEqn}).
It is derived by the same sequence of steps.
There are two basic differences: using $f$ for degree bounds rather than
degree bound 1, and using
\oe.blossoms
(and the family \Ol.)
instead of standard blossoms
(and the family $\Omega^-$).

\subsection{Preparation for the analysis}
\label{PrepSec}
\fD. is executed on the expanded graph $\o G$.
We start by showing
$\o G$ has size similar to $G$:
\begin{eqnarray}
  \bar n &\le& n+2\min\{f(V),m\},\notag\\
\bar m &\le& m+2\min\{f(V),m\},
\label{oGSizeEqn}\\
\bar f(V) &\le& 3f(V).\notag
\end{eqnarray}
Each expanded edge adds 2 vertices, 2 edges, and 2 units of $f$.
So it suffices to show $|I(B)|\le \min\{f(V),m\}$.
$|I(B)|\le m$ is trivial (note this bound is not helpful in bounding $f(V)$,
since the import of our algorithm's time bound is for $f(V)<<m$),
We  complete the proof by showing
\begin{equation*}
\sum_{B\in \Omega^-} 2|I(B)|\le n+f(V)\le 2f(V).
\end{equation*}
Recall $\Omega^-$ is the family of inherited blossoms
in graph $G$. Note  the second inequality is obvious.

Let $M$ be the $f$-factor from the preceding scale.
Clearly $\sum_{B\in \Omega^-} 2|I(B)\cap M|\le 2|M|=f(V)$.
So it suffices to show
$\sum_{B\in \Omega^-} 2|I(B)-M|\le n$.
An unmatched edge in $I(B)$ must be $\eta(B)$.
So the above sum is  $\le 2|\Omega^-|$.
There are $\le n/2$ inherited blossoms of $G$
(they form a laminar family on $V$, each blossom containing
$\ge 3$ maximal subblossoms). Thus $2|\Omega^-|\le n$, as desired.

\bigskip

\begin{figure}[t]
\centering
\input{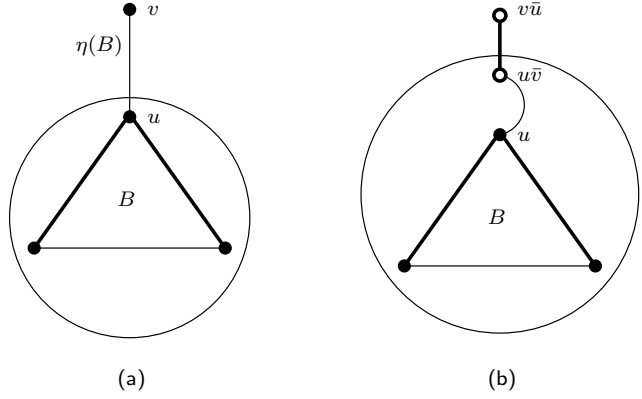}
\caption{Expanding makes $f(B)$ odd.
(a) Heavy blossom in $G$, $f(B)=4$. (b)
Expanding makes $B$ a blossom of $\o G$ with $f(B)=5$.
}
\label{OddBlossomFig}
\end{figure}

The next several lemmas show that e-blossoms are very much like
ordinary matching blossoms. This allows our credit system
to extend, essentially unchanged,  to $f$-factors.

Ordinary matching blossoms have $|B|$ odd. Analogously $f$-factor
blossoms should have $f(B)$ odd. Fig. \ref{OddBlossomFig} illustrates how this 
can fail
for a blossom in the original graph $G$ but it holds after expansion.
The next lemma shows this holds in general.
Consider
the expansion of the $f$-factor of the previous scale
and its matching (in $\o G$) which we denote as \oM..

\begin{lemma}
  \label{lBlossomLemma}
Any \oe.blossom $B\ne V$ has
$|\delta_\oM.(B)|=1$.
The matched edge leaving $B$
is either the $G$-edge $\eta(B)$ or an edge of its expansion.
\end{lemma}

\begin{proof}
  Consider a $G$-edge $uv$ with $u\in B\not\ni v$.
  (Lemma \ref{ExpVertLemma} shows the other possibilities are irrelevant.)
First assume $uv=\eta(B)$.
We will show $uv$ or its expansion has exactly
one edge in $\delta_\oM.(B)$.

\case {$B$ is heavy}
This implies $uv$ is unmatched and $uv\in I(B)$ so it gets expanded.
The expansion has one matched edge,
$(u \bar v,v \bar u)$.
Lemma \ref{ExpVertLemma} shows 
$u \bar v \in B\not\ni v \bar u $.
So  $(u \bar v,v \bar u)\in \delta_\oM.(B)$ and it
is the only such edge.

\case {$B$ is light} This implies
$uv$ is matched and $uv\notin I(B)$.
If $uv$ does not get expanded then $uv\in \oM.$, as desired.
If $uv$ gets expanded then
$(u,u\bar v)$ and
$(v,v\bar u)$ are
matched.
Lemma \ref{ExpVertLemma} shows
$u\bar v,v\bar u \notin B$.
Thus $(u,u\bar v)\in \delta(B)\not\ni
(v,v\bar u)$, as desired.

\bigskip

Now assume $uv\ne \eta(B)$.
We show
$\delta_\oM.(B)$ does not contain
$uv$ or an edge of its expansion.

\case{$uv$ is matched}
By definition $uv\in I(B)$.
Lemma \ref{ExpVertLemma}
shows $u \bar v \in B\not\ni v \bar u$.
So neither of the matched edges $(u, u\bar v),
(v , v\bar u)$ leaves $B$.

\case{$uv$ is unmatched}
We can assume $uv$ gets expanded. Since $uv\not\in I(B)$
Lemma \ref{ExpVertLemma}
shows $u \bar v,v \bar u\not\in B$.
So the matched edge $(u \bar v,v \bar u)$ does not leave $B$.
\end{proof}

  Note the lemma implies $f(B)$ is odd:
  Since $\oM.$ is a perfect matching
  $f(B)= 2|\gamma_\oM.(B)|+|\delta_\oM.(B)|=2|\gamma_\oM.(B)|+1
    $. So any shell with $S^-\ne \emptyset$ is even.

Using the lemma  define
the {\em e-base edge} of an \oe.blossom $B\ne V$
to be the  edge of $\delta_\oM.(B)$, and the
{\em e-base vertex} 
to be the vertex of $B$ on that edge.
We denote these as $\eta(B)$  and
$\o \beta(B)$ respectively, writing the latter as
$\o \beta$ when $B$ is understood.%
\footnote{In contrast to $\beta,\o\beta$, we use the notation
$\eta(B)$ rather than $\o\eta(B)$ since context will
determine that $B$ is an e-blossom.}
A light blossom has $\o\beta=\beta$.
A heavy blossom has $\o\beta\ne\beta$ with
$(\beta,\o\beta)$ an unmatched edge.

Next we show how expanded edges in e-blossoms
modify the internal blossom structure.
First recall the structure of an $f$-factor blossom $B$
(see the review in Appendix \ref{fAppendix}): $B$ is defined
in a graph where the maximal subblossoms of $B$ are contracted.  The
remaining edges of $E(B)$ form a closed path $C(B)$ in the contracted
graph. An {\em atom} of $B$ is a vertex in $B$ that is not in a
maximal subblossom.
The two ends of $C(B)$ are either the atomic vertex $\beta(B)$
or the contracted subblossom containing $\beta(B)$.
In a heavy blossom $B$, $\o\beta(B)$ is an atom not in $C(B)$.

An e-vertex on a cycle $C(B)$ behaves like a $G$-vertex.
Other e-vertices have the following structure
(as illustrated in Fig.\ref{EBlossomFig}).

\begin{lemma}
  \label{eVertexStructureLemma}
  An e-vertex $w \notin C(B_w)$, $B_w\ne V$, is on a matched edge $wx$
  where either
  $x$ is an atomic $G$-vertex of $B_w$, or $wx=\eta(B_w)$ (and $x\not\in B_w$).
\end{lemma}

\begin{proof}
  Applying
  Lemma \ref{ExpVertLemma}
  to blossom $B_w$ shows either $w=u\bar v$ in the middle case
  or, wlog, $w=u\bar v$ in the last case. 

  Consider the middle case. It has $uv\in I(B_w)$.

  If
  $uv$ is unmatched then $uv=\eta(B_w)$ in $G$.
  The expansion has its middle edge $wx$ matched and equal to
  $\eta(B_w)$ in $\o G$, as the lemma claims.
  If
  $uv$ is matched then $uv\ne\eta(B_w)$ in $G$.
  The expansion has $uw$  matched and $B_u=B_w$, i.e.,
  $u$ is atomic, as the lemma with $x=u$ claims.

  Next consider the last case.
  Assume $u$ is not atomic in $B_w$.
  We will show $uv$ is the $\eta$ edge of one of its ends.
  Thus $uv\in C(B_w)$ in $G$. So its image is in $ C(B_w)$ in $\o G$,
  and $w\in C(B_w)$ as the lemma claims.
  The definition of $w=u\bar v$  with $u$ not atomic
  shows $uv \not\in I(B_u)$.

  If $uv$ is matched then $uv = \eta(B_u)$ in $G$.
  If $uv$ is unmatched, then since $uv$ is expanded $uv \in I(B_v)$.
So $uv=\eta(B_v)$.
  \end {proof}

\bigskip

As in ordinary matching we  analyze 
\algcall DismantlePath(Q)
using a matching $M_\omega$ on $Q$
that comes from  the previous scale.
We show how to construct $M_\omega$ for $f$-factors
in Lemmas \ref{PMlemma}--\ref{MOmegaLemma} below.
To set the stage we  first review the construction of $M_\omega$
for ordinary matching. Then we overview the analogous construction
for $f$-factors, and finally we prove the lemmas.

Section \ref{dFSec}  uses the blossom structure
to derive  $M_\omega$ from the algorithm's matching 
at the end of the previous scale. In detail,
let $M$  be the algorithm's matching at the end of the
previous scale. The analysis requires a matching $M_\omega$
that has a certain vertex $\omega$ free. $M_\omega$ is constructed
by rematching $M$ along an alternating path $P(\omega,\beta(Q))$, that
starts with the matched edge incident to $\omega$ and ends at
the base vertex of $Q$. 
\cite{G76} shows how to construct the
$P(\cdot,\cdot)$ paths from the blossom structure.

We proceed similarly for $f$-factors as follows.
Consider an mpr $Q$ and its major path $P(Q)$, both as defined in
$\o G$.
As with ordinary matching we are given  an arbitrary $\o G$-vertex
$\omega\in Q$. The task is to construct an $f$-factor $M_\omega$
that makes $Q$  a light blossom with base vertex $\omega$.
We construct $M_\omega$
by rematching $\oM.$ (the above matching on $\o G$)
along an alternating trail, called a $\o P$ trail.

The $\o P$ trails 
are of two types,
$\o P_M(\cdot,\cdot)$  and $\o P_U(\cdot,\cdot)$.
We use the notation $\o P_\mu(\cdot,\cdot)$, $\mu=M,U$.
By definition the first edge of $\o P_\mu(\cdot,\cdot)$
has M-type $\mu$, i.e., a $\o P_M$ ($\o P_U$) trail
begins with a matched (unmatched) edge.
The $\o P$ trails are in graph $\o G$, so the matching is
$\oM.$.

The $\o P$ trails are analogs of the
trails  $P_i(v,\beta)$ for $f$-factors
from \cite{G18} (see
Appendix \ref{fAppendix}). These $P_i$ trails are in turn analogs of the
$P$ trails of ordinary matching (\cite {G76}). 
However expansion vertices complicate the construction of $\o P$.
For example in Fig. \ref{OddBlossomFig}(b) there is no
possible  trail $\o P_U(u\bar v, \o \beta(B)=u\bar v)$.

The next lemma constructs the $\o P$ trails.
Recall $M_\omega$ is a matching on an mpr $Q$.
We can assume $Q\ne V$. 
(The analysis of mpr $Q=V$ requires no rematching --
the unique shell $(V,\emptyset)$ has both boundaries uncrossed,
so any matching on $V$ can be used.)
Let $\o\beta$ be the e-base vertex of the e-blossom $Q$.

We require the following properties for the 
$\o P$ trails, precise analogs of 
properties of the $P_{i}$ paths:

\bigskip

{\narrower

  {\parindent =0pt
    
    (a) $\o P=\o P_\mu(x,\o\beta), \mu\in \{M,U\}$
    is an alternating $x \o\beta$-trail contained in $B_{x\o\beta}$.


    (b) $\o P$ is composed of $\Omega^-$-blossom edges from $G$ and
    expansion edges from $\o G$. Its last edge is unmatched.

     (c) Any e-blossom $B$ containing a vertex of $\o P$ has
    $|\delta(B,\o P)|\le 2$, and $\o P$ contains either $\o \beta$ or
    $\eta(B)$.
    
  }

}

\bigskip

In property (a) note
there is no danger of interpreting $x\o\beta$ as an e-vertex!
For some examples
let $x$ be either of the two e-vertices
in $S^-$ in Fig. \ref{EBlossomFig}.
The last part of property (b) is needed to rule out
$P_M(x,\beta(S^-))$ being the single matched edge $(x,\beta)$.
(Rematching this trail creates two free vertices, but $M_\omega$
must have only one.)

Regarding property (c), unlike \cite{G18} (c) may be impossible:
In Fig. \ref{EBlossomFig} 
any trail $\o P_U(x,\o \beta)$ must leave $S^-$ on an unmatched edge,
violating (c).  Also in Fig. \ref{OddBlossomFig}(b)
for $x=u\bar v$ a trail 
$P_U(x,x)$ does not even exist.  These examples illustrate the
exceptions in the following lemma.

\begin{lemma}
  \label{PMlemma}
For any e-blossom $B$ and any $\o G$-vertex $x\in B$, 
the alternating trails $\o P_\mu(x,\o \beta(B))$
for $\mu\in \{M,U\}$ exist unless

  \centering
{$x=\o \beta$, or $x$ an e-vertex and $\mu=U$.}
  \end{lemma}

\begin{proof}
We give a recursive definition of the desired trails.
Let the sequence of e-blossoms that contain $v$
and are subsets of the minimal e-blossom
containing both $v$ and $\o\beta$
be $B_i,i=1,\ldots, k$, where $B_1=B_v$, $B_k=B_{v\o\beta}$.
We will define the edges of $\o P$ in each set $E(B_{i})-E(B_{i-1})$,
taking $E(B_0)$ to be $\emptyset$.
We inductively assert that for every $i$,
the  edges of $\o P$ in $E(B_i)$
form a trail $\o P_\mu(v, \o\beta(B_i))$ that
satisfies properties (a)--(c).
Clearly this assertion for  $i=k$ shows the construction is correct. 

We start by giving the inductive step $i>1$.
Then we give the base case $i=1$. (Note that
the base case has some overlap with the inductive step, specifically
it treats the path $CP$
exactly as presented for the inductive step.)

Inductively assume the portion of $\o P$  in $E(B_{i-1})$ ends at
the $\o G$-vertex $x=\o\beta(B_{i-1})$.
$\o P$ follows a trail from $x$ to $\o \beta(B_i)$ in $E(B_i)-E(B_{i-1})$.
If $x=\o\beta(B_i)$ we are done.
Otherwise,  recall that blossoms $B$ have a  closed path
$C(B)$, whose vertices are 
either atoms or maximal subblossoms of $B$
(see Appendix \ref{fAppendix}).

We first define a subset $CP^+$ of edges that belong to $\o P$.
$CP^+$ starts with a path of edges  $CP\con C(B_i)$.
$CP$ ends at the image of $\beta(B_i)$ in $C(B_i)$.
If that image is a blossom then $CP^+=CP$.
Suppose the image is atomic.
If $B_i$ is a light blossom then
$\beta(B_i)=\o\beta(B_i)$ and $CP^+=CP$.
If $B_i$ is heavy
then $\beta(B_i)\ne \o\beta(B_i)$ and
$CP^+$ is
$CP$ extended with the edge
$(\beta(B_i),\o\beta(B_i))$.

$CP$ starts with edge $\eta(B_{i-1})$. This edge
alternates with the last edge of the trail in
$B_{i-1}$ by property (b).

$CP$ consists of the edges of $C(B_i)$ starting with
$\eta(B_{i-1})$ and continuing in the direction that avoids
$B_{i-1}$, ending at the image of $\beta(B_i)$.

Next we  specify
how $\o P$ traverses $CP$ (i.e., how it traverses
the maximal subblossoms of $C(B_i)$ on $CP$).
Consider a maximal subblossom $A$ on $CP$.
First assume $A$ is an interior vertex of $CP$.
So $\delta(A,CP)=\delta(A,C(B_i))=\{\eta(A),\theta(A)\}$.
Let 
$\theta(A) =st$ where $s$ and $t$ are $\o G$-vertices with $s\in A\not \ni t$.
Define $\o P$ to contain the trail $\o P_M (s,\o\beta(A))$.
Since $st=\theta(A)$ is unmatched,
it alternates with  this trail at $s$.
Also the trail exists even if $s$ is an e-vertex
(i.e., it is not a $\o P_U$ trail). It is clear that
properties (a)--(c) are preserved by including
$\o P_M (s,\o\beta(A))$ in $\o P$.

Now assume the subblossom $A$ is the end of $C(B_i)$.
($A$ is the image of $\beta(B_i)$.)
Note that $B_{i-1}$ is disjoint from $A$ (otherwise
$B_{i-1}\con A \pcon B_i$, contradicting the
definition of $B_i$ as the minimal blossom containing $B_{i-1}$.)
Although $A$ has two $\theta(A)$ edges, this implies
only one  belongs to $CP$.
It is treated like $st=\theta(A)$ above.
Again (a)--(c) 
are preserved, by the inductive hypothesis for $A$.
(Note that
(c) always holds for $B=B_i$
since   it has  $\delta(B,\o P)=\emptyset$
and
$\o \beta(B)=\o\beta(B_i)$.) 

It remains to consider the case where the end of $C(B_i)$
is atomic, i.e., $\beta(B_i)$ is atomic.
If $B_i$ is a light blossom then
$\beta(B_i)=\o\beta(B_i)$ and $CP^+=CP$.
The last part of (b) holds since $CP$ ends with an edge
of $e_B$, which is unmatched.
If $B_i$ is heavy
then $\beta(B_i)\ne \o\beta(B_i)$
and $CP^+$ is
$CP$ extended with the edge $(\beta_i,\o\beta_i)$. 
The last edge of $CP$ is an $e_B$ edge, so it is matched.
It alternates with the unmatched edge 
$(\beta_i,\o\beta_i)$. So (a) and (b) hold.

We omit the treatment of a subblossom
$A$ that is the first end of $CP$, since it is $B_{i-1}$.

We turn to the base case
$i=1$. $B_1=B_v$ so $v$ is atomic.
Let $\beta$ and $\o \beta$ be defined for $B_v$.

We first assume $v$ is a vertex ($G$- or e-) in $C(B_v)$ or $v$ is the e-base
of a heavy blossom $B_v$.
Consider the two possible types of blossoms $B_v$.

\case{$B_v$ is light}

\subcase {$v\ne \beta$}
$C(B_v)$ alternates at $v$.  $CP$ starts with the edge of
$\delta(v,C(B_v))$ of M-type $\mu$. The rest of analysis is identical to the
inductive step.

\subcase  {$v=\beta=\o\beta$}
The $\o P_M$ trail for $v$ has no edges in $B_v$.
(In $B_2$ the trail starts with the matched edge $\eta(B_v)$.
There is no trail for $v=\o\beta$ in $B_k$,
an  exceptional case of the lemma,
so we can assume $k>1$.)
The $\o P_U$ trail for $v$ is constructed
by taking $CP^+$ to be the entire closed path $C(B_v)$, starting and ending
at $v$.

\case{$B_v$ is heavy} This case is similar to the previous.

\subcase  {$v\ne \beta,\o\beta$}
$CP$  starts with the edge of
$\delta(v,C(B_v))$
of M-type $\mu$. This path ends at $\beta$, so as before
$CP^+$ adds the edge $(\beta,\o\beta)$.

\subcase{$v=\beta$}
The $\o P_U$ trail for $v$ is edge $(\beta,\o\beta)$.
The $\o P_M$ trail  is constructed
by taking $CP^+$ to be the entire closed path $C(B_v)$ (starting and ending
at $v$), and appending $(\beta,\o\beta)$.

\subcase  {$v=\o\beta$}
The $\o P_M$ trail for $v$ has no edges in $B_v$.
The $\o P_U$ trail does not exist ($v$ is an e-vertex).

\bigskip

It remains to treat the possibility that
$v$ is an e-vertex, $v\notin C(B_v)\cup \o\beta$.
Lemma \ref{eVertexStructureLemma} shows
the matched  edge of $v$ 
joins it to a $G$-vertex  $w\in B_v$.
So  $\o P=\o P_M$ consists of edge $vw$ followed by the $\o P_U$ trail for $w$.
Clearly the latter trail does not contain $vw$
(in $B_v$ that trail consists of edge 
$(\beta,\o\beta)$ if it exists and edges in
$C(B_w)$).
(a)--(c) are preserved.
The $P_U$ trail for $v$ does not exist.
\end{proof}

We can now define the $f$-factor $M_\omega$ for the analysis.
Let $B_\omega$ denote the minimal blossom of $P(Q)$ that contains
a vertex that is free at the chosen instant of time of the analysis.
Let $\omega\in B_\omega$ be such a free vertex.
Let $\o\beta$ be the e-base vertex of $Q$. 
Define
\[M_\omega = \left(\oM. \oplus \o P_M(\omega,\o \beta)\right) \cap \gamma(Q),\]
where we interpret the term $\o P_M$ term to be $\emptyset$ if
$\omega=\o \beta$.

\begin{lemma}
  \label{MOmegaLemma}
  \i Every e-blossom $B$ has $|\gamma(B,M_\omega)|=\f{f(B)/2}$.

  \ii Every  e-blossom $B$ containing $\omega$ has
$\delta(B,M_\omega)=\emptyset$.

  \iii The duals $y_0,z_0$ continue to satisfy
  \eqref{yzScaledEqn} in $\o G$ with matching
  $M_\omega$, i.e.,
\begin{equation*}
    \H{y_0z_0}(e)=w(e)+ \Delta_0(e),
\text{ where }\Delta_0(e)
\begin{cases}
 \ge -2&e\notin M_\omega\\
 \le 12&e\in M_\omega\\
  \in [-2,12]& e \in \cup \set{E(B)}{B \text{ a blossom}} \cup ET.
\end{cases}
\end{equation*}
\end{lemma}

\begin{proof}
  We will use properties (a)--(c) for
$\o P=\o P_M(\omega,\o \beta)$.

  \bigskip
  
  \i   Every vertex of $Q$ is perfectly matched in $M_\omega$ except for
  $\omega$ which lacks one matched edge (properties (a) and (b)).
Any e-blossom
  $B$ has
$f(B)$  odd. So \i holds for $B$
when
either

\bigskip

(1) $\omega\in B$ and $\delta(B,M_\omega)=\emptyset$,
or

(2) $\omega\notin B$ and $|\delta(B,M_\omega)|=1$.

\bigskip

\noindent
Consider the four possibilites for $B$
depending on vertices
$\omega$ and $\o\beta$:

\case {$\omega \in B\not\ni \o\beta$}
We show (1) holds.  (c) shows $|\delta(B,\o P)|\le 2$ and $\o P$
contains $\eta(B)$.

$\o P$ leaves $B$ an odd number of times, so we
get $\delta(B,\o P)= \{\eta(B)\}$.  We conclude (1) holds.

\case {$\o\beta \in B\not\ni \omega$}
We show (2) holds.  (c) shows $|\delta(B,\o P)|\le 2$.
$\o P$ leaves $B$ an odd number of times, so we
get $\delta(B,\o P)= \{\theta(B)\}$.  This gives (2).

\case {$\omega, \o\beta \notin B$}
We show (2) holds.  (c) shows $|\delta(B,\o P)|\le 2$
and $\eta(B)\in \o P$.
$\o P$ leaves $B$ an even number of times, so we
get $\delta(B,\o P)= \{\eta(B), \theta(B)\}$.
Rematching $\o P$ makes $\theta(B)$ matched. This gives (2).

\case {$\omega, \o\beta \in B$}
We show (1) holds.  (a) shows $\delta(B,\o P)=\emptyset$.
Thus $\delta(B, M_\omega)=\delta(B, \oM.)\cap \gamma(Q)$.
The latter is $\emptyset$,
since $\eta(B)\in \delta(Q)$ leaves $\o \beta$.
This gives (1).

\bigskip

\ii $\omega\in B$ iff  the first or last case above holds.
Both satisfy (1).

\bigskip

\iii
Proposition \ref{yzScaledProp} shows \eqref{yzScaledEqn} holds 
  when the \fD. begins.
So we need only consider edges $e\in \o P_M(\omega,\o\beta)$.
Recall the construction of this trail.
Suppose $e$ is in the $CP^+$ set of some blossom $B$.
So $e\in C(B)$ or
$e=(\beta(B),\o \beta(B))$.
Recall  $(\beta(B),\o \beta(B))\in ET$. Thus
$e$ satisfies the last case of \eqref{yzScaledEqn}
(both before and after it gets rematched).

The only other possibility for $e$ is that it is
the first edge $vw$ of a trail $\o P_M(v,\o\beta)$,
$v$ an e-vertex (recall the last case of the construction
of $\o P$ for $i=1$).
This makes $v=\omega$.
We claim $vw$ is not undervalued.
The claim implies $\Delta_0(vw)\ge -2$, so
\eqref{yzScaledEqn} holds 
when $vw$ becomes unmatched as desired.

To prove the claim  assume $vw$ is undervalued.
Recall that
an undervalued edge is matched when
the \fD. begins.
So $v$ is  on a matched
edge throughout  the execution of \fD..
But $v$ is the free vertex $\omega$,
contradiction.
\end{proof}

\bigskip

In the rest of this section we  {\em omit all $\bar\ $ superscripts}, for notational simplicity. For example we write $f$ instead of $\bar f$. Also since the graph is $\o G$
the inherited blossoms are \oe.blossoms.
We remind the reader of this at important points.

At times it is convenient to view sets
of vertices as multisets.
Let $\o S$ denote the shell $S$ where each $v\in S$ has multiplicity $f(v)$.
Let $\o F$ denote the set of free vertices of $S$ where each $v\in F$
has multiplicity equal to its deficiency, i.e., $f(v)-deg(v)$ for
$deg(v)$ the degree of $v$ in the current matching $M$. 

For example,
\begin{equation}
  y(\o S)=\sum_{v\in S} f(v)y(v).
\end{equation}
As another example any set $B\con V$ satisfies
\begin{equation}
  \label{BEqn}
f(S\cap B)=|\o F \cap B| +2|\gamma_M(B)|+|\delta_M(B)|.
\end{equation}
This follows since each of the $f(v)$ copies of $v\in S$ is either
free or matched, and $F\con S, M\con \gamma(S)$.

\paragraph*{The setup}
As for ordinary matching we consider
\algcall DismantlePath(Q) for any mpr $Q$ and
analyze an arbitrary point in the execution of Phase 1 or 2.
The setup is entirely analogous to Section \ref{dFSec}
(although the graph is $\o G$, so $Q\ne V$ is an e-blossom rather than a true blossom).
We describe it in the next several paragraphs.

Let $M_Q$ be the current matching on $Q$.
We analyze an even shell $S=(C,D)$ of $Q$
whose boundaries are uncrossed
by $M_Q$.

Suppose $Q\ne V$, i.e., $Q$ is a blossom.
As before, an $\Omega$-blossom can cross $S^+$ or $S^-$ on
unmatched edges. This includes unmatched $I(B)$ edges.
Let $B_\omega$ be the smallest blossom of $P(Q)$ that is uncrossed
by $M_Q$.
Let $\omega\in B_\omega$ be a free vertex of $M_Q$
in the smallest possible blossom of $P(Q)$.
Here we view $\omega$ as taken from the multiset
of vertices with degree constraint 1, i.e., $\omega$ may have copies
that are free or matched.
Either $\omega\in S^-$ or $\omega=S^-$.

$y$, $z$, and $u$ denote the algorithm's
current duals, $M$ denotes the
restriction of $M_Q$ to
$S$ and $F$ denotes its set of free vertices.
Note that $\omega\notin F$ even though it is free.
$z$ can have positive values on e-blossoms and $\Omega$-blossoms.

$y_0, z_0, u_0$ are the dual functions on entry to
\fD.. Hence they deal with e-blossoms.
The matching $M_\omega$ corresponding to these duals is chosen as follows.
If $S=V$ then $M_\omega$ is the $f$-factor on $V$
after edge expansion.
If $S\ne V$ then $Q$ is a blossom.
Take $M_\omega$ to be the $f$-factor of Lemma \ref{MOmegaLemma}
that has free
vertex $\omega$, restricted to $S$.
Lemma \ref{MOmegaLemma}\ii
shows $M_\omega$ does not cross any
e-blossom  containing $\omega$.
The shell $S$ that we analyze 
either has
$\omega\in S^-$ or $S=(B_\omega,\omega)$.
So neither boundary of $S$ is crossed by $M_\omega$.


Define
\[n=f(S),\]
so $n$ is even.
The shell $(B_\omega,\omega)$ is always somewhat anomalous.
A shell $S$ satisfies $f(S)=f(S^+)-f(S^-)$,
so $f((B_\omega,\omega))=f(B_\omega)-f(\omega)=f(B_\omega)-1$.
Thus $f((B_\omega,\omega))$ is even as claimed above.
The identity
\eqref{BEqn} for $S=(B_\omega,\omega)$ is treated similarly.
Note that $\omega$ is free in the two matchings of interest,
\[\delta(\omega,M_Q)=\delta(\omega,M_\omega)=\emptyset,\]
so $\omega$ is essentially irrelevant to properties concerning matchings.

\bigskip

Define
the same two quantities as
Section \ref{dFSec}:
Let $B\con Q$ be an
inherited  \oe.blossom.
Let $v$ be a vertex that is free at the chosen point of
\algcall DismantlePath(Q).
\begin{eqnarray*}
\tau(B)&=&\text{the number of Phase 1 or 2 unit translations
  of $B$,}\\
&&\text{up to and including the chosen point.}\\
d(v)&=&\text{the number of Phase 1 or 2 dual adjustments 
  made for $v$,}\\
&&\text{up to and including the chosen point.}
\end{eqnarray*}
As before
every free vertex $v$ satisfies
\begin{equation}
\label{FyzPreliminariesEqn}
y(v)=y_0(v)-d(v)+ \sum_{v\in B}\tau(B).
\end{equation}
Similarly $\Ol.$ is a laminar family and so
every \oe.blossom $B$ with $B\cap S\ne \emptyset$
satisfies
\begin{equation}
\label{fRelevantBlossomsEqn}
B\pcon S \text{ or } D\pcon B.
\end{equation}

\bigskip

\paragraph*{Objective functions}
We give the details of the two duals,
$y_0,z_0$ and $y,z$, and their objective functions.

Recall the general discussion of 
$f$-factors duals from Appendix \ref{fAppendix}.
We extend the general definition of objective function
to shells $S$ by defining 
\begin{equation}
  \label{CapDualEqn}
  \H{yz}(S)=y(\bar S) +  \sum_{B} z(B) c(S\cap B),
\end{equation}
where the ``capacity'' function $c$ is the number of
edges $e\in \gamma(S)$ that can include $z(B)$
in their dual value $\H{yz}(e)$.
We specify this capacity function
first for the 
previous duals $y_0,z_0$ and then for the current duals $y,z$.
Note the $u$ function for undervalued edges is not included in
  $\H{yz}$.

\bigskip

The duals $y_0,z_0$ from
the previous scale use \oe.blossoms.
Such a blossom contributes to  $\H{yz}(e)$ if $e\in \gamma(B)\cap
\gamma(S)$ (this set is identical to $\gamma(S\cap B)$).
So the capacity function is
\[ c(S\cap B)=
\f{f(S\cap B)/2}\hskip20pt \forall B\in \Ol..\]
Clearly any matching contains $\le c(S\cap B)$ edges of
$\gamma(S\cap B)$.
(Recall the special treatment of $(B_\omega,\omega)$.)
We show the expanded matching of the previous scale
$M_\omega$ achieves this upper bound, 
i.e.,
\begin{equation}
  \label{ZEqn}
\gamma(S\cap B, M_\omega)= c(S\cap B).
\end{equation}
In proof recall
\eqref{fRelevantBlossomsEqn} and observe that when $S\cap B\ne \emptyset$,
\begin{equation}
  \label{DetailOmegaMEqn}
|\gamma(S\cap B, M_\omega)|=
\begin{cases}
  \f{f(B)/2}&B\pcon S\\
  f(S\cap B)/2&\text{$B$ interior to $S$, or
    $\omega\in B\pcon B_\omega$ if  $S=(B_\omega,\omega)$}\\
f(S)/2&C\con B.
\end{cases}
\end{equation}
(In the middle case for $S=(B_\omega,\omega)$,
$S\cap B=B-\omega$ so $f(S\cap B)$ is even.)
All three quantities equal $\f{f(S\cap B)/2}$.

Next we show the matching $M$ of the \fD. satisfies
\begin{equation}
  \label{fBcapSEqn}
c(S\cap B) =
  |\gamma_M(B)|+
      |\o F\cap B|/2\hskip20pt B\in \Ol.\text{ is undissolved}.
\end{equation}
\noindent As with ordinary matching the case
``$B\in \Ol.\text{ dissolved}$'' is irrelevant.

Suppose $B\in \Ol.$ is undissolved at the chosen instant of time.
    Either $B$ is interior to  $S$
    or $S\con B$. In both cases $S\cap B$ is an even shell.
    Thus $c(S\cap B)=
    f(S\cap B)/2$.
No matched edge crosses $B$ ($B$ undissolved), i.e.,
$\delta_M(B)=\emptyset$. So
\eqref{BEqn} gives
\eqref{fBcapSEqn}.

\bigskip

Now consider the dual functions $y,z$ for the chosen point in time,
with the current matching $M$. An undissolved
e-blossom  has capacity defined above.
An $\Omega$-blossom 
contributes to  $\H{yz}(e)$ if $e\in \gamma(B)\cup I(B)$.
Thus
\[ c(S\cap B)=
\left\lfloor{ \frac{f(S\cap B)+|\gamma(S, I(B))|}{2} }\right\rfloor \hskip20pt \forall B\in \Omega.\]
Recall that an $\Omega$-blossom $B$ can cross $S$ on unmatched edges.
Similarly $I(B)$ may contain edges that relate arbitrarily to $S$.
The capacity function only concerns edges with both ends in $S$,
hence we use the restrictive term $\gamma(S,I(B))$.
Note that $I(B)$ may contain an unmatched edge.

The term $\gamma(S, I(B))$ is not well-defined in the shell
$S=(B_\omega,\omega)$.
Specifically
let
$\omega_V$ be
the  $V$-vertex corresponding to $\omega$.
Assume $\omega_V\notin B$. If
$|\delta(\omega_V,I(B)|=f(\omega_V)$
then $\eta(B)$ must be an unmatched edge in $I(B)$
(since $\omega$ is free), and
$|\delta(\omega_V,I(B)\cap M|=f(\omega_V)-1$.
The latter implies the term $\gamma(S, I(B))$ should not include $\eta(B)$.
On the other hand if $|\delta(\omega_V,I(B)|<f(\omega_V)$ then
$\gamma(S, I(B))$ should include $\eta(B)$.
We make the convention that $\gamma(S, I(B))$ always
includes an unmatched $\eta(B)$ incident to $\omega_V$.
We shall see the capacity function behaves as desired,
i.e., it gives a tight upper bound.

We proceed to show just that.
Clearly any matching contained in $\gamma(S)$ contains $\le
c(S\cap B)$ edges of $\gamma(B)\cup I(B)$. We will show the matching
$M$ achieves this upper bound, i.e.,
\begin{equation}
  \label{capgIEqn}
c(S\cap B) =  |\gamma_M(B)|+|I(B)\cap M|.
\end{equation}
Equivalently we  show
\begin{equation}
  \label{XEqn}
f(S\cap B) +|\gamma(S, I(B))|=2
  \left(|\gamma_M(B)|+|I(B)\cap M|\right)+\epsilon,\hskip20pt 0\le \epsilon \le 1.
\end{equation}
By \eqref{BEqn} and the definition of $I(B)$, the left-hand side
of \eqref{XEqn} is
\[
\left(|\o F \cap B|+2|\gamma_M(B)|+|\delta_M(B)|\right)
+
\bigl(
|I(B)\cap M|+|\eta(B)- M\cap \gamma(S)|
\bigr).
\]
Using $\delta_M(B)=|I(B)\cap M|+|\eta(B)\cap M|$
we rewrite this as
\begin{equation*}
 2\left(|\gamma_M(B)| +|I(B)\cap M|\right)+
\underbrace{
  \bigl(
  |\eta(B)\cap M|+|\eta(B)-M\cap \gamma(S)|
  +|\o F \cap B|
\bigr)
  }
 _{\mbox{$\epsilon$}}.
\end{equation*}
We complete the proof by showing $\epsilon\le1$.
Let $\beta$ be the base vertex of $B$.

If $\beta$ is free then $B$ has no base edge. So $\epsilon=
|\o F \cap B|\le 1$. If $\beta$ is not free then
$\eta(B)$ exists, and is either matched or unmatched.
If it is matched then $\epsilon=1$. If unmatched then
$\epsilon=|\eta(B)\cap \gamma(S)|\le 1$.

\subsection{The analysis}
\label{fdFSec}

The argument is that of
Section \ref{dFSec} with simple modifications
to incorporate e-blossoms and the $f$ function.
As before the first four steps apply the two
dual functions $y_0,z_0$ and $y,z$, and
the last step analyzes
crossings of  e-blossoms.

\bigskip

Every edge $e\in M_\omega$ has
$\H{y_0z_0}(e)\le w(e)+12$ (Lemma \ref{MOmegaLemma}\xiii).
\eqref{ZEqn} shows $M_\omega$ fills each e-blossom to capacity.
So summing these inequalities,
and recalling the definition of $u_0$,
gives
\begin{alignat*}{2}
    \hspace{100pt}
\H{y_0z_0} (S)+u_0(M_\omega)-12(n/2)&\le
w(M_\omega)&\hskip20pt\mbox{\bf 
scaled near tightness of
    \boldmath {$\H{y_0z_0}$}.}
\end{alignat*}

An edge $e\in M_\omega$ is
nearly dominated 
by the $y,z$ duals, unless it is undervalued. 
$M_\omega$ fills each $\Omega$-blossom to at most its capacity. 
So
\begin{xalignat}{2}
\label{LHSEqn}
\hspace{154pt}
w(M_\omega)-2(n/2)&\le \H{yz}(S)+u(M)&\hskip20pt 
\mbox{\bf near domination of \boldmath {$\H{yz}$}.}\notag     
\end{xalignat}

Every edge $e\in M$ is underrated by the $y,z$ duals.
\eqref{capgIEqn} shows $M$ fills each $\Omega$-blossom to its capacity.
So  
\begin{xalignat}{2}
\hspace{61pt}
\H{yz}(S)+u(M)&\le y(\bar F) + w(M) +
\sum_{D\pcon B\in \Oe.} z(B)|\bar F\cap B|/2&\hspace{20pt}\mbox{\bf near tightness of \boldmath 
{$\H{yz}$}.}\notag
\end{xalignat}
The range of the summation is justified by
\eqref{fRelevantBlossomsEqn} and the fact that e-blossoms $B\pcon S$ 
have dissolved.
(As before we refrain from assuming $D\ne \omega$ in the summation
in order to simplify notation.)

To summarize at this point, combine the last 3 inequalities to get
\begin{equation}
\label{SummaryEqn}
\H{y_0z_0} (S)+u_0(M_\omega)-7n \le y(\bar F) + w(M) +
\sum_{D\pcon B\in \Oe.} z(B)|\bar F\cap B|/2.
\end{equation}

To upper bound the right-hand side of 
\eqref{SummaryEqn}
first sum \eqref{FyzPreliminariesEqn}
for every $v\in \bar F$:
\begin{equation}
y(\bar F)=y_0(\bar F)-d(\bar F)+\sum_{B} |\bar F\cap B|\tau(B),
\notag
\end{equation}
and then bound the matched edges by
\begin{eqnarray*}
  \hspace{0pt}
w(M)\le y_0(\bar S \cap V(M))+ \sum_{B}  z_0(B)|\gamma_M(B)|+u_0(M_\omega)+2(n/2)
&\text{\hspace{16pt}
  {\bf near domination of \boldmath{$\H{y_0z_0}$}.}}
\end{eqnarray*}
Combining the last two inequalities gives
\begin{align}
\label{Fyz1Toyz0Eqn}
y(\bar F)+w(M)\le  
y_0(\bar S) -d(\bar F)+
\underbrace{\sum_{B} \Big(  |\bar F\cap B|\tau(B) +z_0(B)|\gamma_M(B)|\Big)}_{SUM}
+u_0(M_\omega)+n.
\end{align}

\subsubsection*{Analysis of blossom crossings}
\begin{figure}[t]
\centering
\input{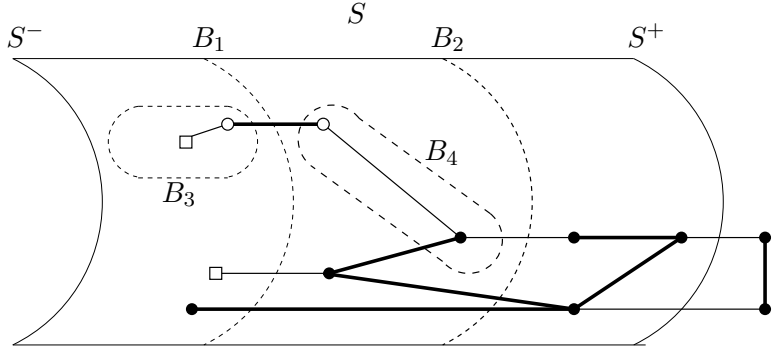}
\caption{Crossings of e-blossoms by matched edges.
  The two e-blossoms $B_3,B_4$ are joined by an expanded edge
  in $I(B_3)\cap I(B_4)$.
      $CRI(S)=z_0(B_1)+z_0(B_2)$, 
  $CRB(SUB(S))=z_0(B_4)/2$.
}
 \label{FbcrFig}
\end{figure}
This section is almost identical to the case of ordinary matching,
with simple changes for the $f$ function. (This is unsurprising since
the analysis  concerns e-blossoms, which are simple versions of
ordinary matching blossoms.)
We include the details
for completeness.

As before we will show
\begin{equation}
  \label{fUpperboundedByEqn}
SUM\le \sum_{B} z_0(B)\f{f(B\cap S)/2}
-\sum_{D\pcon B\in \Oe.}|\bar F\cap B|z(B)/2
+\Delta
\end{equation}
where
$\Delta$
consists of $\tau, CRI$ and $CRB$ terms.
Fig.\ref{FbcrFig} illustrates
the latter two for $f$-factors.
The blossoms $B$ contributing to $SUM$
belong to $\Oe.$
with $B\cap S\ne \emptyset$, so they satisfy
\eqref{fRelevantBlossomsEqn}.
We  consider the two  corresponding possibilities,
choosing the possibility $B\pcon S$ when $S=Q=V$.

\case{$D\pcon B$}

\subcase{$D\ne \omega$}
With $S\ne V$, i.e., $Q$ a blossom, this implies
either $B\in P(Q)$ or $Q\pcon B$.
In the first case the 
unit translations
of \algcall DismantlePath(Q)
maintain the invariant $\tau(B)+z(B)/2 = z_0(B)/2$.
This holds trivially in the second case, since $\tau(B)=0$,
$z(B) = z_0(B)$. Rearranging the invariant to
$\tau(B)= z_0(B)/2-z(B)/2$
and using 
\eqref{BEqn}
shows
$B$'s term in $SUM$ is
\begin{eqnarray*}
|\bar F\cap B|\tau(B) +z_0(B)|\gamma_M(B)|
&=&(z_0(B)/2)\big(|\bar F\cap B|+2|\gamma_M(B)|\big)-(z(B)/2)|\bar F\cap B|\\
&=& z_0(B)\big(f(S\cap B)-|\delta_M(B)|\big)/2 -(z(B)/2)|\bar F\cap B|.
\end{eqnarray*}
A blossom $B$ of this case has   $f(S\cap B)$ even
(recall the last two cases of \eqref{DetailOmegaMEqn}).
So $f(S\cap B)/2=\f{f(S\cap B)/2}$. Using this  and rearranging terms changes the last line to
\[z_0(B)\f{f(S\cap B)/2}-(z(B)/2)|\bar F\cap B|
 -|\delta_M(B)|z_0(B)/2.\]
The first
two terms match the
terms for $B$ in the two summations of
\eqref{fUpperboundedByEqn}. We include the third term
\[ -|\delta_M(B)|z_0(B)/2\]
in $\Delta$. 
It is a nonpositive quantity,
nonzero only on  crossed interior blossoms of $S$.

\subcase{$D= \omega$}
Using $\tau(B)\le z_0(B)/2$ the term for $B$ 
in $SUM$ is at most
\begin{equation*}
  z_0(B)(|\bar F\cap B|+2|\gamma_M(B)|)/2=
  z_0(B)(f(S\cap B)-|\delta_M(B)|)/2
  \le z_0(B)f(S\cap B)/2.
\end{equation*}
Since $S=(B_\omega,\omega)$, $S\cap B=B-\omega$.
Thus $f(S\cap B)=f(B)-1$ is even. So
the above right-hand side is 
$z_0(B)\f{ f(S\cap B)/2}$. This  matches the
term for $B$ in the first summation of
\eqref{fUpperboundedByEqn}.
As mentioned above there is no contribution to the
second summation ($z(B)=0$).

\case{$B\pcon S$}

\subcase{$B$ crossed}
Again 
using $\tau(B)\le z_0(B)/2$, the term for $B$ 
in $SUM$ is at most
\begin{eqnarray*}
z_0(B)(|\bar F\cap B|+2|\gamma_M(B)|)/2&=& z_0(B)(f(B)-|\delta_M(B)|)/2\\
&=&z_0(B)\f{f(B\cap S)/2} +z_0(B)( 1-|\delta_M(B)|)/2.
\end{eqnarray*}
The first term 
of the last line
 matches the
term for $B$ in the first summation of
\eqref{fUpperboundedByEqn}.  Recalling the assumption
$|\delta_M(B)|\ge 1$, we include the second term
\[ z_0(B)( 1-|\delta_M(B)|)/2\]
in $\Delta$ when $|\delta_M(B)|>1$.
It is a nonpositive quantity.

\subcase{$B$ uncrossed}
This implies
$B$ contains a free vertex. Using
\eqref{BEqn}
$B$'s term in $SUM$ is
\[|\bar F\cap B|\tau(B) +z_0(B)|\gamma_M(B)|\le
\tau(B) +z_0(B)(f(B)-1)/2=
\tau(B) +z_0(B)\f{f(B\cap S)/2}.\]
In the rightmost bound the second term matches
term for $B$ in the first summation of
\eqref{fUpperboundedByEqn}.
We include the nonnegative term
\[\tau(B)\]
in $\Delta$.

\bigskip

We conclude
\eqref{Fyz1Toyz0Eqn}
and \eqref{fUpperboundedByEqn}
give
\begin{xalignat*}{2}
y(\o F)+w(M)\le
&y_0(S)-d(\o F) 
+\sum_{B} z_0(B)\f{f(B\cap S)/2}
-\sum_{D\pcon B\in \Omega^-}|\o F\cap B|z(B)/2+\Delta+u_0(M_\omega)+n
\notag\\
=&\H{y_0z_0}(S) -d(\o F)
-\sum_{D\pcon B\in \Omega^-}|\o F\cap B|z(B)/2+\Delta+u_0(M_\omega)+n. 
\end{xalignat*}
Combining this with
\eqref{SummaryEqn}
gives
\[\H{y_0z_0} (S)+u_0(M_\omega)-7n 
\le \H{y_0z_0}(S) -d(\o F)+\Delta+u_0(M_\omega)+n.
\] 
Hence
\begin{equation}
\label{fGoalEqn}
d(\o F)\le 8n+\Delta.
\end{equation}

We use the same notation as before for the terms of $\Delta$.
Let $S$ denote a shell in \algcall DismantlePath(Q),
$M$ a matching on $S$,
$BS$ an arbitrary collection of
inherited e-blossoms. $B$ always denotes an inherited e-blossom.
\begin{eqnarray*}
{\U.}    &=& 
\set {B} {\text{$B\pcon S$ an inherited e-blossom not crossed by $M$}},\\
  INT(S) &=& \set {B} {\text{$B$ interior to $S$}},\\
  SUB(S) &=& \set {B} {\text{$B\pcon S$}},\\
  CRI(S)&=& \sum_{B\in INT(S)}|\delta_M(B)| z_0(B)/2,\\
  CRB(BS)&=&\sum_{B\in BS,\; |\delta_M(B)|>1} (|\delta_M(B)|-1)z_0(B)/2. 
\end{eqnarray*}
As before we comment when necessary that \U.
and the crossed blossom functions $CRI,CRB$ depend on the matching $M$.
\eqref{fGoalEqn} becomes
\begin{equation}
\label{fGoalCrossedEqn}
d(\o F)+ CRI(S)+ CRB(SUB(S)) \le cn+\tau(\U.)
\end{equation}
for the constant $c=8$ ($c$ used in the credit system).
Again we
reiterate the setting for \eqref{fGoalCrossedEqn}.
\eqref{fGoalCrossedEqn} applies 
at any chosen point of Phase 1 or 2 of \algcall DismantlePath(Q),
$Q$ an mpr of the expanded graph $\o G$.
The matching is $M$,
$S$ is an
even shell of $P(Q)$  uncrossed by $M$, 
$\o F$ is the multiset of free vertices of $M$
contained in  $\o S$,
and $M$ is used to define $CRI$ and $CRB$.

The extended $d$ function is defined exactly as in
Section \ref{dFSec}.



\setcounter{section}{0}
\renewcommand{\thesection}{\Alph{section}}
\renewcommand{\thetheorem}{\Alph{section}.\arabic{theorem}}

\setcounter{equation}{0}
\renewcommand{\theequation}{\Alph{section}.\arabic{equation}}

\def\Myfor #1{{\bf for} {\em #1} {\bf do}}
\def\Myif #1{{\bf if} {\em #1} {\bf then}}
\def\Myelseif #1{{\bf else if} {\em #1} {\bf then}}
\def\Myelse #1{{\bf else} {#1}}

\section{Edmonds' algorithm for matching}
\label{EdAppendix}

For background we first review the
linear program for  maximum weight perfect matching.
The  variables are given by
the function $x:E \to \mathbb {R_+}$ which
 indicates whether or not an edge is matched. 
Recall
our summing convention, e.g.,
$x(\delta(v))=\sum_{e\in \delta(v)}x(e)$.

\bigskip

\hskip60pt maximize $\sum_{e\in E} w(e)x(e)$ subject to

\[\begin{array}{llll}
x(\delta(v))&=&1&\forall v\in V\\
x(\gamma(B))&\le& \f{|B|\over 2}&\forall B\con V\\ 
x(e)&\ge& 0&\forall  e\in E
\end{array}
\]

The dual LP uses dual functions
$y:V \to \mathbb {R}$,
$z:2^V\to \mathbb {R_+}$.
Define
$\H{yz}:E\to \mathbb {R}$ by
\begin{equation}
\H{yz}(e) = y(e)  + z\set {B} {e \con B}.
\end{equation}
\noindent
(Note for  $e=uv$,  $y(e)$ denotes
$y(u)+y(v)$ and
$z\set {B} {e \con B}$
 denotes  $\sum_{e \con B} z(B)$.) 

\bigskip

\hskip60pt minimize 
$y(V) +  \sum_{B\con V} \f{|B|\over 2}\, z(B) $
subject to

\[\begin{array}{llll}
\H{yz}(e) &\ge& w(e)&\forall e\in E\\
z(B)&\ge& 0&\forall  B\con V
\end{array}
\]

\bigskip

\setlength{\figwidth}{\textwidth}
\addtolength{\figwidth}{-1in}
\setlength{\figindent}{.5in}
\setlength{\vmargin}{.1in}

   \setlength{\Efigwidth}{\textwidth}
   \addtolength{\Efigwidth}{-.3in}

\begin{figure}
\begin{center}
  \fbox{
\begin{minipage}{\Efigwidth}
\setlength{\parindent}{.2in}

\narrower{
\setlength{\parindent}{0pt}
\vspace{\vmargin}
\setlength{\parindent}{20pt}

\noindent
Make every free vertex or blossom the  root of an \os.-tree.
Then repeat the following statement until an augmenting path is found.

\smallskip

\Myif {$\exists$  eligible edge $e=xy$, $x\in \S., y\notin \S.$, alternating with $B_x$}

{\hi

$/*$ {grow step} $*/$

add $xy, B_y$ to \S.
}

\Myelseif {$\exists$  eligible edge $e=xy$, $x,y\in \S.$, alternating with both $B_x$ and $B_y$}
 
{\hi

\Myif  {$x$ and $y$ are in the same search tree}

{\hi

$/*$ {blossom step} $*/$

merge all blossoms in the fundamental cycle of $e$ in \os.
}

$/*$ if $x$ and $y$ are in different  trees  an augmenting path has been found $*/$ 

}

\Myelseif{$\exists$ nonsingleton inner blossom $B$ with $z(B)=0$}
Expand$(B)$

\Myelse AdjustDuals

\bigskip

\noindent
{\bf Algorithm} Expand$(B)$:

in \os. replace $B$ by the even length alternating path of subblossoms
$B_0,\ldots, B_k$, where

{\hi
the unmatched edge entering $B$ enters $B_0$

the matched edge entering $B$ enters $B_k$

}

$/*$ the remaining subblossoms of $B$ are no longer in \S. $*/$

\bigskip

\noindent
{\bf Algorithm} AdjustDuals:

$\delta_1\gets\min \set{|y(e)-t(e)|}{e=xy \mbox{ with $x\in \S., y\notin \S.$,
$xy$ alternates with $B_x$}}$

$\delta_2=\min \set{|y(e)-t(e)|/2}{e=xy \mbox{ with $x,y\in \S.$,
$xy$ alternates with both $B_x$ and $B_y$}}$

$\delta_3=\min \set{z(B)/2}{B \mbox{ an inner blossom of }\os.}$

$\delta=\min \{\delta_1,\delta_2,\delta_3\}$

\smallskip

For  {every vertex $v\in \S.$}

{\hi

$y(v)\gets y(v) + $ $($\Myif {$B_v$ is inner} {$\delta$} 
\Myelse{$-\delta)$}

}

For {every blossom $B$ in \os.}

{\hi

$z(B) \gets z(B) +$  $($\Myif {$B$ is inner}  {$-2\delta$} \Myelse 
{$2\delta)$}

}

}

\vspace{\vmargin}

\end{minipage}

}

\caption{Pseudocode for an Edmonds' search, for perfect matching
with near-optimum duals.}
\label{EdSrchFig}
\end{center}
\end{figure}

A precise statement of Edmonds' algorithm is given in Fig. \ref{EdSrchFig}.
We briefly summarize the 
algorithm, although we assume the reader is familiar with a complete
treatment.

The search structure 
\os. is a forest. Its nodes are contracted 
blossoms or vertices that are not in blossoms.
For simplicity in this section we call the latter ``blossoms'' as well.

The 
roots of \os. are the {\em free} blossoms, i.e., the blossoms that contain 
an 
unmatched vertex. The forest is {\em alternating}, i.e., the edges
from a node to its root alternate between being matched and 
unmatched. The last edge (incident to the root) is necessarily 
unmatched.
If the first edge is unmatched
the node is {\em inner}.
Otherwise (i.e., the first 
edge
is matched or the node is a root) the node is {\em outer}.

For $B$  a blossom in \os., an edge $e \notin \os.$ 
{\em alternates with $B$} if $e$ is unmatched for $B$ outer
and matched for $B$ inner. (Note 
if $e$ alternates with $B$ 
then $B$ can be replaced by an 
appropriate alternating path
so there is an alternating path of edges of $G$ from $e$ to its forest 
root.)

For any vertex $v$ of $G$, $B_v$ denotes the maximal blossom containing 
$v$ (or $v$ itself if no bigger blossom exists). Let $M$ denote the current matching.
In a grow step,
if $B_x$ is outer then $xy\notin M$ and $B_y$ becomes inner;
if $B_x$ is inner then $xy\in M$ and $B_y$ becomes outer.
A blossom step necessarily creates a new outer node.
When an augmenting path is discovered, the algorithm proceeds to
augment a maximal collection of disjoint augmenting paths
of eligible edges.

In the dual adjustment step,
define the "target" $t(e)$ of edge $e$ 
that is not in the search structure to be the value of $yz(e)$ that 
makes $e$ 
eligible,
i.e., $t(e)$ is the value of
\eqref{EligibleEqn}. 
(When we execute Edmonds' algorithm in Phases 2 or 3
the definition of eligibility changes to  
\eqref{EligibleTwoEqn} and the target becomes the value
closest to the current value of $yz(e)$.
For instance an unmatched edge with 
$yz(e)>w(e)$ has target
$t(e)=w(e)$.)

\def\ce #1.{\lceil #1 \rceil}
\def\Bce #1.{\Big\lceil #1 \Big\rceil}
\section{Implementation of the \D.}
\label{Phase2Appendix}
This section presents the data structures and implementation details for
Phases 1 and 2 of \algcall DismantlePath(Q).
We show the total time is $O(\sqrt{|Q|\log |Q|} m(Q))$ for matching on graphs.
The discussion applies equally well to $f$-factors (since the \fD. is identical
to the \D.) making obvious textual changes, e.g., the time is
$O(\sqrt{f(Q)\log f(Q)} m(Q))$.

Edmonds' search is implemented with known data structures.
(These include nontrivial data structures for set merging
\cite{GT85} and split-findmin \cite{Th}.)
We use a data structure for set merging to track the current partition
of $V$ into atomic shells.
It suffices to
use the simple ``relabel-the-smaller-half'' strategy:
Over the course of an entire scale, the atomic shell containing
a given vertex $v$ changes by merging operations (even as 
\algname DismantlePath. transitions from one
major path to the next). So $v$ can be examined
every time its shell is merged into a larger shell.
Thus the total overhead for set-merging, over an  entire scale, is
$O(n\log n)$.

Each current atomic shell $S$ records various items.
The size $|S|$ is known from the set-merging data structure.
$S$ has  a list of its free vertices.
$S$ has pointers to its boundary blossoms.
Each inherited blossom $B$ records the initial dual 
value $z_0(B)$ and the current value $\tau(B)$. All current atomic shells of $P(Q)$ are maintained in
a linked list that is ordered as in $P(Q)$. The list has a pointer to
the maximal undissolved blossom of $P(Q)$.

The graph is represented using a simple adjacency structure.  When
scanning the adjacencies of a vertex in shell $S$ it is easy to
distinguish edges that remain in $S$ from edges that leave $S$.

The current dual functions $y,z$ are maintained using the quantities
\[y'(v)=y(v)+z\set{B}{v\in B\in \Omega^-}/2.\]
These quantities are initialized at the start of the scale
using $y_0$ and $z_0$. They are maintained in Edmonds searches, i.e.,
a dual adjustment changing $y$ by $\pm 1$ makes the same change to $y'$.
$y'(v)$ does not change in any unit translation of an inherited blossom
(unlike $y'$ of \cite{GT}).

We  compute $\H{yz}$ values (as needed in Edmonds'algorithm)
by 
\begin{equation*}
\label{yzComputedByYPrimeEqn}
\H{yz}(uv)=y'\{u,v\} + z\set{B}{u,v\in B\in \Omega}.
\end{equation*}
Here $uv$ is an edge in a current shell $S$.
To see this is correct consider a blossom $B\in \Omega^-$.
If $B$ contains both $u$ and $v$ 
$y'(u)$ and $y'(v)$ contribute a total of
$z(B)/2 +z(B)/2=z(B)$ to the right-hand side, as desired.
If $B$ contains only one of $u,v$, say $u$, then blossom $B$
dissolves before  \algcall ShellSearch(S), so $y'(u)$ correctly includes 
the contribution of $z(B)/2$
to $y(u)$.

\bigskip

Further details depend on the phase.

\bigskip

\subsection*{Phase 1}
The list \A. is implemented using buckets 
$B[i]$, $1\le i\le |Q|$, where $B[i]$ contains the atomic shells of
size $i$ that contain $\ge 1$ free
vertex.

The rematch step of
Phase 1 computes a maximal set of disjoint augmenting paths
using the algorithm of Gabow and Tarjan \cite{G17,GT} for ordinary matching,
and Huang and Pettie \cite{HP} for $f$-factors.
Both run in linear time $O(m)$.

\subsection*{Phase 2}

We use a priority queue $PQ$ of $c|Q|\log |Q|$ buckets.
An entry $PQ[t]$ is a list of  the events of Phase 2
that are predicted to occur in the $t$th dual adjustment
of the current shell $S$ being searched.

To keep the space linear $O(n)$,
$PQ$ is divided into $c\log |n|$ ``pages'' of $n$ buckets.
Only the current page is implemented as an array of $n$ buckets.
The other pages are simply
lists of events scheduled for time units  in that page.

An event in $PQ[t]$
is either a step of Edmonds' search in the current shell $S$, or
a shell boundary $S^-$ or $S^+$ that will dissolve in time unit $t$.
When such a boundary dissolves, merging $S$ with an adjacent shell $S'$,
we scan the adjacency list of each vertex $v\in S'$. This
entails adding new Edmonds
events to $PQ$, and possibly executing Edmonds events that occur in the current
time unit. In particular new search trees are added for free vertices in
$S'$.

The adjacency lists scans use total time $O(m)$ per augment. 
as desired for our time bound. (In addition a scan may occur in a search whose
shell
eventually gets deactivated, again within our time bound.)

\section{The $f$-factor algorithm}
\label{fAppendix}

\bigskip

The LP for $f$-factors is similar to matching (Appendix \ref{EdAppendix}),
incorporating the degree constraint function
$f$ and  $I$-sets of blossoms.
(It is derived in \cite{G18}. Alternatively
it follows from
\cite[Theorem 33.2]{S} by increasing edge weights by a large amount).
We allow the graph to have parallel edges.
Each copy of an edge has its own $x$ variable
indicating membership in the $f$-factor.

\hskip60pt maximize $\sum_{e\in E} w(e)x(e)$ subject to

\[\begin{array}{llll}
x(\delta(v))+2x(\gamma(v))&=&f(v)&\forall v\in V\\
x(\gamma(B) \cup I )&\le& \f{f(B)+|I|\over 2}&\forall B\con V,\,
I\con \delta(B)\\ 
x(e)&\le& 1&\forall  e\in E\\
x(e)&\ge& 0&\forall  e\in E
\end{array}
\]

We call $e$  {\em dominated, tight,} or {\em underrated}
depending on whether
$\H{yz}(e)$ is  $\ge w(e)$, $= w(e)$, or $\le w(e)$, respectively;
{\em strictly dominated} and {\em strictly underrated}
refer to the possibilities  $>w(e)$ and $< w(e)$ respectively.

The dual LP uses dual functions
$y:V \to \mathbb {R}$,
$z:2^V \times 2^E\to \mathbb {R_+}$.
Define
$\H{yz}:E\to \mathbb {R}$ by
\begin{equation}
\label{fHyzEqn}
\H{yz}(e) = y(e)  + z\set {(B,I)} {e \in \gamma(B)\cup I}.
\end{equation}

\hskip60pt minimize 
$\sum_{v\in V} f(v)y(v) +  
\sum_{B\con V,I\con \delta(B)} \f{f(B)+|I|\over2}\, z(B,I) +u(E)$
subject to

\[\begin{array}{llll}
\H{yz}(e) +u(e)&\ge& w(e)&\forall e\in E\\
u(e)&\ge& 0&\forall e\in E\\
z(B,I)&\ge& 0&\forall B\con V,\,I\con \delta(B)
\end{array}
\]

In
the $f$-factor algorithm
every nonzero $z$ value has the form $z(B,I(B))$ for $B$ a mature blossom.
So we write $z(B)$ as a shorthand for $z(B,I(B))$.

Gabow \cite{G18} treats $f$-factors
by first generalizing algorithmic concepts from ordinary matching, and then
modifying the matching procedure of
Fig. \ref{EdSrchFig} to apply to $f$-factors.
We briefly review this presentation, referring the reader to
\cite{G18} for a complete development.

We begin by defining $f$-factor
blossoms. We start with some notation before
giving the complete  definition.
We are given a graph $G$ with degree constraints $f(v)$, and
a subgraph $M$ (called a ``matching'') with
each $d(v,M)\le f(v)$.%
\footnote{Throughout this appendix $d$ denotes the degree function,
not to be confused with the $d$ counting dual adjustments
in the analysis of the scaling algorithm.
Note $d(v)=|\delta(v)|+2|\gamma(v)|$.}
A blossom $B$ is a subgraph of $G$.
$B$ has a {\em base vertex} $\beta(B)\in V(B)$, denoted $\beta$
if the blossom is clear.
The subgraph $B$ contains various matched edges but there may also be
matched edges in $(\gamma(B) \cup \delta(B))-E(B)$.
Every vertex $v\in V(B)-\beta$ is perfectly matched, i.e.,
$d(v,M)=f(v)$.
The base vertex has
$d(\beta,M)\ge f(\beta)-1$.

If $d(\beta,M)=f(v)$ then $B$ has an associated
{\em base edge} $\eta(B)=\beta \beta'$, $\beta'\notin V(B)$.
$B$ is a {\em light} blossom if $\eta(B)\in M$ and  a {\em heavy}
blossom if
$\eta(B)\notin M$. (The light/heavy designations come from the
$e_B$ edges in the definition below.)

If $d(\beta,M)=f(\beta)-1$ the blossom is {\em free}.
A free blossom has $\eta(B)=\emptyset$ in the formal definition.
But for the purpose of defining blossoms it is convenient
to add an artificial edge $\beta \beta'$ incident to $B$,
call it matched, and set $\eta(B)=\beta\beta'$.
So a free blossom is light.

The following inductive definition  of a blossom 
is taken from \cite{G18}, with minor changes that incorporate the above notation.
Fig. \ref{InOutIFig} partially illustrates  the definition.

\begin{definition}
\label{ABlossomDefn}
Let $\o G$ be a graph derived from $G$ by contracting
a family \A. of zero or more vertex-disjoint blossoms.
A vertex  of $\o G$ but not \A. is an {\em atom}.
Let $C$ be a
closed path in $\o G$ that starts and ends at
a vertex $\alpha$ of $\o G$.
  The preimage of $C$ in $G$ is
a {\em blossom} $B$
with {\em base vertex}
$\beta$ if $C$ has the following properties:

\bigskip

If $\alpha$ is an atom then $\alpha=\beta(B)$.
$C$ starts and ends with two edges $e_B$ incident to $\beta$
that are both matched in a heavy blossom, both unmatched in a light blossom.

If $\alpha\in \A.$ then $\beta(B)=\beta(\alpha)$.
The $e_B$ edges are also those of $\alpha$.

\bigskip

If $v$ is an atom of $C-\beta$ then the
two edges of $\delta_C(v) $ alternate.

If $v\in \A.\cap C-\alpha$ then 
$\eta(v)\in \delta_C(\beta(v))$.
\end{definition}

The closed path $C$ is denoted $C(B)$ if the blossom needs to
be specified. The 
{\em subgraph of $B$}, denoted $(V(B),E(B))$,
is a subgraph of $G$
consisting of
the subgraphs of all subblossoms \A. of $B$
plus the atoms of $B$ and the
 edges of $C(B)$.
We sometimes treat $B$ as just the vertex set $V(B)$.

A key property of blossoms is that every vertex
$v\in V(B)$ has two alternating trails
from $v$ to $\beta$: $P_0(v,\beta)$ which has even length
and $P_1(v,\beta)$ which has  odd length.
Each $P_i$ is contained in $B$'s subgraph, $E(P_i)\con E(B)$.
The $P_i$'s are recursively defined:
For every blossom $A\in \A.$ containing a vertex of $P_i$,
$E(P_i)\cap E(A)$ is some trail $P_{i'}(v', \beta(A))$, and
$P_i$ contains $\eta(A)$ unless $\beta(A)= \beta$.
($P_{i'}$ may occur in reverse order in $P_i$.)
The definition easily gives this generalization:

{\hi\hi
For every blossom $A\not\ni \beta$, $P_i(v,\beta)$
contains $\eta(A)$
if it contains any vertex of $A$.}

\noindent
One of the trails $P_0, P_1$ starts
with a matched edge, the other starts with an unmatched edge.
Both trails end with an $e_B$ edge
with one exception,
$P_0(\beta,\beta)$ has length 0.
$P_1(\beta,\beta)$ is nonsimple, it contains both $e_B$ edges.
So any $P_i(v,\beta)$ extended with the
edge $\eta(B)$ is still an alternating trail.
(In ordinary matching, blossoms are always light.
The  $P_0$ are required to be paths, and $P_1$ paths need not exist.)

A blossom $B$ has an associated set 
\[I(B)=\delta_M(B)\oplus \eta(B).\]
(A free $B$ has $I(B)=\delta_M(B)\oplus \emptyset=\delta_M(B)$.)
These sets play the role of $I$ in the above LP, i.e.,
instead of writing $z(B,I)$ we denote $z$ duals as $z(B)$ and define
\[\H{yz}(e) = y(e)  + z\set {B} {e \in \gamma(B)\cup I(B)}.\]
Although Fig. \ref{InOutIFig} draws $I(B)$ edges as directed,
an edge $xy$ may belong to both $I(B_x)$ and $I(B_y)$.

The $f$-factor algorithm constructs a search forest analogous to matching.
The roots of the forest are the free blossoms and free atoms
(vertices with $d(v,M)<f(v)$).
For a nonroot node $v$ of the forest, $\tau(v)$ denotes
the first edge of the path from $v$ to its root.
As in matching we say an edge $xy\notin \os.$ {\em alternates} with  node
$x\in \os.$ if $xy$ can be added to the forest, i.e., there is an alternating
path of edges of $G$ that starts with $xy$. If $x$ is in a contracted
blossom the path includes
$P_i(x,\beta)$,  where $i$ is chosen so $P_i$ starts with an edge alternating with $xy$. The classification of nodes as {\em inner} and {\em outer}
is illustrated in Fig. \ref{InOutIFig}. (In ordinary matching
possibilities (c) and (f) do not exist and there are no $I(B)$ edges.)

\begin{figure}[th]
\centering
\input{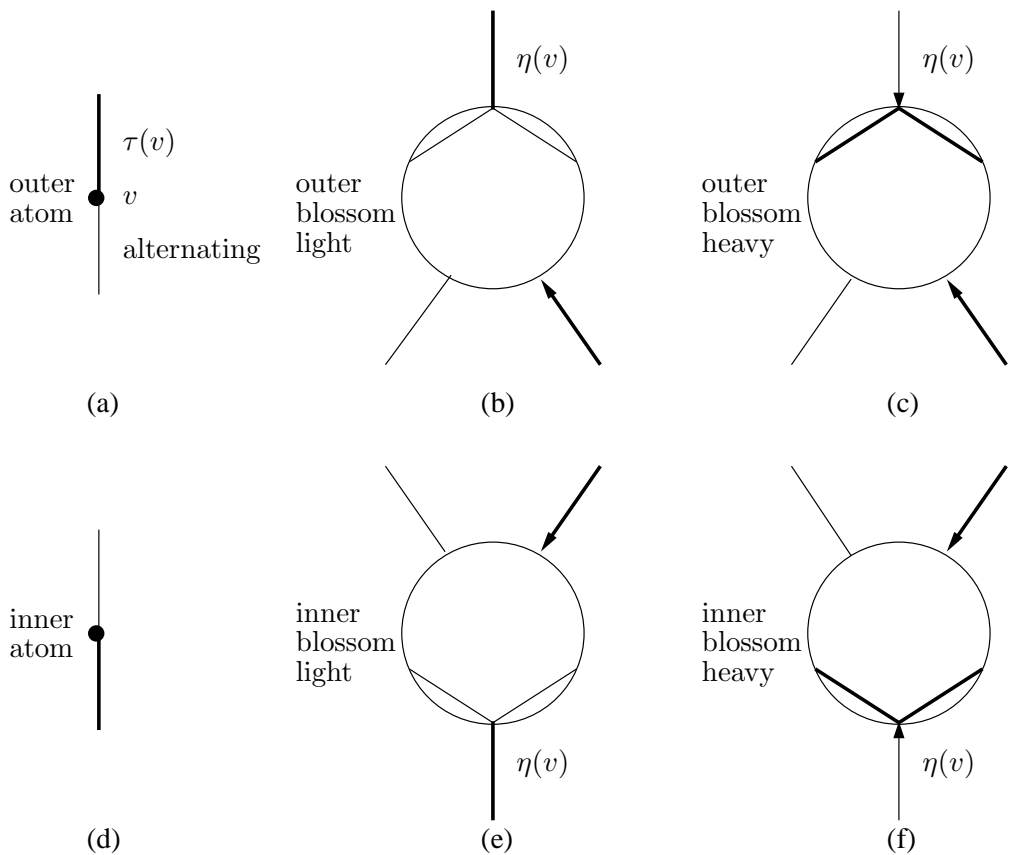}
\caption{The six types of nodes $v$ in a search forest.
  (a)--(c) are outer nodes and (d)--(f) are inner nodes.
  Edges drawn heavy are matched, light edges are unmatched.
  As labelled in (a),
  $\tau(v)$ edges are always drawn above $v$ and
edges alternating with $v$ are always drawn below $v$.
(They need not be in the search forest.)
$\tau(v)$ does not exist for
a search forest root, which has the form of either (a) or (b). 
For blossoms, the two $e_B$ edges
  are always shown. Edges in $I(B)$ sets are drawn directed to blossom $B$.
    In (e) and (f) $\tau(v)$ can be matched or unmatched.
  }
 \label{InOutIFig}
 \end{figure}

We turn to the $f$-factor algorithm.
As in matching the search forest is rooted
at every free atom and free blossom.
Eligibility is defined exactly as in matching.
Pseudocode for a search is the same as matching Fig. \ref{EdSrchFig} except for
new versions of the blossom step, the Expand routine, and
minor changes to AdjustDuals, all given in Fig. \ref{fFactorSrchFig}.

\begin{figure}
\begin{center}
  \fbox{     
\begin{minipage}{\Efigwidth}
\setlength{\parindent}{.2in}

\narrower{
\setlength{\parindent}{0pt}
\vspace{\vmargin}
\setlength{\parindent}{20pt}

\def\dotter{.\hskip 20pt}
\def\bkup{\vspace{-7pt}}

\long\def\vellipsis{\begin{center}.\\
\bkup.\\
\bkup.
\end{center}
\vspace{-10pt}}


\vellipsis
\Myelseif
{$\exists$  eligible edge $e=xy$, $x,y\in \S.$, alternating with both $B_x$ 
and $B_y$}

{\hi

$\alpha\gets$ the nca of $B_x$ and $B_y$ in \os., if it exists

{\bf unless}

{\hi

$/*$ an augmenting path has been found  in the next two possibilities: $*/$

$\alpha$ does not exist $/*$ $B_x$ and $B_y$ are in different search 
trees $*/$

{\bf or}  
$\alpha$ is atomic and $d(\alpha,M)\le f(\alpha)-2$
$/*$ $\alpha$ is a search tree root $*/$

{\hi

$/*$ {blossom step} $*/$

$C\gets$ the fundamental cycle of $e$

contract  $C$ to an outer blossom with $\eta(C)=\tau(\alpha)$  

$/*$ now $B_x=B_y=C$ $*/$
}
}
}
\vspace{-20pt}
\vellipsis
\vspace{-10pt}
\noindent {\bf Algorithm} Expand$(B)$:

let $e=\tau(B)$, $f=\eta(B)$, $v=e\cap V(B)$,  $\beta(B)=f\cap V(B)$

let $C_e$ be the subtrail of $C(B)$ traversed by the
alternating trail $P_i(v,\beta(B))$,

{\hi  for $i\in \{0,1\}$   chosen so  $P_i(v,\beta(B))$ 
alternates with $e$ at $v$
}

{\bf if} {$C_e=C(B)$} {\bf then}
{make $B$ an outer blossom by assigning $\eta(B)\gets e$}

{\bf else} {replace $B$ by $C_e$}
$/*$ atoms and subblossoms of $C(B)-C_e$ leave \S. $*/$

\smallskip

\noindent
{\bf Algorithm} AdjustDuals:

$\delta_1\gets\min \set{|\H{yz}(e)-t(e)|}{e=xy \mbox{ with $x\in \S., y\notin \S.$,
$xy$ alternates with $B_x$}}$

$\delta_2=\min \set{|\H{yz}(e)-t(e)|/2}{e=xy \mbox{ with $x,y\in \S.$,
$xy$ alternates with both $B_x$ and $B_y$}}$

\vspace{-20pt}
\vellipsis

}

\vspace{\vmargin}

\end{minipage}
}

\end{center}
\caption{Search of the $f$-factor algorithm:
The pseudocode that differs from Edmonds' search.}
\label{fFactorSrchFig}

\end{figure}

Fig.\ref{ExceptionalSearchFig} illustrates
a simple execution of the algorithm.
It shows that the ill-formed blossoms
treated in the compression step of our algorithm
(Fig.\ref{fFactorAlgFig})
can actually be
formed.

\begin{figure}[th]
\centering
\input{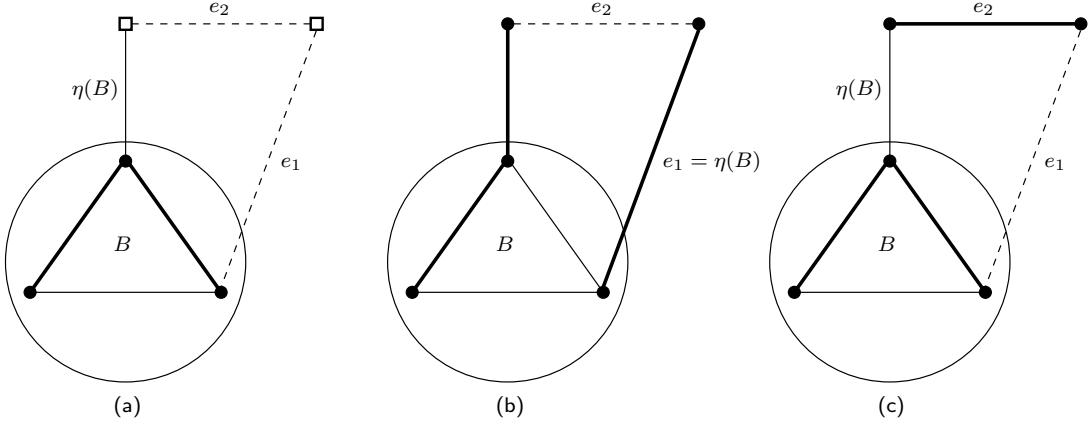}
\caption{Search steps leading to an ill-formed blossom. 
At the start of the search the solid edges are eligible. The ineligible
edges $e_i$ have $\H{yz}(e_i)=y(e_i)=w(e_i)+\Delta(e_i)$.
(a) The search begins by forming outer blossom $B$ with unmatched
base edge $\eta(B)$. (b) If $0<\Delta(e_1)<\Delta(e_2)$,
duals are adjusted by $1+\Delta(e_1)/2$. An augmenting path through $B$
is found and the matching is augmented.
(c) If $0<\Delta(e_2)<\Delta(e_1)$,
duals are adjusted by $1+\Delta(e_2)/2$. An augmenting path
of one edge 
is found and the matching is augmented.}
 \label{ExceptionalSearchFig}
 \end{figure}

We conclude with a property  closely related to
the tightening step of our algorithm
(Fig.\ref{fFactorAlgFig}).
Consider a blossom $B_u$ with base edge $\eta(B_u)=uv$.
$\eta(B_u)$ is eligible if $B_u$ is in a larger blossom
with a different base vertex. This need not hold
if $u$ is the base of a maximal blossom, say $B$.
To illustrate assume $B$ is not in the  search tree.
If $uv$ is unmatched and $v$ is an inner vertex,
dual adjustments increase $\H{yz}(uv)$.
If $uv$ is matched and $v$ is an outer atom or $v$ is in
an inner blossom,
dual adjustments decrease $\H{yz}(uv)$.

The following lemma 
identifies another configuration where $uv$ is always eligible. 
We do not use the lemma in the main body of the paper -- it is included for
completeness, as well as illustrating the execution of the algorithm. Also
for completeness we prove the lemma for duals that
are optimum
as well as near optimum.
An {\em $\eta$ pair} is an
edge $uv$ with $uv=\eta(B_u)=\eta(B_v)$.

\begin{lemma}
\label{MutualEtasTightLemma}
At any point in the $f$-factor algorithm, an $\eta$ pair $e$
has $\H{yz}(e)=w(e)+\Delta$ where
\[
\Delta\ 
\begin{cases}
=0 &\text{for optimum duals}\\
\in [-2,0] &\text{for near-optimum duals.}
\end{cases}
\]
\end{lemma}

\begin{proof}
The condition  $uv=\eta(B_u)=\eta(B_v)$ becomes satisfied
in a blossom step that creates one of the blossoms $B_u$, $B_v$,
or in an expand step that changes $B$ from inner to outer,
or in an augment step that makes $uv$ an $\eta$ pair.
In all cases $uv$ is eligible, so $\Delta$ has the claimed value.

Suppose a dual adjustment step changes
$\Delta$.
Clearly $u$ and $v$ must be
in different maximal blossoms, and at least one
of those blossoms is in the search structure.
Let $B$ ($B'$) be the maximal blossom
containing $u$ ($v$) respectively, and wlog assume
$B$ enters  the search structure
before $B'$.
$B$ enters in a grow
step for an edge $xy$ with $y\in B\not\ne x$, $xy\ne uv$.
So $B$ is an inner vertex (Fig.\ref{InOutIFig}(e) or (f)).

We will show the dual adjustment maintains
$\Delta$ as claimed in the lemma.
First
suppose the algorithm is using optimum duals. $uv$ is eligible, so a
grow step for $uv$ adds $B'$ to the search structure as an outer
vertex (Fig.\ref{InOutIFig}(b) or (c)).
The subsequent dual
adjustment changes duals by
\begin{eqnarray*}
y(u)\gets y(u)+\delta, & &z(B)\gets z(B)-\delta\\
y(v)\gets y(v)-\delta, & &z(B')\gets z(B')+\delta.
\end{eqnarray*}
The changes in $y$ cancel each other.
The changes in $z$ do not 
change $\H{yz}(uv)$ when
$uv$ is matched (since $uv\notin
I(B)\cup I(B')$ and $z$ is irrelevant)
or unmatched (since $uv\in
I(B)\cap I(B')$ and the changes in $z$ cancel).

Suppose the algorithm is using near optimum duals.
If $uv$ is eligible the analysis for optimal duals applies.
Suppose $uv$ is ineligible.
The dual adjustment changes $y(u)$ and $z(B)$ as above.
If $uv$ is matched then $uv\notin I(B)$ so
$\H{yz}(uv)$ increases. $\Delta$ increases from an initial value
$<0$, so the lemma's condition is preserved.
If $uv$ is unmatched then $uv\in I(B)$ so
$\H{yz}(uv)$ decreases. $\Delta$ decreases from an initial value
$>-2$, so the lemma's condition is preserved.
\end{proof}

\or
\input exit
\fi

\ifcase 1 
\or

\section*{Acknowledgments}
The author thanks Seth Pettie for helpful
conversations regarding reference \cite{DPS}.

\fi
\end{document}